\UseRawInputEncoding
\documentclass{aastex631}

\usepackage{amsmath}
\usepackage{ifthen}

\newboolean{refresponse}
\setboolean{refresponse}{false}
\ifthenelse{\boolean{refresponse}}
{
\newcommand{\addedtext}[1]{{\bf {#1}}}
}
{
\newcommand{\addedtext}[1]{{#1}}
}

\newcommand{\cfa}{\affiliation{Center for Astrophysics \textbar\ Harvard \& Smithsonian \\
60 Garden Street \\
Cambridge, MA 02138, USA}}
\newcommand{\mki}{\affiliation{MIT Kavli Institute for Astrophysics and Space Research \\
77 Massachusetts Avenue \\
Cambridge, MA 02139, USA}}

\newcommand{\Chandra}{{\em Chandra\/}}
\newcommand{\paperI}{{Paper~I}}
\newcommand{\canon}[1]{\ensuremath{\langle #1 \rangle}}
\newcommand{\likelihood}{{\mathcal{L}}}
\newcommand{\rhat}{{\hat{R}}}
\newcommand{\truevalue}{{\tt TRUE}}
\newcommand{\marginalvalue}{{\tt MARGINAL}}
\newcommand{\falsevalue}{{\tt FALSE}}

\newcommand{\ape}{{\tt acis\_process\_events}}
\newcommand{\hpe}{{\tt hrc\_process\_events}}
\newcommand{\mkvtbkg}{{\tt mkvtbkg}}
\newcommand{\sherpa}{{\tt Sherpa}}
\newcommand{\wavdetect}{{\tt wavdetect}}
\newcommand{\obsid}{{ObsID}}

\newcommand{\Mean}{\operatorname{Mean}}
\newcommand{\Var}{\operatorname{Var}}

\DeclareMathOperator{\Gammafunc}{\Gamma}
\DeclareMathOperator{\gammafunc}{\gamma}

\shorttitle{Chandra Source Catalog Release 2}
\shortauthors{Evans et al.}

\begin{document}

\title{THE CHANDRA SOURCE CATALOG RELEASE 2 SERIES}

\correspondingauthor{Ian N. Evans}
\email{ievans@cfa.harvard.edu}

\author[0000-0003-0494-743X]{Ian~N.~Evans}
\cfa

\author[0000-0003-3509-0870]{Janet D. Evans}
\cfa

\author[0000-0002-5069-0324]{J. Rafael Mart\'inez-Galarza}
\cfa

\author[0009-0009-9067-2030]{Joseph B. Miller}
\cfa

\author[0000-0002-6077-0643]{Francis A. Primini}
\cfa

\author[0000-0001-6004-9728]{Mojegan Azadi}
\cfa

\author[0000-0003-4428-7835]{Douglas J. Burke}
\cfa

\author[0000-0002-2115-1137]{Francesca M. Civano}
\altaffiliation{Present address: NASA Goddard Space Flight Center, Mail Code 660, Greenbelt, MD 20771, USA.}
\cfa

\author[0000-0003-3073-0605]{Raffaele D'€™Abrusco}
\cfa

\author[0000-0002-3554-3318]{Giuseppina Fabbiano}
\cfa

\author[0000-0003-0920-482X]{Dale E. Graessle}
\cfa

\author[0009-0009-1096-574X]{John D. Grier}
\cfa

\author[0000-0002-6761-6796]{John C. Houck}
\altaffiliation{Present address: Center for Astrophysics \textbar\ Harvard \& Smithsonian, 60 Garden Street, Cambridge, MA 02138, USA.}
\mki

\author[0009-0007-9475-512X]{Jennifer Lauer}
\cfa

\author[0000-0002-8384-3374]{Michael L. McCollough}
\cfa

\author[0000-0001-6923-1315]{Michael A. Nowak}
\altaffiliation{Present address: Washington University in St. Louis, CB 1105, One Brookings Drive, St. Louis, MO 63130, USA.}
\mki

\author[0009-0008-1560-2318]{David A. Plummer}
\cfa

\author[0000-0003-2377-2356]{Arnold H. Rots}
\cfa

\author[0000-0002-0905-7375]{Aneta Siemiginowska}
\cfa

\author[0009-0002-0861-240X]{Michael S. Tibbetts}
\cfa

\begin{abstract}
The \Chandra\ Source Catalog (CSC) is a virtual X-ray astrophysics facility that enables both detailed individual source studies and statistical studies of large samples of X-ray sources detected in ACIS and HRC-I imaging observations obtained by the {\em Chandra X-ray Observatory\/}.  The catalog provides carefully-curated, high-quality, and uniformly calibrated and analyzed tabulated positional, spatial, photometric, spectral, and temporal source properties, as well as science-ready X-ray data products.  The latter includes multiple types of source- and field-based FITS format products that can be used as a basis for further research, significantly simplifying followup analysis of scientifically meaningful source samples.  We discuss in detail the algorithms used for the CSC Release~2 Series, including CSC 2.0, which includes 317,167 unique X-ray sources on the sky identified in observations released publicly through the end of 2014, and CSC 2.1, which adds \Chandra\ data released through the end of 2021 and expands the catalog to 407,806 sources.  Besides adding more recent observations, the CSC Release~2 Series includes multiple algorithmic enhancements that provide significant improvements over earlier releases.  The compact source sensitivity limit for most observations is $\sim\!5$ photons over most of the field of view, which is $\sim\!2\times$ fainter than Release~1, achieved by co-adding observations and using an optimized source detection approach.  A Bayesian X-ray aperture photometry code produces robust fluxes even in crowded fields and for low count sources.  The current release, CSC 2.1, is tied to the Gaia-CRF3 astrometric reference frame for the best sky positions for catalog sources.
\end{abstract}


\section{Introduction} \label{sec:intro}
Launched in 1999, NASA's {\em Chandra X-ray Observatory\/} \citep{2000SPIE.4012....2W,2002PASP..114....1W} continues to observe the soft X-ray universe in the 0.1--10\thinspace keV energy band. \Chandra\ is a pointed observation mission and obtains roughly 800 imaging observations per year using the Advanced CCD Imaging Spectrometer \citep[ACIS;][]{1998SPIE.3444..210B,2003SPIE.4851...28G} or High Resolution Camera \citep[HRC;][]{2000SPIE.4012...68M} instruments.  \Chandra\ provides the highest angular resolution of any X-ray telescope to date with a sub-arcsecond on-axis point spread function (PSF), a field of view up to several hundred $\hbox{\rm arcmin}^2$ (depending on the instrument), and low instrumental background.  These combined capabilities yield a high detectable-source density with low confusion and good astrometry, typically resulting in many serendipitous sources per field.

The aim of the \Chandra\ Source Catalog (CSC) is to facilitate a wide range of X-ray and multi-wavelength studies by providing homogeneously analyzed properties and science ready data products for X-ray sources detected in uniformly calibrated \Chandra\ imaging observations.  X-ray sources from many different astrophysical environments, both Galactic and extragalactic, are represented in the catalog, including hot stars, X-ray binary systems, accreting black holes, supernova remnants, shocked and ionized nebulae, star forming and active galaxies, clusters of galaxies, and the intra-cluster medium.  A wide variety of uniformly calibrated properties measured across a broad range of source fluxes make the CSC a highly valuable resource for many diverse scientific investigations, ranging from single source studies to statistical analyses of large samples of objects in each of these classes, while removing the need for detailed data reductions for each \Chandra\ observation.

The first release of the CSC \citep[][hereafter Paper~I]{2010ApJS..189...37E} provided properties for $\sim\!95,000$ X-ray sources identified in roughly 4,000 imaging observations obtained by \Chandra\ prior to the end of 2008\null.  The {\em CSC Release~2 Series \/} currently includes the CSC 2.0 release published in October 2019 and the more recent CSC 2.1 release published in April 2024\null. CSC 2.0 is a full release, meaning that all of the observation were processed through all of the catalog production pipelines using the latest algorithms at the time the data were processed.  CSC 2.1 is an incremental release that includes, but does not in general reprocess, data from regions of the sky that were included in CSC 2.0 and for which there are no new added observations.  There are some exceptions to this rule, as the CSC 2.1 production pipelines use some new or updated algorithms and correct a few known issues that are identified as caveats to CSC 2.0\null.  CSC Release~2 Series  data are available from the \Chandra\ X-ray Center:\dataset[doi:10.25574/csc2]{\doi{10.25574/csc2}}. Throughout this paper we use terms such as ``CSC~2'' or ``release~2'' when discussing topics that apply to the generic CSC Release~2 Series of releases. Topics that apply to a specific release are described using the release number, for example ``CSC 2.0" or ``release 2.1". 

\begin{figure}
\epsscale{0.9}
\plotone{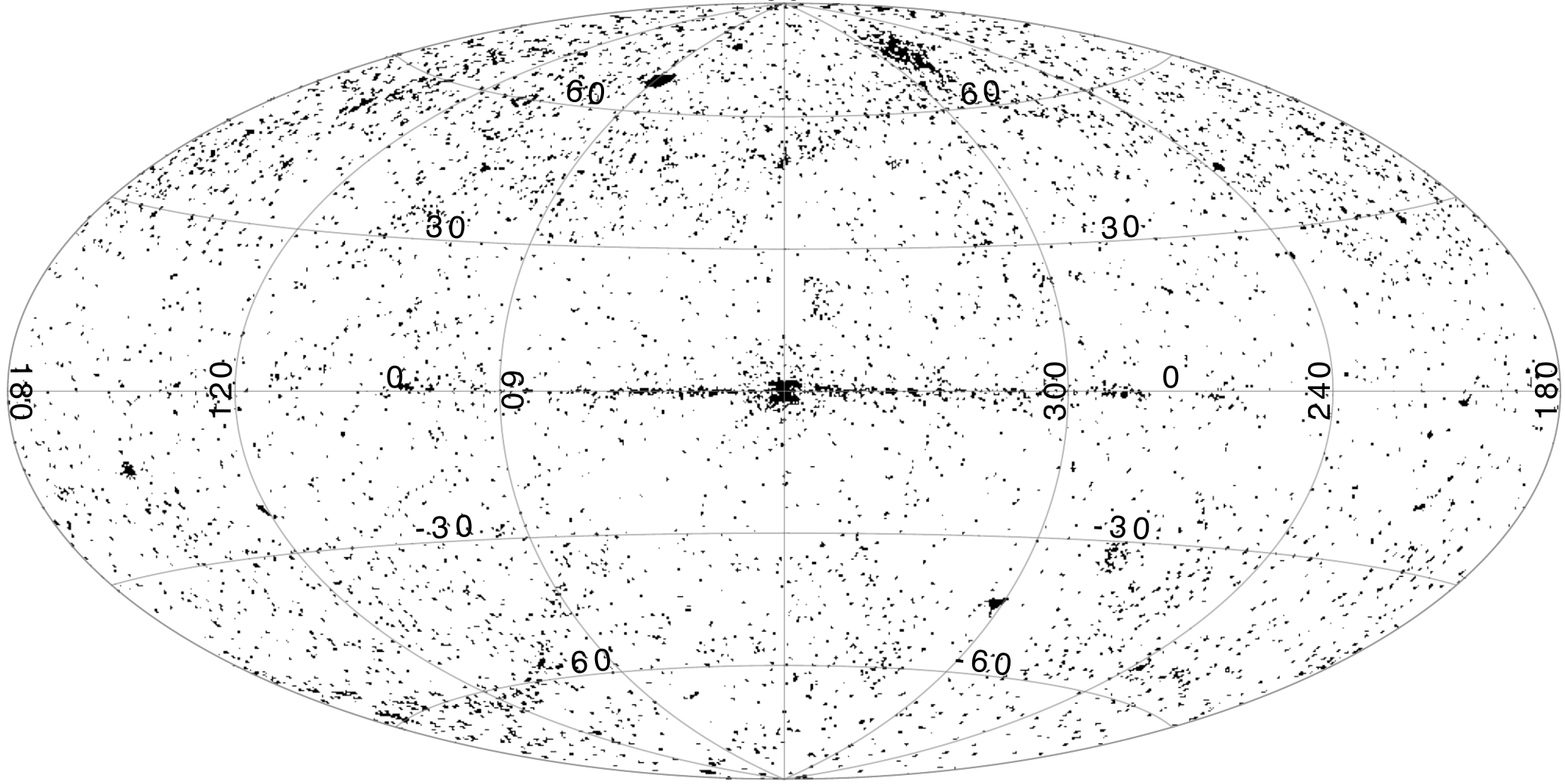}
\caption{\label{fig:skydistribution}
Distribution of CSC release 2.1 master sources on the sky, in Galactic coordinates.}
\end{figure}

Release 2.0 more than tripled the size of the original release 1 catalog to 317,167 distinct X-ray sources on the sky detected in 10,382 ACIS and HRC-I imaging observations released publicly prior to the end of 2014.  Release 2.1 further extends CSC 2.0 by adding observations released publicly through the end of 2021, with the resulting catalog containing 407,806 X-ray sources detected in 15,533 observations. The distribution of CSC 2.1 sources on the sky is shown in Figure~\ref{fig:skydistribution}.

The increased number of sources in the CSC~2 is achieved in part by stacking (co-adding) multiple observations with similar pointings for higher signal-to-noise, as well as an improved background determination technique combined with a new method of source detection that can reliably detect compact sources down to roughly 5 net counts over much of the field-of-view.  The enhancements to the source detection approach allow the inclusion of sky regions containing bright, extended emission that were excluded from release~1 because of the excessive false source rate they generated. Better handling of field and detector edges lowers the false source rate in those areas also.  Evaluation of source and detection properties, including X-ray aperture photometry, spectral fitting, and temporal variability analyses, are significantly improved in release 2, especially for sources detected in multiple observations.  Finally, CSC~2 includes enhanced limiting sensitivity information.  This paper describes in detail the new and updated algorithms used to produce the CSC~2 catalogs. 

\begin{figure}
\epsscale{0.65}
\plotone{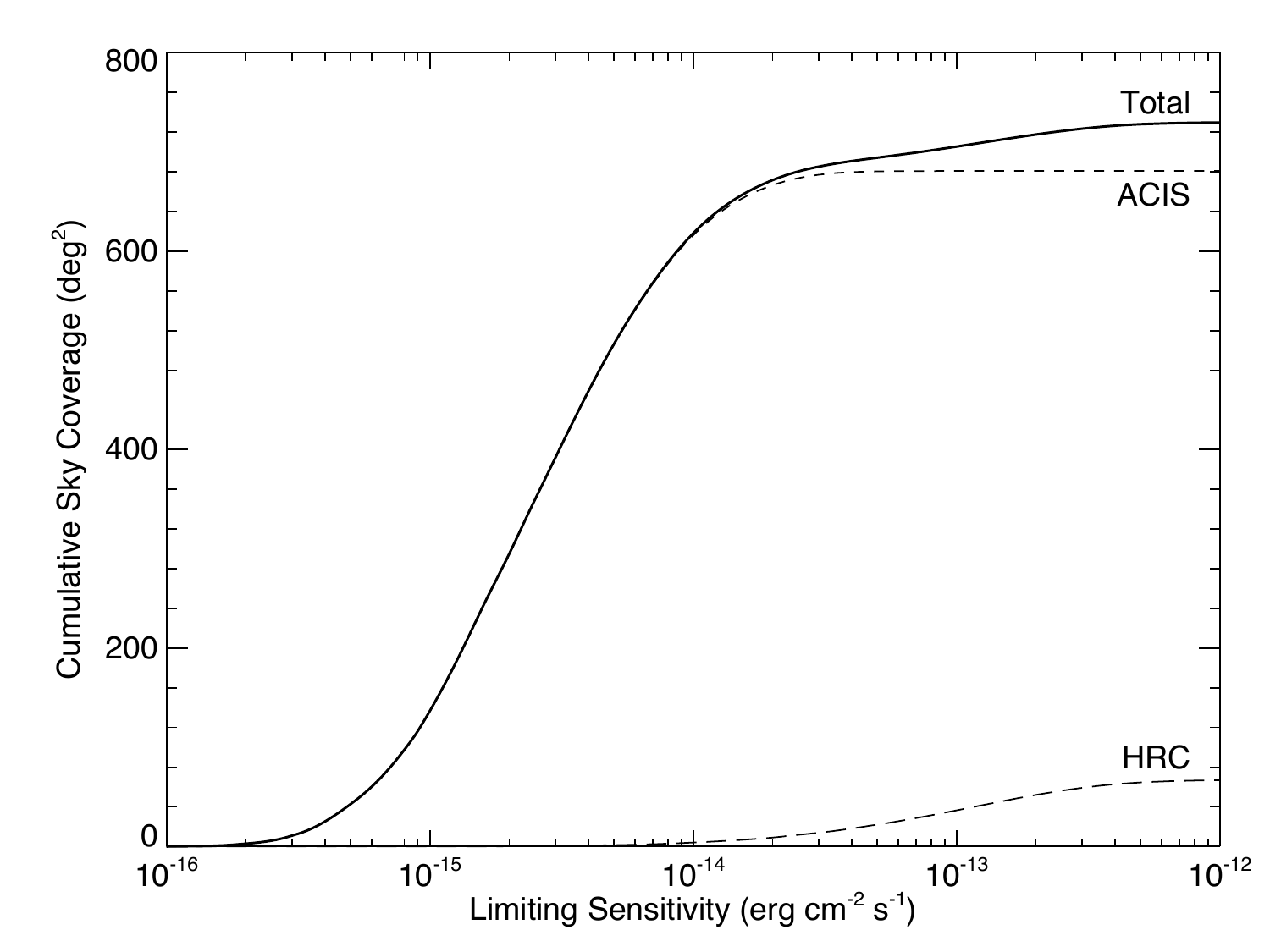}
\caption{\label{fig:skycoverage}
Sky coverage of release 2.1 of the CSC\null. The ordinate value is the total sky area included in the CSC that is sensitive to point sources with fluxes in any energy band at least as large as the corresponding value on the abscissa.}
\end{figure}

Like release~1, the CSC~2 catalogs are constructed from pointed observations obtained using \Chandra, and are neither all-sky nor uniform in depth.  The sky coverage of CSC 2.1 totals $\sim\!730\,\hbox{\rm deg}^2$, with $\sim\!705\,\hbox{\rm deg}^2$ fainter than $1.0\times10^{-13}\,\hbox{\rm erg}\,\hbox{\rm cm}^2\,\hbox{\rm s}^{-1}$ in one or more energy bands, decreasing to  $\sim\!618\,\hbox{\rm deg}^2$ fainter than $1.0\times10^{-14}\,\hbox{\rm erg}\,\hbox{\rm cm}^2\,\hbox{\rm s}^{-1}$ and $\sim\!137\,\hbox{\rm deg}^2$ fainter than $1.0\times10^{-15}\,\hbox{\rm erg}\,\hbox{\rm cm}^2\,\hbox{\rm s}^{-1}$ (see Figure~\ref{fig:skycoverage}).  The sky coverage of observations obtained using the ACIS instrument is $\sim\!681\,\hbox{\rm deg}^2$, and for observations obtained using HRC the coverage is $\sim\!67\,\hbox{\rm deg}^2$, implying $\sim\!18\,\hbox{\rm deg}^2$ sky coverage overlap by both instruments.  These values compare to a total of $\sim\!559\,\hbox{\rm deg}^2$ for CSC 2.0 ($\sim\!517\,\hbox{\rm deg}^2$ for ACIS, $\sim\!56\,\hbox{\rm deg}^2$ for HRC, and a $\sim\!15\,\hbox{\rm deg}^2$ overlap).

\section{CATALOG DESCRIPTION}
The design goals of the CSC and the characteristics of the data recorded by the CXO instruments are described in \S~2 of \paperI, to which the reader is referred for more detail.  For each X-ray source, the catalog tables record many commonly used measured and derived properties, including source position, significance, spatial extent, aperture photometry, hardness ratios, spectral model fit parameters, and inter- and intra-observation temporal variability measures.  The tabulated quantities are computed using robust, state-of-the-art algorithms and will directly address many scientific queries.  Most properties have associated independent lower and upper confidence limits, and are evaluated in 5 energy bands covering the energy range 0.2--$7.0\,\hbox{\rm keV}$ for ACIS and a single energy band covering $\sim\!0.1$--$10.0\,\hbox{\rm keV}$ for HRC-I\null.  The names, designations, energy ranges, and effective monochromatic energies for the  energy bands, are unchanged from release~1, and are presented in Table~\ref{tab:ebands} for convenience.
\begin{deluxetable}{llccc}
\tabletypesize{\small}
\tablecolumns{4}
\tablecaption{CSC Energy Bands\label{tab:ebands}}
\tablehead{
\colhead{Instrument} &  \colhead{Name}	& \colhead{Designation}	& \colhead{Range (keV)}	& \colhead{Mono Energy (keV)} }
\startdata
ACIS & Ultrasoft	& {\it u\/}	& 0.2--0.5	& 0.4\phn  \\
ACIS & Soft		& {\it s\/}	& 0.5--1.2	& 0.92  \\
ACIS & Medium	& {\it m\/}	& 1.2--2.0	& 1.56 \\
ACIS & Hard		& {\it h\/}	& 2.0--7.0	& 3.8\phn  \\
ACIS & Broad		& {\it b\/}	& 0.5--7.0	& 2.3\phn \\
HRC & Wide		& {\it w\/}	& 0.1--10	& 1.5\phn \\
\enddata
\end{deluxetable}

The catalog also includes 37 different types of uniformly processed and calibrated FITS file-based data products that can immediately be used as a starting point for further detailed analyses.  These data products include not only the observation and individual source region photon event lists and products derived from them, such as images, pulse-invariant spectra (with associated responses), and light curves, but also higher-level products such as Bayesian marginalized probability density functions for aperture photometry fluxes, extended source polygons, and limiting sensitivity maps.

\subsection{Catalog Organization}
Because the spatial extent of the \Chandra\  PSF varies by roughly a factor of 50 between the center and edge of the field of view, the CSC differentiates between {\em detections\/} and {\em sources\/}.  Detections are blobs of photon counts on the detector image that are identifiable by the catalog source detection algorithms, while sources are the interpretation of detections in terms of distinct X-ray sources on the sky.  There is potentially a many-to-many relationship between detections and sources.  A single source may correspond to multiple detections because the source is detected in multiple observations obtained at different epochs that include the same region of the sky.  Conversely, a single detection may correspond to multiple sources.  This typically occurs when a single far off-axis detection (with a large PSF) in one observation is resolved into multiple sources by another observation where the same location on the sky is positioned closer to the telescope optical-axis (with a small PSF).

To improve the detection limiting sensitivity of CSC~2, observations from the same instrument (ACIS or HRC-I) with telescope optical axis pointing directions co-aligned within $60''$ are co-added as described in \S~\ref{sec:stackgen} to form an ``observation stack'' prior to source detection.  The limitation on the maximum misalignment of the individual observation pointing directions ensures that their PSFs are comparably sized at any location within the field of view.  This permits source detection to be performed on the observation stack and allows detection properties to be extracted from both the observation stack and the individual observations that comprise the stack.  The former typically yields higher S/N for faint detections that might otherwise be undetectable in a single observation, while for brighter sources the latter enables studies of properties that vary temporally between the individual observation epochs. Table~\ref{tab:releases} summarizes the source and detection content of the CSC 2.0 and CSC 2.1 releases.
\begin{deluxetable}{p{185pt}cc}
\tabletypesize{\small}
\tablecolumns{3}
\tablecaption{CSC Release 2 Series Summary\label{tab:releases}}
\tablehead{
\colhead{} &  \colhead{Release 2.0}	& \colhead{Release 2.1} }
\startdata
Number of Master Sources	& \phn317167		& \phn407806 \\
Number of Stacked Observation Detections	& \phn376343		& \phn493236 \\
\hbox{Number of Stacked Observation Detections} {\ \ Including}~Photometric Upper Limits	& \phn620555		& \phn855402\\
Number of Individual Observation Detections	& \phn928280		& 1304376 \\
\hbox{Number of Individual Observation Detections} {\ \ Including}~Photometric Upper Limits	& 1420545	& 2143847\\
Total Sky Coverage ($\hbox{\rm deg}^2$)	& \phn\phn\phn\phn559 & \phn\phn\phn\phn730 \\
Sky Coverage (ACIS; $\hbox{\rm deg}^2$)	& \phn\phn\phn\phn517	& \phn\phn\phn\phn681 \\
Sky Coverage (HRC-I; $\hbox{\rm deg}^2$)	& \phn\phn\phn\phn\phn56	& \phn\phn\phn\phn\phn67 \\
\enddata
\end{deluxetable}

The measured and derived source and detection properties included in CSC~2 are organized into three principal and six auxiliary tables listed in Table~\ref{tab:csctab}.
{\newlength{\tablecolwidth}
\setlength{\tablecolwidth}{160pt}
\newlength{\descripcolwidth}
\setlength{\descripcolwidth}{290pt}
\begin{deluxetable*}{p{\tablecolwidth}cp{\descripcolwidth}}
\tabletypesize{\footnotesize}
\tablewidth{0pt}
\tablecolumns{3}
\tablecaption{CSC Tables\label{tab:csctab}}
\tablehead{
\colhead{Table} & \colhead{Type} & \colhead{Description}
}
\startdata
Master Sources & Principal & Extracted properties for each distinct X-ray source on the sky \\
Stacked Observation Detections & Principal & Extracted properties for each detection identified in an observation stack \\
Per-Observation Detections & Principal & Extracted properties for each stacked observation detection from each observation that includes the detection \\
Master Source/Stacked Observation Detection Associations & Auxiliary & Mapping between master sources and stacked observation detections \\
Stacked Observation Detection/Per-Obser\-vation Detection Associations & Auxiliary & Mapping between stacked observation detections and per-observation detections \\
Detect Stack & Auxiliary & Mapping between observation stacks and observations that comprise the stack \\
Valid Stack & Auxiliary & Mapping that identifies which observations are valid ({\em i.e.\/} the detection falls on the valid pixel mask within the field of view) for each stacked observation detection \\
Likely Stack & Auxiliary & Mapping that identifies which observations were used to compute the detection likelihood for each stacked observation detection \\
Limiting Sensitivity & Auxiliary & Table of limiting photon and energy flux sensitivity values for a point source to be included in the catalog as a function of position on the sky (on a $3\farcs 22$ angular resolution HEALPIX pixel grid) \\
\enddata
\end{deluxetable*}}

The primary catalog table is the {\em Master Sources\/} table, which records the best estimate properties for each distinct X-ray source on the sky, determined by combining data from the individual detections and observations that are uniquely associated with the source, as described below.  The contents of the Master Sources table are described in Table~\ref{tab:masterproperties}.  The {\em Stacked Observation Detections\/} table (see Table~\ref{tab:stackobsproperties}) records the best estimate properties for each stacked observation detection, computed by combining data from the complete set of observations that comprise the observation stack in which the detection is identified.  The third principal catalog table is the {\em Per-Observation Detections\/} table (see Table~\ref{tab:perobsproperties}).  For each stacked observation detection, this table records the properties extracted separately from each observation included in the observation stack.
\begin{figure}[t]
\epsscale{0.8}
\plotone{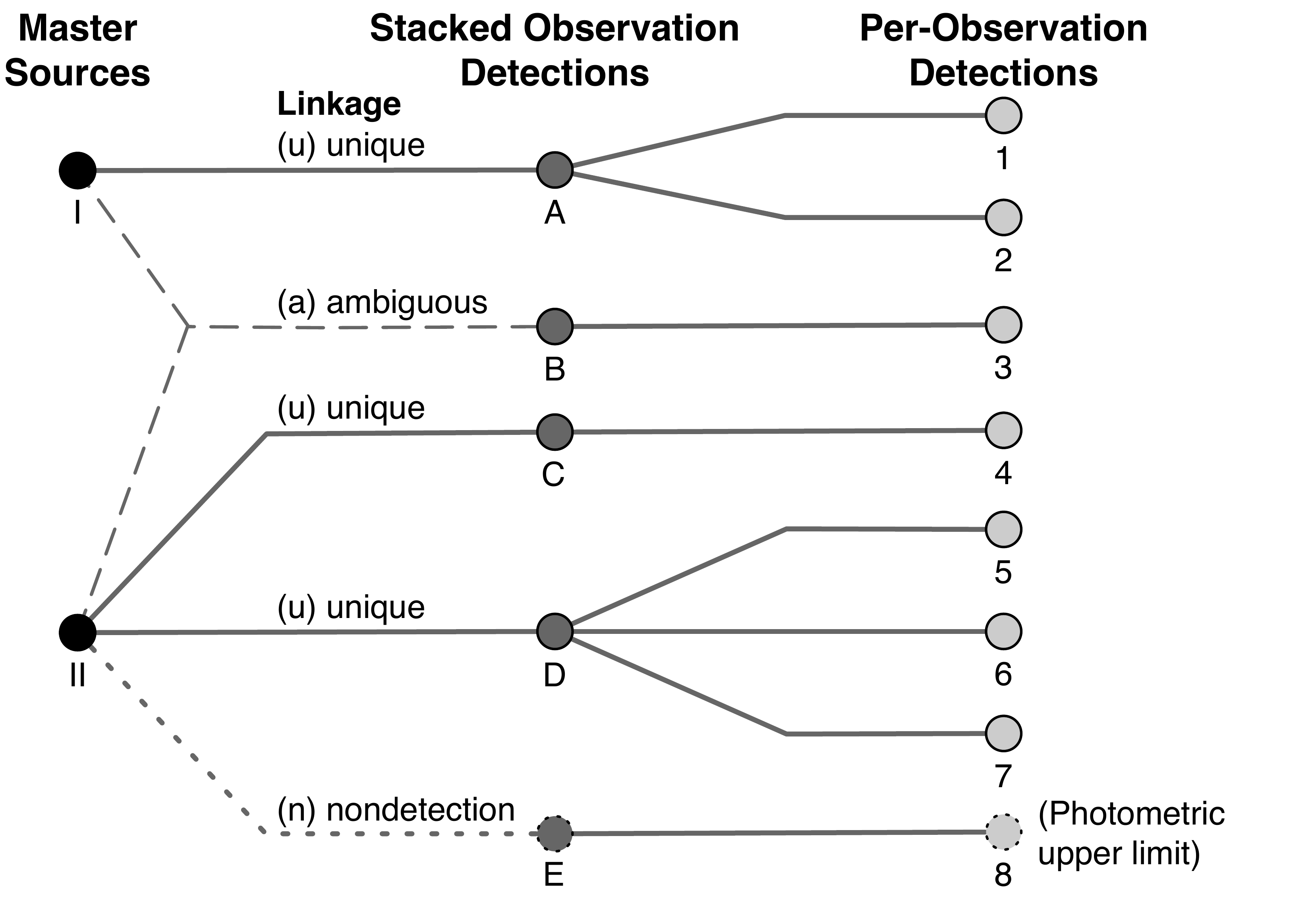}
\caption{\label{fig:linkage}
Linkage between master sources, stacked observation detections, and per-observation detections in CSC 2\null.  \addedtext{Master sources and stacked observation detections are linked uniquely when the stacked observation detection is associated with a single master source, whereas they are linked ambiguously if the stacked observation detection can be associated with more than one master source.  Non-detection linkage implies that no source was detected at the location of a known master source in the associated stacked observation.}  Source detection is performed at the stacked observation detections level.  \addedtext{In this example, master source~I is linked uniquely to stacked observation detection~A and linked ambiguously to stacked observation detection~B\null. Per-observation (individual observation) detections 1 and 2 will contribute to master source~I source properties, whereas per-observation detection~3 will provide a photometric upper limit at the epoch of that observation.  Master source~II is linked uniquely to stacked observation detections C and D, linked ambiguously to stacked observation detection~B, and has a nondetection linkage to stacked observation detection~E\null. Per-observation detections 4, 5, 6, and 7 will contribute to master source~II source properties, and per-observation detection~3 will provide a photometric upper limit at the epoch of that observation. Since no source was detected at the location of stacked observation detection~E, per-observation detection~8 is defined to be a local PSF-sized region that will be used to compute a photometric upper limit at the epoch of that observation.  For more information on linkage, see the details in the description of the Master Source/Stacked Observation Detection Associations table in the text.  The algorithm used to match stacked observation detections to identify master sources is described in \S~\ref{sec:matchdets}.}}
\end{figure}

With the hierarchy of three principal catalog tables described above, each distinct X-ray source is represented by a single record in the Master Sources table and one or more associated detection records in the Stacked Observation Detections table (one for each observation stack in which the source is identified).  Each record in the Stacked Observation Detections table will further be associated with one or more records in the Per-Observation Detections table, corresponding to the individual detections in the observations that comprise the observation stack for which the detection falls within the observation field of view.  The mappings between the table records are shown schematically in Figure~\ref{fig:linkage}.  They are managed transparently by the catalog user interfaces, so that a scientist can access all observation data for a single source seamlessly.

Within all the catalog tables, a {\em null\/} (missing) value should be interpreted as meaning the property is undetermined.  Null values may arise because either (a)~the property is meaningless in the context of the current source or detection, (b)~the property was intentionally not computed, or (c)~the algorithm used to compute the property failed to converge or generated an error when applied to the source or detection data in question.  The null value is distinct from the value zero, which is a properly computed quantity ({\em i.e.\/}, the algorithm used to compute the value executed without error).  Since most numeric quantities have associated independent lower and upper confidence limits, computed {\em upper limits\/} are indicated by the combination of a zero value, zero lower confidence limit, and non-zero upper confidence limit.  If the lower confidence limit is zero but the value is non-zero, this indicates a computed quantity that is consistent with zero but that is not an upper limit.

The {\em Master Source/Stacked Observation Detection Associations\/} table (Table~\ref{tab:masterstackassoc}) is the first of the auxiliary tables and maps Master Sources table records to Stacked Observation Detections table records.  Within this table, the {\tt match\_type} value defines the type of linkage that connects the detection and source.  If a detection in the Stacked Observation Detections table can be related unambiguously to a single X-ray source in the Master Sources table, then the corresponding table entries will be associated by ``unique'' linkage (${\tt match\_type} = \null${\tt `u'}).  A source must always have at least one uniquely linked detection.  Detections that cannot be related uniquely to a single X-ray source will have entries associated by ``ambiguous'' linkage (${\tt match\_type} = \null${\tt `a'}).  A detection with ambiguous linkage must always be associated with at least two sources.  Newly added in CSC~2 is the ``non-detection'' linkage (${\tt match\_type} = \null${\tt `n'}).  Artificial detections are created for each observation stack in which an X-ray source (that is detected in one or more other observation stacks) is not detected (see \S~\ref{sec:nondet})\null, and provide photometric upper limits for the source flux at the epochs of the observations comprising the stack.  Non-detections created in this manner are associated with the X-ray source using non-detection linkage.

Similar to the Master Source/Stacked Observation Detection Associations table, the {\em Stacked Observation Detection/Per-Observation Detection Associations\/} table (Table~\ref{tab:stackobiassoc}) records the mappings between Stacked Observation Detections table records and Per-Observation Detections table records.  

Data from stacked observation detections with ambiguous or non-detection linkage are generally not used when computing the best estimates of source properties included in the Master Sources table (the exception being that non-detections {\em are\/} used when evaluating inter-observation variability properties).  When computing stacked observation detection properties or source properties based on ACIS detections, per-observation detections for which the estimated photon pile-up fraction \citep{Davis_pileup} exceeds $\sim\!10\%$ are not used unless all ACIS per-observation detections exceed this threshold.

{
\newlength{\descripwidth}
\setlength{\descripwidth}{254pt}
\renewcommand\arraystretch{0.9}
\startlongtable
\begin{deluxetable*}{lcccp{\descripwidth}}
\tabletypesize{\scriptsize}
\tablecolumns{5}
\tablecaption{Master Sources Table Properties\label{tab:masterproperties}}
\tablehead{
\noalign{\vskip 2.0ex} \colhead{Property} & \colhead{Multi-\tablenotemark{\scriptsize a}} & \colhead{Conf.\tablenotemark{\scriptsize b}} & \colhead{Units} & \colhead{Description} \\
\noalign{\vskip -1.5ex} & \colhead{Band} & \colhead{Lim.} & &
}
\startdata
\cutinhead{Source Designation}	
name								& No  & No  &							& Source name in the format ``2CXO J${\it hhmmss.s}\pm{\it ddmmss}$[X]'' \\
\cutinhead{Source Position}	
ra									& No  & No  & deg						& Source position, ICRS right ascension \\
dec									& No  & No  & deg						& Source position, ICRS declination \\
gal\_l								& No  & No  & deg 						& Source position, Galactic longitude (equinox J2000.0, epoch J2000.0) \\
gal\_b								& No  & No  & deg						& Source position, Galactic latitude (equinox J2000.0, epoch J2000.0) \\
err\_ellipse\_r0							& No  & No  & arcsec					& Major radius of the 95\% confidence level position error ellipse \\
err\_ellipse\_r1							& No  & No  & arcsec					& Minor radius of the 95\% confidence level position error ellipse \\
err\_ellipse\_ang						& No  & No  & deg						& Position angle\tablenotemark{\scriptsize c} of the major axis of the 95\% confidence level position error ellipse \\
\cutinhead{Source Significance}
significance							& No  & No  &							& Highest source flux significance across all stacked observations and energy bands \\
likelihood								& No  & No  &							& Highest detection log likelihood\tablenotemark{\scriptsize d} across all stacked observations and energy bands \\
likelihood\_class						& No  & No  &							& Highest detection likelihood classification across all stacked observations and energy bands \\
\cutinhead{Source Flags}
conf\_flag								& No  & No  &							& Source may be confused (source regions overlap in one or more contributing stacked observations) \\
dither\_warning\_flag						& No  & No  &							& Highest statistically significant peak in the power spectrum of the source region count rate occurs at the dither frequency or at a beat frequency of the dither frequency in one or more observations \\
extent\_flag							& No  & No  &							& Source is extended, or deconvolved source extent is inconsistent with a point source ($a_{\rm rss}>5\,\Delta a_{\rm rss}$) in one or more observations and energy bands \\
pileup\_flag							& No  & No  &							& ACIS pile-up fraction exceeds $\sim\!10\%$ in all observations; source properties may be affected \\
sat\_src\_flag							& No  & No  &							& Source is saturated in all observations; source properties are unreliable \\
streak\_src\_flag						& No  & No  &							& Source is located on an ACIS readout streak in all observations; source properties may be affected \\
var\_flag								& No  & No  &							& Source displays flux variability within one or more observations (intra-observation variability index $\ge 3$), in one or more energy bands\tablenotemark{\scriptsize e} \\
var\_inter\_hard\_flag					& No  & No  &							& Source hardness ratios are statistically inconsistent between two or more observations \\
man\_add\_flag							& No  & No  &							& Source was manually added to the catalog via human review (passes automated validation criteria)\\
man\_inc\_flag							& No  & No  &							& Source was manually included in the catalog via human review (may not pass automated validation criteria) \\
man\_match\_flag						& No  & No  &							& Source detections were manually matched between overlapping stacked observations via human review \\
man\_pos\_flag                                                   & No  & No  &                                                     &  Best fit source position was manually modified via human review \\
man\_reg\_flag							& No  & No  &							& Source region parameters (dimensions, initial guess position input to the Maximum Likelihood Estimator fit) were manually modified via human review \\
\cutinhead{Source Extent\tablenotemark{\scriptsize f}}	
major\_axis					 		& Yes & Yes & arcsec 					& $1\,\sigma$ radius along the major axis of the ellipse defining the deconvolved source extent \\
minor\_axis					 		& Yes & Yes & arcsec				        & $1\,\sigma$ radius along the minor axis of the ellipse defining the deconvolved source extent \\
pos\_angle					 		& Yes & Yes & deg						& Position angle\tablenotemark{\scriptsize c} of the major axis of the ellipse defining the deconvolved source extent \\
src\_area								& Yes & No  & $\rm arcsec^2$				& Area of the deconvolved source extent ellipse, or area of the source polygon for extended sources \\
\cutinhead{Aperture Photometry}	
photflux\_aper 							& Yes & Yes & $\rm photon\,cm^{-2}\,s^{-1}$	& Aperture-corrected net photon flux inferred from the source region aperture, best estimate derived from the longest block of a multi-band, flux-ordered Bayesian Block analysis of the contributing observations, and calculated by counting X-ray events \\
photflux\_aper\_avg						& Yes & Yes & $\rm photon\,cm^{-2}\,s^{-1}$	& Aperture-corrected net photon flux inferred from the source region aperture, averaged over all contributing observations, and calculated by counting X-ray events \\
flux\_aper 								& Yes & Yes & $\rm erg\,cm^{-2}\,s^{-1}$		& Aperture-corrected net energy flux inferred from the source region aperture, best estimate derived from the longest block of a multi-band, flux-ordered Bayesian Block analysis of the contributing observations, and calculated by counting X-ray events \\
flux\_aper\_avg							& Yes & Yes & $\rm erg\,cm^{-2}\,s^{-1}$		& Aperture-corrected net energy flux inferred from the source region aperture, averaged over all contributing observations, and calculated by counting X-ray events \\
photflux\_aper90 						& Yes & Yes & $\rm photon\,cm^{-2}\,s^{-1}$	& Aperture-corrected net photon flux inferred from the Point Spread Function 90\% Enclosed Counts Fraction aperture, calculated by counting X-ray events \\
photflux\_aper90\_avg					& Yes & Yes & $\rm photon\,cm^{-2}\,s^{-1}$	& Aperture-corrected net photon flux inferred from the PSF 90\% ECF aperture, averaged over all contributing observations, and calculated by counting X-ray events \\
flux\_aper90 							& Yes & Yes & $\rm erg\,cm^{-2}\,s^{-1}$		& Aperture-corrected net energy flux inferred from the PSF 90\% ECF aperture, calculated by counting X-ray events \\
flux\_aper90\_avg						& Yes & Yes & $\rm erg\,cm^{-2}\,s^{-1}$		& Aperture-corrected net energy flux inferred from the PSF 90\% ECF region aperture, averaged over all contributing observations, and calculated by counting X-ray events \\
phot\_nsrcs							& No  & No  & 	   						& Number of sources fit simultaneously to compute aperture photometry quantities \\
flux\_powlaw\_aper 						& Yes & Yes & $\rm erg\,cm^{-2}\,s^{-1}$		& Source region aperture model energy flux inferred from the canonical absorbed power law model ($N_{\rm H} = N_{\rm H(Gal)}$; $\Gamma = 2.0$) \\
flux\_bb\_aper 							& Yes & Yes & $\rm erg\,cm^{-2}\,s^{-1}$  		& Source region aperture model energy flux inferred from the canonical absorbed black body model ($N_{\rm H} = N_{\rm H(Gal)}$; $kT = 0.75\,\rm keV)$ \\
flux\_brems\_aper						& Yes & Yes & $\rm erg\,cm^{-2}\,s^{-1}$	  	& Source region aperture model energy flux inferred from the canonical absorbed bremsstrahlung model ($N_{\rm H} = N_{\rm H(Gal)}$; $kT = 3.5\,\rm keV)$ \\
flux\_apec\_aper						& Yes & Yes & $\rm erg\,cm^{-2}\,s^{-1}$		& Source region aperture model energy flux inferred from the canonical absorbed APEC model ($N_{\rm H} = N_{\rm H(Gal)}$; $kT = 6.5\,\rm keV)$ \\
flux\_powlaw\_aper90 					& Yes & Yes & $\rm erg\,cm^{-2}\,s^{-1}$		& PSF 90\% ECF region aperture model energy flux inferred from the canonical absorbed power law model ($N_{\rm H} = N_{\rm H(Gal)}$; $\Gamma = 2.0$) \\
flux\_bb\_aper90						& Yes & Yes & $\rm erg\,cm^{-2}\,s^{-1}$  		& PSF 90\% ECF region aperture model energy flux inferred from the canonical absorbed black body model ($N_{\rm H} = N_{\rm H(Gal)}$; $kT = 0.75\,\rm keV)$ \\
flux\_brems\_aper90						& Yes & Yes & $\rm erg\,cm^{-2}\,s^{-1}$	  	& PSF 90\% ECF region aperture model energy flux inferred from the canonical absorbed bremsstrahlung model ($N_{\rm H} = N_{\rm H(Gal)}$; $kT = 3.5\,\rm keV)$ \\
flux\_apec\_aper90						& Yes & Yes & $\rm erg\,cm^{-2}\,s^{-1}$		& PSF 90\% ECF region aperture model energy flux inferred from the canonical absorbed APEC model ($N_{\rm H} = N_{\rm H(Gal)}$; $kT = 6.5\,\rm keV)$ \\
nh\_gal 								& No  & No  & $\rm 10^{20}\,cm^{-2}$ 		& Galactic neutral Hydrogen column density, $N_{\rm H}({\rm Gal})$, in the direction of the source determined from \citet{1990ARAA..28..215D} \\
\cutinhead{Spectral Hardness Ratios}
hard\_$\canon{x}\canon{y}$				& No  & Yes &							& Photon flux hardness ratio measured between ACIS energy bands $\canon{x}$ and $\canon{y}$, $\rm hard\_\canon{\mathit{x}}\canon{\mathit{y}} = (photflux\_aper\_\canon{\mathit{x}} - photflux\_aper\_\canon{\mathit{y}}) / (photflux\_aper\_\canon{\mathit{x}} + photflux\_aper\_\canon{\mathit{y}})$\tablenotemark{\scriptsize g} \\
var\_inter\_hard\_prob\_$\canon{x}\canon{y}$	& No  & Yes &							& Inter-observation photon flux hardness ratio variability probability \\
var\_inter\_hard\_sigma\_$\canon{x}\canon{y}$	& No  & Yes &							& Inter-observation photon flux hardness ratio variability standard deviation \\
\cutinhead{Model Spectral Fits\tablenotemark{\scriptsize h}}
flux\_powlaw 							& No  & Yes & $\rm erg\,cm^{-2}\,s^{-1}$ 		& Net integrated $0.5$--$7.0\rm\,keV$ energy flux of the best fitting absorbed power-law model spectrum to the source region aperture PI spectrum \\
powlaw\_gamma						& No  & Yes &							& Photon index ($\Gamma$, defined as $F_{\rm E}\propto\!E^{-\Gamma}$) of the best fitting absorbed power-law model spectrum to the source region aperture PI spectrum \\
powlaw\_gamma\_rhat					& No  & No  &							& Photon index convergence criterion, $\rhat$, of the best fitting absorbed power-law model spectrum to the source region aperture PI spectrum\tablenotemark{\scriptsize i} \\
powlaw\_nh 							& No  & Yes & $\rm 10^{20}\,cm^{-2}$ 		& Total neutral Hydrogen column density, $N_{\rm H}$, of the best fitting  absorbed power-law model spectrum to the source region aperture PI spectrum \\
powlaw\_nh\_rhat						& No  & No  & 			   				& Total neutral Hydrogen column density convergence criterion, $\rhat$, of the best fitting absorbed power-law model spectrum to the source region aperture PI spectrum\tablenotemark{\scriptsize i} \\
powlaw\_ampl							& No  & Yes &							& Amplitude of the best fitting absorbed power-law model spectrum to the source region aperture PI spectrum \\
powlaw\_ampl\_rhat						& No  & No  &							& Amplitude convergence criterion, $\rhat$, of the best fitting absorbed power-law model spectrum to the source region aperture PI spectrum\tablenotemark{\scriptsize i} \\
powlaw\_stat							& No  & No  &							& $\chi^2$ statistic per degree of freedom of the best fitting absorbed power-law model spectrum to the source region aperture PI spectrum \\
flux\_bb 								& No  & Yes & $\rm erg\,cm^{-2}\,s^{-1}$	 	& Net integrated $0.5$--$7.0\rm\,keV$ energy flux of the best fitting absorbed black body model spectrum to the source region aperture PI spectrum \\
bb\_kt								& No  & Yes & keV						& Temperature ($kT$) of the best fitting absorbed black body model spectrum to the source region aperture PI spectrum \\
bb\_kt\_rhat							& No  & No  &							& Temperature convergence criterion, $\rhat$, of the best fitting absorbed black body model spectrum to the source region aperture PI spectrum\tablenotemark{\scriptsize i} \\
bb\_nh 								& No  & Yes & $\rm 10^{20}\,cm^{-2}$ 		& Total neutral Hydrogen column density, $N_{\rm H}$, of the best fitting absorbed black body model spectrum to the source region aperture PI spectrum \\
bb\_nh\_rhat							& No  & No  & 			   				& Total neutral Hydrogen column density convergence criterion, $\rhat$, of the best fitting absorbed black body model spectrum to the source region aperture PI spectrum\tablenotemark{\scriptsize i} \\
bb\_ampl								& No  & Yes &							& Amplitude of the best fitting absorbed black body model spectrum to the source region aperture PI spectrum \\
bb\_ampl\_rhat							& No  & No  &							& Amplitude convergence criterion, $\rhat$, of the best fitting absorbed black body model spectrum to the source region aperture PI spectrum\tablenotemark{\scriptsize i} \\
bb\_stat								& No  & No  &							& $\chi^2$ statistic per degree of freedom of the best fitting absorbed black body model spectrum to the source region aperture PI spectrum \\
flux\_brems 							& No  & Yes & $\rm erg\,cm^{-2}\,s^{-1}$ 		& Net integrated $0.5$--$7.0\rm\,keV$ energy flux of the best fitting absorbed bremsstrahlung model spectrum to the source region aperture PI spectrum \\
brems\_kt								& No  & Yes & keV						& Temperature ($kT$) of the best fitting absorbed bremsstrahlung model spectrum to the source region aperture PI spectrum \\
brems\_kt\_rhat							& No  & No  &							& Temperature convergence criterion, $\rhat$, of the best fitting absorbed bremsstrahlung model spectrum to the source region aperture PI spectrum\tablenotemark{\scriptsize i} \\
brems\_nh 							& No  & Yes & $\rm 10^{20}\,cm^{-2}$ 		& Total neutral Hydrogen column density, $N_{\rm H}$, of the best fitting absorbed bremsstrahlung model spectrum to the source region aperture PI spectrum \\
brems\_nh\_rhat						& No  & No  & 			   				& Total neutral Hydrogen column density convergence criterion, $\rhat$, of the best fitting absorbed bremsstrahlung model spectrum to the source region aperture PI spectrum\tablenotemark{\scriptsize i} \\
brems\_norm							& No  & Yes &							& Normalization of the best fitting absorbed bremsstrahlung model spectrum to the source region aperture PI spectrum \\
brems\_norm\_rhat						& No  & No  &							& Normalization convergence criterion, $\rhat$, of the best fitting absorbed bremsstrahlung model spectrum to the source region aperture PI spectrum\tablenotemark{\scriptsize i} \\
brems\_stat							& No  & No  &							& $\chi^2$ statistic per degree of freedom of the best fitting absorbed bremsstrahlung model spectrum to the source region aperture PI spectrum \\
flux\_apec 							& No  & Yes & $\rm erg\,cm^{-2}\,s^{-1}$ 		& Net integrated $0.5$--$7.0\rm\,keV$ energy flux of the best fitting absorbed APEC model spectrum to the source region aperture PI spectrum \\
apec\_kt								& No  & Yes & keV						& Temperature ($kT$) of the best fitting absorbed APEC model spectrum to the source region aperture PI spectrum \\
apec\_kt\_rhat							& No  & No  &							& Temperature convergence criterion, $\rhat$, of the best fitting absorbed APEC model spectrum to the source region aperture PI spectrum\tablenotemark{\scriptsize i} \\
apec\_abund							& No  & Yes &							& Abundance of the best fitting absorbed APEC model spectrum to the source region aperture PI spectrum \\
apec\_abund\_rhat						& No  & No  &							& Abundance convergence criterion, $\rhat$, of the best fitting absorbed APEC model spectrum to the source region aperture PI spectrum\tablenotemark{\scriptsize i} \\
apec\_z								& No  & Yes &							& Redshift of the best fitting  absorbed APEC model spectrum to the source region aperture PI spectrum \\
apec\_z\_rhat							& No  & No  &							& Redshift convergence criterion, $\rhat$, of the best fitting absorbed APEC model spectrum to the source region aperture PI spectrum\tablenotemark{\scriptsize i} \\
apec\_nh 								& No  & Yes & $\rm 10^{20}\,cm^{-2}$ 		& Total neutral Hydrogen column density, $N_{\rm H}$, of the best fitting absorbed APEC model spectrum to the source region aperture PI spectrum \\
apec\_nh\_rhat							& No  & No  & 	  		 				& Total neutral Hydrogen column density convergence criterion, $\rhat$, of the best fitting absorbed APEC model spectrum to the source region aperture PI spectrum\tablenotemark{\scriptsize i} \\
apec\_norm							& No  & Yes &							& Normalization of the best fitting absorbed APEC model spectrum to the source region aperture PI spectrum \\
apec\_norm\_rhat						& No  & No  &							& Normalization convergence criterion, $\rhat$, of the best fitting absorbed APEC model spectrum to the source region aperture PI spectrum\tablenotemark{\scriptsize i} \\
apec\_stat								& No  & No  &							& $\chi^2$ statistic per degree of freedom of the best fitting absorbed APEC model spectrum to the source region aperture PI spectrum \\
\cutinhead{Temporal Variability}
var\_intra\_index 						& Yes & No  &			 	 	  	 	& Intra-observation Gregory-Loredo variability index in the range $[0, 10]$; indicates whether the source region photon flux is constant within an observation (highest value across all observations) \\
var\_intra\_prob 						& Yes & No  &	 			 	 	 	& Intra-observation Gregory-Loredo variability probability (highest value across all observations) \\
ks\_intra\_prob 							& Yes & No  &	 	 	  			 	& Intra-observation Kolmogorov-Smirnov test variability probability (highest value across all observations) \\
kp\_intra\_prob 							& Yes & No  &			 	 	 	 	& Intra-observation Kuiper's test variability probability (highest value across all observations) \\
var\_inter\_index 						& Yes & No  &	 	 			  	 	& Inter-observation variability index in the range $[0, 10]$; indicates whether the source region photon flux is constant between observations \\
var\_inter\_prob 						& Yes & No  &	 						& Inter-observation variability probability, calculated from the $\chi^2$ distribution of the photon fluxes of the individual observations \\
var\_inter\_sigma						& Yes & No  & $\rm photon\,cm^{-2}\,s^{-1}$	& Inter-observation flux variability standard deviation; the spread of the individual observation photon fluxes about the error weighted mean \\
\cutinhead{Observation Summary\tablenotemark{\scriptsize j}}
acis\_num 							& No  & No  &	 			 	 	 	& Total number of ACIS imaging observations contributing to the Master Sources Table record of the source \\
acis\_hetg\_num 						& No  & No  &	 	 			 	 	& Total number of ACIS/HETG observations contributing to the Master Sources Table record of the source \\
acis\_letg\_num 						& No  & No  &	 	 	 	 			& Total number of ACIS/LETG observations contributing to the Master Sources Table record of the source \\
hrc\_num 								& No  & No  &			 	 	 	 	& Total number of HRC imaging observations contributing to the Master Sources Table record of the source \\
hrc\_letg\_num 							& No  & No  &	 			 	 	 	& Total number of HRC/LETG observations contributing to the Master Sources Table record of the source \\
hrc\_hetg\_num 						& No  & No  &	 	 	 			 	& Total number of HRC/HETG observations contributing to the Master Sources Table record of the source \\
acis\_time 							& No  & No  & s 			 	 		& Total livetime for all ACIS imaging observations contributing to the Master Sources Table record of the source \\
acis\_hetg\_time 						& No  & No  & s 	 			 		& Total livetime for all ACIS/HETG observations contributing to the Master Sources Table record of the source \\
acis\_letg\_time 						& No  & No  & s 	 	 				& Total livetime for all ACIS/LETG observations contributing to the Master Sources Table record of the source \\
hrc\_time 								& No  & No  & s		 	 	 			& Total livetime for all HRC imaging observations contributing to the Master Sources Table record of the source \\
hrc\_letg\_time 							& No  & No  & s 			 	 		& Total livetime for all HRC/LETG observations contributing to the Master Sources Table record of the source \\
hrc\_hetg\_time 						& No  & No  & s 	 	 				& Total livetime for all HRC/HETG observations contributing to the Master Sources Table record of the source \\
\enddata
\tablenotetext{a}{Indicates that tabulated properties include separate entries
  for each energy band.  The individual band entries are identified by the
  suffix ``\_$\canon{x}$'', where $\canon{x}$ is one of the energy band
  designations.}
\vskip -6pt
\tablenotetext{b}{Indicates that tabulated properties include separate entries
  for $\sim\!68\%$ lower and upper confidence limits.  The data value is
  tabulated using the indicated property name, while the lower and upper
  confidence limits are identified by the suffixes ``\_lolim'' and ``\_hilim,''
  respectively.  If a property includes both confidence limits {\em and\/}
  separate entries for each band, then the confidence limit suffix {\em
    precedes\/} the band designation suffix.} 
\vskip -6pt
\tablenotetext{c}{Position angles are defined relative to {\em local\/} true
North.}
\vskip -6pt
\tablenotetext{d}{Set to $+\infty$ if the source is saturated in all observations.}
\vskip -6pt
\tablenotetext{e}{Starting with CSC release 2.1: Source displays flux variability within or between one or more observations (intra- or inter-observation variability index $\ge 6$), in one or more energy bands.}
\vskip -6pt
\tablenotetext{f}{In the CSC Release 2 Series, the source extent ellipse
  is approximated by a circle.}
\vskip -6pt
\tablenotetext{g}{This definition is changed from CSC Release 1.}
\vskip -6pt
\tablenotetext{h}{Spectral model fits are only performed if the source has at
  least 150 net counts in the ACIS broad energy band.}
\vskip -6pt
\tablenotetext{i}{In the CSC Release 2 Series, the $\rhat$ values are not populated.}
\vskip -6pt
\tablenotetext{j}{The CSC Release 2 Series do not include observations
  obtained using the High or Low Energy Transmission Gratings.}
\end{deluxetable*}
}

{
\setlength{\descripwidth}{254pt}
\renewcommand\arraystretch{0.9}
\startlongtable
\begin{deluxetable*}{lcccp{\descripwidth}}
\tabletypesize{\scriptsize}
\tablecolumns{5}
\tablecaption{Stacked Observation Detections Table Properties\label{tab:stackobsproperties}}
\tablehead{
\noalign{\vskip 2.0ex} \colhead{Property} & \colhead{Multi-\tablenotemark{\scriptsize a}} & \colhead{Conf.\tablenotemark{\scriptsize b}} & \colhead{Units} & \colhead{Description} \\
\noalign{\vskip -1.5ex} & \colhead{Band} & \colhead{Lim.} & &
}
\startdata
\cutinhead{Stack-Specfic Information --- Stack Identification}	
detect\_stack\_id						& No  & No  &							& Detect stack identifier (designation of observation stack used for source detection) in the format ``\{acis|hrc\}fJ{\it hhmmsss}\{p|m\}{\it ddmmss\_nnn}'' \\
\cutinhead{Stack-Specfic Information --- Stack Reference Position}	
ra\_stack								& No  & No  & deg						& Detect stack tangent plane reference position, ICRS right ascension \\
dec\_stack							& No  & No  & deg						& Detect stack tangent plane reference position, ICRS declination \\
\cutinhead{Stack-Specfic Information --- Instrument Configuration}	
instrument								& No  & No  & 	 						& Instrument used for the stacked observations, 'ACIS' or 'HRC' \\
grating								& No  & No  & 							& Transmission grating used for the stacked observations, 'NONE', 'HETG', or 'LETG' \\
\cutinhead{Stack-Specfic Information --- Aspect Information\tablenotemark{\scriptsize c}}
deltax								& No  & No  & arcsec					& SKY coordinate system $X$ translation correction required to co-align the observation stack astrometric frame with the Gaia-CRF3 realization of the ICRF \\
deltay								& No  & No  & arcsec					& SKY coordinate system $Y$ translation correction required to co-align the observation stack astrometric frame with the Gaia-CRF3 realization of the ICRF \\
deltarot								& No  & No  & deg						& SKY coordinate system roll angle rotation correction required to co-align the observation stack astrometric frame with the Gaia-CRF3 realization of the ICRF \\
dscale								& No  & No  & 							& SKY coordinate system scale factor correction required to co-align the observation stack astrometric frame with the Gaia-CRF3 realization of the ICRF \\
man\_astrom\_flag						& No  & No  &							& Observation stack astrometric transform was manually modified via human review \\
\cutinhead{Stack-Specfic Information --- Processing Information}	
ascdsver								& No  & No  &							& Software version used to create the Level 3 detect stack event data file \\
caldbver								& No  & No  &							& Calibration database version used to calibrate the Level 3 detect stack event data file \\
crdate								& No  & No  &							& Creation date/time of the Level 3 detect stack event data file, UTC \\
\cutinhead{Stack Detection Properties --- Stack-Specific Detection Identification}
region\_id								& No  & No  &							& Detection region identifier (component number) \\
\cutinhead{Stack Detection Properties --- Detection Position}
ra									& No  & No  & deg						& Detection position, ICRS right ascension \\
dec									& No  & No  & deg						& Detection position, ICRS declination \\
err\_ellipse\_r0							& No  & No  & arcsec					& Major radius of the 95\% confidence level position error ellipse \\
err\_ellipse\_r1							& No  & No  & arcsec					& Minor radius of the 95\% confidence level position error ellipse \\
err\_ellipse\_ang						& No  & No  & deg						& Position angle\tablenotemark{\scriptsize d} of the major axis of the 95\% confidence level position error ellipse \\
theta\_mean							& No  & No  & arcmin					& Mean source region aperture off-axis angle from all stacked observations \\
\cutinhead{Stack Detection Properties --- Detection Significance}
flux\_significance						& Yes & No  &							& Significance of the stacked observation detection determined from the ratio of the stacked observation detection photon flux to the estimated error in the photon flux \\
detect\_significance						& Yes & No  &							& Significance of the stacked observation detection computed by the stacked observation detection algorithm \\
likelihood								& Yes & No  &							& log likelihood\tablenotemark{\scriptsize e} of the stacked observation detection computed by the Maximum Likelihood Estimator fit to the photon counts distribution \\
likelihood\_class						& No  & No  &							& Highest detection likelihood classification across all energy bands \\
\cutinhead{Stack Detection Properties --- Detection Codes and Flags}
conf\_code							& No  & No  &							& Compact detection may be confused (bit encoded: 1: background region overlaps another background region; 2: background region overlaps another source region; 4: source region overlaps another background region; 8: source region overlaps another source region; 256: compact detection is overlaid on an extended detection) \\
dither\_warning\_flag						& No  & No  &							& Highest statistically significant peak in the power spectrum of the detection source region count rate occurs at the dither frequency or at a beat frequency of the dither frequency in one or more of the stacked observations \\
edge\_code							& No  & No  &							& Detection position, or source or background region dithered off a detector boundary (chip pixel mask) during one or more of the stacked observations (bit encoded: 1: background region dithers off detector boundary; 2: source region dithers off detector boundary; 4: detection position dither off detector boundary) \\
extent\_code							& No  & No  &							& Detection is extended, or deconvolved compact detection extent is inconsistent with a point source ($a_{\rm rss}>5\,\Delta a_{\rm rss}$) in one or more of the stacked observations and energy bands (bit encoded: 1, 2, 4, 8, 16, 32: deconvolved compact detection extent is not consistent with a point source in the ACIS ultrasoft, soft, medium, hard, broad, or HRC wide ($\sim\!0.1$--$10.0\,$keV) energy band, respectively; 256: extended detection) \\
multi\_chip\_code						& No  & No  &							& Detection position, or source or background region dithered multiple detector chips during one or more of the stacked observations (bit encoded: 1: background region dithers across 2 chips; 2: background region dithers across $>2$ chips; 4: source region dithers across 2 chips; 8: source region dithers across $>2$ chips; 16: detection position dithers across 2 chips; 32: detection position dither across $>2$ chips) \\
pileup\_flag							& No  & No  &							& ACIS pile-up fraction exceeds $\sim\!10\%$ in any stacked observation; detection properties may be affected \\
sat\_src\_flag							& No  & No  &							& Detection is saturated in all stacked observations; detection properties are unreliable \\
streak\_src\_flag						& No  & No  &							& Detection is located on an ACIS readout streak in all stacked observations; detection properties may be affected \\
var\_flag								& No  & No  &							& Detection displays flux variability within one or more observations (intra-observation variability index $\ge 3$), in one or more energy bands\tablenotemark{\scriptsize f}  \\
var\_inter\_hard\_flag					& No  & No  &							& Detection hardness ratios are statistically inconsistent between two or more of the stacked observations \\
man\_add\_flag							& No  & No  &							& Detection was manually added to the catalog via human review (passes automated validation criteria) \\
man\_inc\_flag							& No  & No  &							& Detection was manually included in the catalog via human review (may not pass automated validation criteria) \\
man\_pos\_flag							& No  & No  &							& Best fit detection position was manually modified via human review \\
man\_reg\_flag							& No  & No  &							& Detection region parameters (dimensions, initial guess position input to the Maximum Likelihood Estimator fit) were manually modified via human review \\
\cutinhead{Stack Detection Properties --- Detection Extent\tablenotemark{\scriptsize g}}	
major\_axis					 		& Yes & Yes & arcsec					& $1\,\sigma$ radius along the major axis of the ellipse defining the deconvolved detection extent \\
minor\_axis					 		& Yes & Yes & arcsec				        	& $1\,\sigma$ radius along the minor axis of the ellipse defining the deconvolved detection extent \\
pos\_angle					 		& Yes & Yes & deg						& Position angle\tablenotemark{\scriptsize d} of the major axis of the ellipse defining the deconvolved detection extent \\
src\_area								& Yes & No  & $\rm arcsec^2$				& Area of the deconvolved detection extent ellipse, or area of the detection polygon for extended detections \\
\cutinhead{Stack Detection Properties --- Aperture Photometry}	
src\_cnts\_aper							& Yes & Yes & counts						& Aperture-corrected detection net counts inferred from the source region aperture \\
src\_rate\_aper							& Yes & Yes & $\rm counts\,s^{-1}$				& Aperture-corrected detection net count rate inferred from the source region aperture \\
photflux\_aper 							& Yes & Yes & $\rm photon\,cm^{-2}\,s^{-1}$		& Aperture-corrected detection net photon flux inferred from the source region aperture, calculated by counting X-ray events \\
flux\_aper 								& Yes & Yes & $\rm erg\,cm^{-2}\,s^{-1}$			& Aperture-corrected detection net energy flux inferred from the source region aperture, calculated by counting X-ray events \\
src\_cnts\_aper90						& Yes & Yes & counts						& Aperture-corrected detection net counts inferred from the PSF 90\% ECF aperture \\
src\_rate\_aper90						& Yes & Yes & $\rm counts\,s^{-1}$				& Aperture-corrected detection net count rate inferred from the PSF 90\% ECF aperture \\
photflux\_aper90 						& Yes & Yes & $\rm photon\,cm^{-2}\,s^{-1}$		& Aperture-corrected detection net photon flux inferred from the PSF 90\% ECF aperture, calculated by counting X-ray events \\
flux\_aper90 							& Yes & Yes & $\rm erg\,cm^{-2}\,s^{-1}$			& Aperture-corrected detection net energy flux inferred from the PSF 90\% ECF aperture, calculated by counting X-ray events \\
ra\_aper								& No  & No  & deg  							& Center of the source region and background region apertures, ICRS right ascension \\
dec\_aper								& No  & No  & deg							& Center of the source region and background region apertures, ICRS declination \\
mjr\_axis\_aper							& No  & No  & arcsec						& Semi-major axis of the elliptical source region aperture \\
mnr\_axis\_aper						& No  & No  & arcsec						& Semi-minor axis of the elliptical source region aperture \\
pos\_angle\_aper						& No  & No  & deg							& Position angle\tablenotemark{\scriptsize d} of the major axis of the elliptical source region aperture \\
mjr\_axis1\_aperbkg						& No  & No  & arcsec						& Semi-major axis of the inner ellipse of the annular background region aperture \\
mnr\_axis1\_aperbkg						& No  & No  & arcsec						& Semi-minor axis of the inner ellipse of the annular background region aperture \\
mjr\_axis2\_aperbkg						& No  & No  & arcsec						& Semi-major axis of the outer ellipse of the annular background region aperture \\
mnr\_axis2\_aperbkg						& No  & No  & arcsec						& Semi-minor axis of the outer ellipse of the annular background region aperture \\
pos\_angle\_aperbkg						& No  & No  & deg							& Position angle\tablenotemark{\scriptsize d} of the major axis of the annular background region aperture \\
phot\_nsrcs							& No  & No  & 	   							& Number of detections fit simultaneously to compute aperture photometry quantities \\
\cutinhead{Stack Detection Properties --- Spectral Hardness Ratios}
hard\_$\canon{x}\canon{y}$					& No  & Yes &							& Photon flux hardness ratio measured between ACIS energy bands $\canon{x}$ and $\canon{y}$, $\rm hard\_\canon{\mathit{x}}\canon{\mathit{y}} = (photflux\_aper\_\canon{\mathit{x}} - photflux\_aper\_\canon{\mathit{y}}) / (photflux\_aper\_\canon{\mathit{x}} + photflux\_aper\_\canon{\mathit{y}})$\tablenotemark{\scriptsize h} \\
var\_inter\_hard\_prob\_$\canon{x}\canon{y}$	& No  & Yes &								& Inter-observation photon flux hardness ratio variability probability \\
var\_inter\_hard\_sigma\_$\canon{x}\canon{y}$	& No  & Yes &								& Inter-observation photon flux hardness ratio variability standard deviation \\
\cutinhead{Stack Detection Properties --- Temporal Variability}
var\_intra\_index 						& Yes & No  &			 	 	  	 		& Intra-observation Gregory-Loredo variability index in the range $[0, 10]$; indicates whether the source region photon flux is constant within an observation (highest value across all stacked observations) \\
var\_intra\_prob 						& Yes & No  &	 			 	 	 		& Intra-observation Gregory-Loredo variability probability (highest value across all stacked observations) \\
ks\_intra\_prob 							& Yes & No  &	 	 	  			 		& Intra-observation Kolmogorov-Smirnov test variability probability (highest value across all stacked observations) \\
kp\_intra\_prob 							& Yes & No  &			 	 	 	 		& Intra-observation Kuiper's test variability probability (highest value across all stacked observations) \\
var\_inter\_index 						& Yes & No  &	 	 			  	 		& Inter-observation variability index in the range $[0, 10]$; indicates whether the source region photon flux is constant between observations \\
var\_inter\_prob 						& Yes & No  &	 							& Inter-observation variability probability, calculated from the $\chi^2$ distribution of the photon fluxes of the individual observations \\
var\_inter\_sigma						& Yes & No  & $\rm photon\,cm^{-2}\,s^{-1}$		& Inter-observation flux variability standard deviation; the spread of the individual observation photon fluxes about the error weighted mean \\
\cutinhead{Stack Detection Properties --- Detection-Specific Stacked Observation Summary}
livetime 								& No  & No  & s	 			 	 	 		& Effective stacked observation exposure time, after applying the good time intervals and the deadtime correction factor (vignetting and dead area corrections are {\em not\/} applied) \\
\enddata
\tablenotetext{a}{Indicates that tabulated properties include separate entries
  for each energy band.  The individual band entries are identified by the
  suffix ``\_$\canon{x}$'', where $\canon{x}$ is one of the energy band
  designations.}
\vskip -6pt
\tablenotetext{b}{Indicates that tabulated properties include separate entries
  for $\sim\!68\%$ lower and upper confidence limits.  The data value is
  tabulated using the indicated property name, while the lower and upper
  confidence limits are identified by the suffixes ``\_lolim'' and ``\_hilim,''
  respectively.  If a property includes both confidence limits {\em and\/}
  separate entries for each band, then the confidence limit suffix {\em
    precedes\/} the band designation suffix.} 
\vskip -6pt
\tablenotetext{c}{These five properties are present starting in CSC release 2.1.}
\vskip -6pt
\tablenotetext{d}{Position angles are defined relative to {\em local\/} true
North.}
\vskip -6pt
\tablenotetext{e}{Set to $+\infty$ if the source is saturated in all observations.}
\vskip -6pt
\tablenotetext{f}{Starting with CSC release 2.1: Source displays flux variability within or between one or more observations (intra- or inter-observation variability index $\ge 6$), in one or more energy bands.}
\vskip -6pt
\tablenotetext{g}{In the CSC Release 2 Series, the source extent ellipse
  is approximated by a circle.}
\vskip -6pt
\tablenotetext{h}{This definition is changed from CSC Release 1.}
\end{deluxetable*}
}

{
\setlength{\descripwidth}{277pt}
\renewcommand\arraystretch{0.9}
\startlongtable
\begin{deluxetable*}{lcccp{\descripwidth}}
\tabletypesize{\scriptsize}
\tablecolumns{5}
\tablecaption{Per-Observation Detections Table Properties\label{tab:perobsproperties}}
\tablehead{
\noalign{\vskip 2.0ex} \colhead{Property} & \colhead{Multi-\tablenotemark{\scriptsize a}} & \colhead{Conf.\tablenotemark{\scriptsize b}} & \colhead{Units} & \colhead{Description} \\
\noalign{\vskip -1.5ex} & \colhead{Band} & \colhead{Lim.} & &
}
\startdata
\cutinhead{Observation-Specfic Information --- Observation Identification}	
obsid								& No  & No  &								& Observation identifier \\
obi									& No  & No  &								& Observation interval identifier \\
\cutinhead{Observation-Specfic Information --- Observation Target and Pointing}
targname		       		       		   		& No  & No  &								& Observation target name specified by the observer \\
ra\_targ								& No  & No  & deg							& Target position specified by the observer, ICRS right ascension \\
dec\_targ								& No  & No  & deg							& Target position specified by the observer, ICRS declination \\
ra\_pnt								& No  & No  & deg							& Mean spacecraft pointing during the observation, ICRS right ascension \\
dec\_pnt								& No  & No  & deg							& Mean spacecraft pointing during the observation, ICRS declination \\
roll\_pnt								& No  & No  & deg							& Mean spacecraft roll angle during the observation \\
chipx\_pnt								& No  & No  & pixel							& Detector (chip coordinates) Cartesian $x$ position corresponding to $({\rm ra\_pnt}, {\rm dec\_pnt})$ \\
chipy\_pnt								& No  & No  & pixel							& Detector (chip coordinates) Cartesian $y$ position corresponding to $({\rm ra\_pnt}, {\rm dec\_pnt})$ \\
chip\_id\_pnt							& No  & No  & 								& Detector (chip coordinates) identifier used to define $({\rm chipx\_pnt}, {\rm chipy\_pnt})$ \\
ra\_nom								& No  & No  & deg							& Observation tangent plane reference position, ICRS right ascension \\
dec\_nom								& No  & No  & deg							& Observation tangent plane reference position, ICRS declination \\
roll\_nom								& No  & No  & deg							& Observation tangent plane roll angle (used to determine tangent plane North)\\
\cutinhead{Observation-Specfic Information --- Observation Timing}
gti\_start								& No  & No  & s								& Start time of valid observation data (MET: seconds since 1998 Jan 01 00:00:00 TT) \\
gti\_stop								& No  & No  & s								& Stop time of valid observation data (MET: seconds since 1998 Jan 01 00:00:00 TT) \\
gti\_elapse							& No  & No  & s								& Total elapsed time of the observation $({\rm gti\_stop} - {\rm gti\_start})$ \\
gti\_obs								& No  & No  & 								& Start time of valid observation data (TT), ISO 8601 format (yyyy-mm-ddThh:mm:ss) \\
gti\_end								& No  & No  &								& Stop time of valid observation data (TT), ISO 8601 format (yyyy-mm-ddThh:mm:ss) \\
gti\_mjd\_obs							& No  & No  &								& MJD corresponding to start time of valid observation data (TT) \\
mjd\_ref								& No  & No  &								& MJD corresponding to $\rm 0\,s$ MET \\
\cutinhead{Observation-Specfic Information --- Observating Cycle}
ao									& No  & No  &								& Chandra observing cycle in which the observation was scheduled \\
\cutinhead{Observation-Specfic Information --- Instrument Configuration}
instrument								& No  & No  &								& Instrument used for the observation, 'ACIS' or 'HRC' \\
grating								& No  & No  &								& Transmission grating used for the observation, 'NONE', 'HETG', or 'LETG \\
datamode								& No  & No  &								& Instrument data mode used for the observation \\
readmode								& No  & No  &								& ACIS readout mode used for the observation \\
cycle									& No  & No  &								& ACIS readout cycle for the observation, 'P' (primary) or 'S' (secondary) for alternating exposure (interleaved) mode observations, or 'P' for other ACIS modes \\
exptime								& No  & No  &								& ACIS CCD frame time \\
timing\_mode							& No  & No  &								& HRC precision timing mode \\
\cutinhead{Observation-Specfic Information --- Aspect Information}
sim\_x								& No  & No  & mm							& SIM focus stage position during observation \\
sim\_z								& No  & No  & mm							& SIM translation stage position during observation \\
dy									& No  & No  & mm							& Mean aspect $dy$ offset during observation \\
dz									& No  & No  & mm							& Mean aspect $dz$ offset during observation \\
dtheta								& No  & No  & deg							& Mean aspect $d\theta$ offset during observation \\
deltax								& No  & No  & arcsec						& SKY coordinate system $X$ translation correction required to co-align the observation astrometric frame within observation stack \\
deltay								& No  & No  & arcsec						& SKY coordinate system $Y$ translation correction required to co-align the observation astrometric frame within observation stack \\
deltarot								& No  & No  & deg							& SKY coordinate system roll angle rotation correction required to co-align the observation astrometric frame within observation stack \\
dscale								& No  & No  & 								& SKY coordinate system scale factor correction required to co-align the observation astrometric frame within observation stack \\
man\_astrom\_flag						& No  & No  &								& Observation astrometric transform was manually modified via human review \\
\cutinhead{Observation-Specfic Information --- Processing Information}
ascdsver								& No  & No  &								& Software version used to create the Level 3 observation event data file \\
caldbver								& No  & No  &								& Calibration database version used to calibrate the Level 3 observation event data file \\
crdate								& No  & No  &								& Creation date/time of the Level 3 observation event data file, UTC \\
\cutinhead{Observation Detection Properties --- Observation-Specific Detection Identification}
region\_id	       		 	    				& No  & No  &								& Detection region identifier (component number) \\
\cutinhead{Observation Detection Properties --- Detection Position}	
theta									& No  & No  & arcmin						& Detection position, PSF 90\% ECF aperture off-axis angle \\
phi									& No  & No  & deg							& Detection position, PSF 90\% ECF aperture azimuthal angle \\
chipx									& No  & No  & pixel 							& Detector (chip coordinates) Cartesian $x$ position corresponding to $({\rm theta}, {\rm phi})$ \\
chipy									& No  & No  & pixel							& Detector (chip coordinates) Cartesian $y$ position corresponding to $({\rm theta}, {\rm phi})$ \\
chip\_id								& No  & No  & 								& Detector (chip coordinates) identifier used to define $({\rm chipx}, {\rm chipy})$ \\
\cutinhead{Observation Detection Properties --- Detection Significance}
flux\_significance						& Yes & No  &								& Significance of the single-observation detection determined from the ratio of the single-observation detection photon flux to the estimated error in the photon flux \\
likelihood								& Yes & No  &								& log likelihood of the single-observation detection computed by the Maximum Likelihood Estimator fit to the photon counts distribution \\
likelihood\_class						& No  & No  &								& Highest detection likelihood classification across all stacked observations and energy bands \\
\cutinhead{Observation Detection Properties --- Detection Codes and Flags}
conf\_code							& No  & No  &								& Compact detection may be confused (bit encoded: 1: background region overlaps another background region; 2: background region overlaps another source region; 4: source region overlaps another background region; 8: source region overlaps another source region; 256: compact detection is overlaid on an extended detection) \\
dither\_warning\_flag						& No  & No  &								& Highest statistically significant peak in the power spectrum of the detection source region count rate occurs at the dither frequency or at a beat frequency of the dither frequency of the observation \\
edge\_code							& No  & No  &								& Detection position, or source or background region dithered off a detector boundary (chip pixel mask) during the observation (bit encoded: 1: background region dithers off detector boundary; 2: source region dithers off detector boundary; 4: detection position dither off detector boundary) \\
extent\_code							& No  & No  &								& Detection is extended, or deconvolved compact detection extent is inconsistent with a point source ($a_{\rm rss}>5\,\Delta a_{\rm rss}$) in one or more energy bands (bit encoded: 1, 2, 4, 8, 16, 32: deconvolved compact detection extent is not consistent with a point source in the ACIS ultrasoft, soft, medium, hard, broad, or HRC wide ($\sim\!0.1$--$10.0\,$keV) energy band, respectively; 256: extended detection) \\
multi\_chip\_code						& No  & No  &								& Source position, or source or background region dithered multiple detector chips during the observation (bit encoded: 1: background region dithers across 2 chips; 2: background region dithers across $>2$ chips; 4: source region dithers across 2 chips; 8: source region dithers across $>2$ chips; 16: detection position dithers across 2 chips; 32: detection position dither across $>2$ chips) \\
pileup\_warning							& No  & No  & count/frame					& ACIS observed per-frame count rate of the brightest $3\times 3$ pixel island; ${\rm estimated pileup fraction} = {\rm pileup\_warning} / 2$ for pileup\_warning values $leq 0.2$ \\
sat\_src\_flag							& No  & No  &								& Detection is saturated; detection properties are unreliable \\
streak\_src\_flag						& No  & No  &								& Detection is located on an ACIS readout streak; detection properties may be affected \\
var\_code								& No  & No  &								& Detection displays flux variability within one or more observations (intra-observation variability index $\ge 3$), in one or more energy bands\tablenotemark{\scriptsize e} (bit encoded: 1, 2, 4, 8, 16, 32: intra-observation variability detected in the ACIS ultrasoft, soft, medium, hard, broad, or HRC wide ($\sim\!0.1$--$10.0\,$keV) energy band, respectively) \\
\cutinhead{Observation Detection Properties --- Detection Extent\tablenotemark{\scriptsize f}}	
mjr\_axis\_raw	       		 	    	       		& Yes & Yes & arcsec						& $1\,\sigma$ radius along the major axis of the ellipse defining the observed detection extent \\
mnr\_axis\_raw							& Yes & Yes & arcsec						& $1\,\sigma$ radius along the minor axis of the ellipse defining the observed detection extent \\
pos\_angle\_raw						& Yes & Yes & deg							& Position angle of the major axis of the ellipse defining the observed detection extent \\
psf\_mjr\_axis\_raw	       		 	    	       	& Yes & Yes & arcsec						& $1\,\sigma$ radius along the major axis of the ellipse defining the local PSF extent \\
psf\_mnr\_axis\_raw						& Yes & Yes & arcsec						& $1\,\sigma$ radius along the minor axis of the ellipse defining the local PSF extent \\
psf\_pos\_angle\_raw					& Yes & Yes & deg							& Position angle of the major axis of the ellipse defining the local PSF extent \\
major\_axis					 		& Yes & Yes & arcsec 						& $1\,\sigma$ radius along the major axis of the ellipse defining the deconvolved detection extent \\
minor\_axis					 		& Yes & Yes & arcsec				        		& $1\,\sigma$ radius along the minor axis of the ellipse defining the deconvolved detection extent \\
pos\_angle					 		& Yes & Yes & deg							& Position angle\tablenotemark{\scriptsize c} of the major axis of the ellipse defining the deconvolved detection extent \\
src\_area								& Yes & No  & $\rm arcsec^2$					& Area of the deconvolved detection extent ellipse, or area of the detection polygon for extended sources \\
\cutinhead{Observation Detection Properties --- Aperture Photometry}
cnts\_aper								& Yes & No  & counts						& Total counts observed in the modified\tablenotemark{\scriptsize g} source region aperture \\
cnts\_aperbkg							& Yes & No  & counts						& Total counts observed in the modified\tablenotemark{\scriptsize g} background region aperture \\
src\_cnts\_aper							& Yes & Yes & counts						& Aperture-corrected detection net counts inferred from the source region aperture \\
src\_rate\_aper							& Yes & Yes & $\rm counts\,s^{-1}$				& Aperture-corrected detection net count rate inferred from the source region aperture \\
photflux\_aper 							& Yes & Yes & $\rm photon\,cm^{-2}\,s^{-1}$		& Aperture-corrected detection net photon flux inferred from the source region aperture, calculated by counting X-ray events \\
flux\_aper 								& Yes & Yes & $\rm erg\,cm^{-2}\,s^{-1}$			& Aperture-corrected detection net energy flux inferred from the source region aperture, calculated by counting X-ray events \\
cnts\_aper90							& Yes & No  & counts						& Total counts observed in the modified\tablenotemark{\scriptsize g} PSF 90\% ECF aperture \\
cnts\_aper90bkg						& Yes & No  & counts						& Total counts observed in the modified\tablenotemark{\scriptsize g} PSF 90\% ECF background aperture \\
src\_cnts\_aper90						& Yes & Yes & counts						& Aperture-corrected detection net counts inferred from the PSF 90\% ECF aperture \\
src\_rate\_aper90						& Yes & Yes & $\rm counts\,s^{-1}$				& Aperture-corrected detection net count rate inferred from the PSF 90\% ECF aperture \\
photflux\_aper90 						& Yes & Yes & $\rm photon\,cm^{-2}\,s^{-1}$		& Aperture-corrected detection net photon flux inferred from the PSF 90\% ECF aperture, calculated by counting X-ray events \\
flux\_aper90 							& Yes & Yes & $\rm erg\,cm^{-2}\,s^{-1}$			& Aperture-corrected detection net energy flux inferred from the PSF 90\% ECF aperture, calculated by counting X-ray events \\
ra\_aper90							& No  & No  & deg  							& Center of the PSF 90\% ECF and PSF 90\% ECF background apertures, ICRS right ascension \\
dec\_aper90							& No  & No  & deg							& Center of the PSF 90\% ECF and PSF 90\% ECF background apertures, ICRS declination \\
mjr\_axis\_aper90						& No  & No  & arcsec						& Semi-major axis of the elliptical PSF 90\% ECF aperture \\
mnr\_axis\_aper90						& No  & No  & arcsec						& Semi-minor axis of the elliptical PSF 90\% ECF aperture \\
pos\_angle\_aper90						& No  & No  & deg							& Position angle\tablenotemark{\scriptsize c} of the major axis of the elliptical PSF 90\% ECF aperture \\
mjr\_axis1\_aper90bkg					& No  & No  & arcsec						& Semi-major axis of the inner ellipse of the annular PSF 90\% ECF background aperture \\
mnr\_axis1\_aper90bkg					& No  & No  & arcsec						& Semi-minor axis of the inner ellipse of the annular PSF 90\% ECF background aperture \\
mjr\_axis2\_aper90bkg					& No  & No  & arcsec						& Semi-major axis of the outer ellipse of the annular PSF 90\% ECF background aperture \\
mnr\_axis2\_aper90bkg					& No  & No  & arcsec						& Semi-minor axis of the outer ellipse of the annular PSF 90\% ECF background aperture \\
pos\_angle\_aper90bkg					& No  & No  & deg							& Position angle\tablenotemark{\scriptsize c} of the major axis of the annular PSF 90\% ECF background aperture \\
area\_aper							& No  & No  & $\rm arcsec^2$					& Area of the modified\tablenotemark{\scriptsize g} elliptical source region aperture \\
area\_aperbkg							& No  & No  & $\rm arcsec^2$					& Area of the modified\tablenotemark{\scriptsize g} annular background region aperture \\
area\_aper90							& Yes & No  & $\rm arcsec^2$					& Area of the modified\tablenotemark{\scriptsize g} elliptical PSF 90\% ECF aperture \\
area\_aper90bkg						& Yes & No  & $\rm arcsec^2$					& Area of the modified\tablenotemark{\scriptsize g} annular PSF 90\% ECF background aperture \\
psf\_frac\_aper							& Yes & No  & 	   							& Fraction of the PSF included in the modified\tablenotemark{\scriptsize g} elliptical source region aperture \\
psf\_frac\_aperbkg						& Yes & No  &								& Fraction of the PSF included in the modified\tablenotemark{\scriptsize g} annular background region aperture \\
psf\_frac\_aper90						& Yes & No  & 	   							& Fraction of the PSF included in the modified\tablenotemark{\scriptsize g} elliptical PSF 90\% ECF aperture \\
psf\_frac\_aper90bkg					& Yes & No  &								& Fraction of the PSF included in the modified\tablenotemark{\scriptsize g} annular PSF 90\% ECF background aperture \\
phot\_nsrcs							& No  & No  & 	   							& Number of detections fit simultaneously to compute aperture photometry quantities \\
flux\_powlaw\_aper 						& Yes & Yes & $\rm erg\,cm^{-2}\,s^{-1}$			& Source region aperture model energy flux inferred from the canonical absorbed power law model ($N_{\rm H} = N_{\rm H(Gal)}$; $\Gamma = 2.0$) \\
flux\_bb\_aper 							& Yes & Yes & $\rm erg\,cm^{-2}\,s^{-1}$  			& Source region aperture model energy flux inferred from the canonical absorbed black body model ($N_{\rm H} = N_{\rm H(Gal)}$; $kT = 0.75\,\rm keV)$ \\
flux\_brems\_aper						& Yes & Yes & $\rm erg\,cm^{-2}\,s^{-1}$	  		& Source region aperture model energy flux inferred from the canonical absorbed bremsstrahlung model ($N_{\rm H} = N_{\rm H(Gal)}$; $kT = 3.5\,\rm keV)$ \\
flux\_apec\_aper						& Yes & Yes & $\rm erg\,cm^{-2}\,s^{-1}$			& Source region aperture model energy flux inferred from the canonical absorbed APEC model ($N_{\rm H} = N_{\rm H(Gal)}$; $kT = 6.5\,\rm keV)$ \\
flux\_powlaw\_aper90 					& Yes & Yes & $\rm erg\,cm^{-2}\,s^{-1}$			& PSF 90\% ECF region aperture model energy flux inferred from the canonical absorbed power law model ($N_{\rm H} = N_{\rm H(Gal)}$; $\Gamma = 2.0$) \\
flux\_bb\_aper90						& Yes & Yes & $\rm erg\,cm^{-2}\,s^{-1}$  			& PSF 90\% ECF region aperture model energy flux inferred from the canonical absorbed black body model ($N_{\rm H} = N_{\rm H(Gal)}$; $kT = 0.75\,\rm keV)$ \\
flux\_brems\_aper90						& Yes & Yes & $\rm erg\,cm^{-2}\,s^{-1}$	  		& PSF 90\% ECF region aperture model energy flux inferred from the canonical absorbed bremsstrahlung model ($N_{\rm H} = N_{\rm H(Gal)}$; $kT = 3.5\,\rm keV)$ \\
flux\_apec\_aper90						& Yes & Yes & $\rm erg\,cm^{-2}\,s^{-1}$			& PSF 90\% ECF region aperture model energy flux inferred from the canonical absorbed APEC model ($N_{\rm H} = N_{\rm H(Gal)}$; $kT = 6.5\,\rm keV)$ \\
\cutinhead{Observation Detection Properties --- Spectral Hardness Ratios}
hard\_$\canon{x}\canon{y}$				 & No  & Yes &								& Photon flux hardness ratio measured between ACIS energy bands $\canon{x}$ and $\canon{y}$, $\rm hard\_\canon{\mathit{x}}\canon{\mathit{y}} = (photflux\_aper\_\canon{\mathit{x}} - photflux\_aper\_\canon{\mathit{y}}) / (photflux\_aper\_\canon{\mathit{x}} + photflux\_aper\_\canon{\mathit{y}})$\tablenotemark{\scriptsize h} \\
\cutinhead{Observation Detection Properties --- Model Spectral Fits\tablenotemark{\scriptsize i}}
flux\_powlaw 							& No  & Yes & $\rm erg\,cm^{-2}\,s^{-1}$			& Net integrated $0.5$--$7.0\rm\,keV$ energy flux of the best fitting absorbed power-law model spectrum to the source region aperture PI spectrum \\
powlaw\_gamma						& No  & Yes &								& Photon index ($\Gamma$, defined as $F_{\rm E}\propto\!E^{-\Gamma}$) of the best fitting absorbed power-law model spectrum to the source region aperture PI spectrum \\
powlaw\_gamma\_rhat					& No  & No  &								& Photon index convergence criterion, $\rhat$, of the best fitting absorbed power-law model spectrum to the source region aperture PI spectrum\tablenotemark{\scriptsize j} \\
powlaw\_nh 							& No  & Yes & $\rm 10^{20}\,cm^{-2}$ 			& Total neutral Hydrogen column density, $N_{\rm H}$, of the best fitting  absorbed power-law model spectrum to the source region aperture PI spectrum \\
powlaw\_nh\_rhat						& No  & No  & 			   					& Total neutral Hydrogen column density convergence criterion, $\rhat$, of the best fitting absorbed power-law model spectrum to the source region aperture PI spectrum\tablenotemark{\scriptsize j} \\
powlaw\_ampl							& No  & Yes &								& Amplitude of the best fitting absorbed power-law model spectrum to the source region aperture PI spectrum \\
powlaw\_ampl\_rhat						& No  & No  &								& Amplitude convergence criterion, $\rhat$, of the best fitting absorbed power-law model spectrum to the source region aperture PI spectrum\tablenotemark{\scriptsize j} \\
powlaw\_stat							& No  & No  &								& $\chi^2$ statistic per degree of freedom of the best fitting absorbed power-law model spectrum to the source region aperture PI spectrum \\
flux\_bb 								& No  & Yes & $\rm erg\,cm^{-2}\,s^{-1}$	 		& Net integrated $0.5$--$7.0\rm\,keV$ energy flux of the best fitting absorbed black body model spectrum to the source region aperture PI spectrum \\
bb\_kt								& No  & Yes & keV							& Temperature ($kT$) of the best fitting absorbed black body model spectrum to the source region aperture PI spectrum \\
bb\_kt\_rhat							& No  & No  &								& Temperature convergence criterion, $\rhat$, of the best fitting absorbed black body model spectrum to the source region aperture PI spectrum\tablenotemark{\scriptsize j} \\
bb\_nh 								& No  & Yes & $\rm 10^{20}\,cm^{-2}$ 			& Total neutral Hydrogen column density, $N_{\rm H}$, of the best fitting absorbed black body model spectrum to the source region aperture PI spectrum \\
bb\_nh\_rhat							& No  & No  & 			   					& Total neutral Hydrogen column density convergence criterion, $\rhat$, of the best fitting absorbed black body model spectrum to the source region aperture PI spectrum\tablenotemark{\scriptsize j} \\
bb\_ampl								& No  & Yes &								& Amplitude of the best fitting absorbed black body model spectrum to the source region aperture PI spectrum \\
bb\_ampl\_rhat							& No  & No  &								& Amplitude convergence criterion, $\rhat$, of the best fitting absorbed black body model spectrum to the source region aperture PI spectrum\tablenotemark{\scriptsize j} \\
bb\_stat								& No  & No  &								& $\chi^2$ statistic per degree of freedom of the best fitting absorbed black body model spectrum to the source region aperture PI spectrum \\
flux\_brems 							& No  & Yes & $\rm erg\,cm^{-2}\,s^{-1}$ 			& Net integrated $0.5$--$7.0\rm\,keV$ energy flux of the best fitting absorbed bremsstrahlung model spectrum to the source region aperture PI spectrum \\
brems\_kt								& No  & Yes & keV							& Temperature ($kT$) of the best fitting absorbed bremsstrahlung model spectrum to the source region aperture PI spectrum \\
brems\_kt\_rhat							& No  & No  &								& Temperature convergence criterion, $\rhat$, of the best fitting absorbed bremsstrahlung model spectrum to the source region aperture PI spectrum\tablenotemark{\scriptsize j} \\
brems\_nh 							& No  & Yes & $\rm 10^{20}\,cm^{-2}$ 			& Total neutral Hydrogen column density, $N_{\rm H}$, of the best fitting absorbed bremsstrahlung model spectrum to the source region aperture PI spectrum \\
brems\_nh\_rhat						& No  & No  & 			   					& Total neutral Hydrogen column density convergence criterion, $\rhat$, of the best fitting absorbed bremsstrahlung model spectrum to the source region aperture PI spectrum\tablenotemark{\scriptsize j} \\
brems\_norm							& No  & Yes &								& Normalization of the best fitting absorbed bremsstrahlung model spectrum to the source region aperture PI spectrum \\
brems\_norm\_rhat						& No  & No  &								& Normalization convergence criterion, $\rhat$, of the best fitting absorbed bremsstrahlung model spectrum to the source region aperture PI spectrum\tablenotemark{\scriptsize j} \\
brems\_stat							& No  & No  &								& $\chi^2$ statistic per degree of freedom of the best fitting absorbed bremsstrahlung model spectrum to the source region aperture PI spectrum \\
flux\_apec 							& No  & Yes & $\rm erg\,cm^{-2}\,s^{-1}$ 			& Net integrated $0.5$--$7.0\rm\,keV$ energy flux of the best fitting absorbed APEC model spectrum to the source region aperture PI spectrum \\
apec\_kt								& No  & Yes & keV							& Temperature ($kT$) of the best fitting absorbed APEC model spectrum to the source region aperture PI spectrum \\
apec\_kt\_rhat							& No  & No  &								& Temperature convergence criterion, $\rhat$, of the best fitting absorbed APEC model spectrum to the source region aperture PI spectrum\tablenotemark{\scriptsize j} \\
apec\_abund							& No  & Yes &								& Abundance of the best fitting absorbed APEC model spectrum to the source region aperture PI spectrum \\
apec\_abund\_rhat						& No  & No  &								& Abundance convergence criterion, $\rhat$, of the best fitting absorbed APEC model spectrum to the source region aperture PI spectrum\tablenotemark{\scriptsize j} \\
apec\_z								& No  & Yes &								& Redshift of the best fitting  absorbed APEC model spectrum to the source region aperture PI spectrum \\
apec\_z\_rhat							& No  & No  &								& Redshift convergence criterion, $\rhat$, of the best fitting absorbed APEC model spectrum to the source region aperture PI spectrum\tablenotemark{\scriptsize j} \\
apec\_nh 								& No  & Yes & $\rm 10^{20}\,cm^{-2}$ 			& Total neutral Hydrogen column density, $N_{\rm H}$, of the best fitting absorbed APEC model spectrum to the source region aperture PI spectrum \\
apec\_nh\_rhat							& No  & No  & 	  		 					& Total neutral Hydrogen column density convergence criterion, $\rhat$, of the best fitting absorbed APEC model spectrum to the source region aperture PI spectrum\tablenotemark{\scriptsize j} \\
apec\_norm							& No  & Yes &								& Normalization of the best fitting absorbed APEC model spectrum to the source region aperture PI spectrum \\
apec\_norm\_rhat						& No  & No  &								& Normalization convergence criterion, $\rhat$, of the best fitting absorbed APEC model spectrum to the source region aperture PI spectrum\tablenotemark{\scriptsize j} \\
apec\_stat								& No  & No  &								& $\chi^2$ statistic per degree of freedom of the best fitting absorbed APEC model spectrum to the source region aperture PI spectrum \\
\cutinhead{Observation Detection Properties --- Temporal Variability}
var\_index								& Yes & No  &			 	 	  	 		& Intra-observation Gregory-Loredo variability index in the range $[0, 10]$; indicates whether the source region photon flux is constant within an observation \\
var\_prob 								& Yes & No  &	 			 	 	 		& Intra-observation Gregory-Loredo variability probability \\
ks\_prob 								& Yes & No  &	 	 	  			 		& Intra-observation Kolmogorov-Smirnov test variability probability \\
kp\_prob 								& Yes & No  &			 	 	 	 		& Intra-observation Kuiper's test variability probability \\
var\_sigma							& Yes & No  & $\rm counts\,s^{-1}$				& Flux variability standard deviation, calculated from the optimally-binned light curve \\
var\_mean							& Yes & No  & $\rm counts\,s^{-1}$				& Flux variability mean value, calculated from the optimally-binned light curve \\
var\_min								& Yes & No  & $\rm counts\,s^{-1}$				& Flux variability minimum value, calculated from the optimally-binned light curve \\
var\_max								& Yes & No  & $\rm counts\,s^{-1}$				& Flux variability maximum value, calculated from the optimally-binned light curve \\
\cutinhead{Observation Detection Properties --- Observation Summary}
livetime 								& No  & No  & s	 			 	 	 		& Effective observation exposure time, after applying the good time intervals and the deadtime correction factor (vignetting and dead area corrections are {\em not\/} applied) \\
detector								& No  & No  & 								& Detector elements over which the background region bounding box dithers during the observation \\
\enddata
\tablenotetext{a}{Indicates that tabulated properties include separate entries
  for each energy band.  The individual band entries are identified by the
  suffix ``\_$\canon{x}$'', where $\canon{x}$ is one of the energy band
  designations.}
\vskip -6pt
\tablenotetext{b}{Indicates that tabulated properties include separate entries
  for $\sim\!68\%$ lower and upper confidence limits.  The data value is
  tabulated using the indicated property name, while the lower and upper
  confidence limits are identified by the suffixes ``\_lolim'' and ``\_hilim,''
  respectively.  If a property includes both confidence limits {\em and\/}
  separate entries for each band, then the confidence limit suffix {\em
    precedes\/} the band designation suffix.} 
\vskip -6pt
\tablenotetext{c}{Position angles are defined relative to {\em local\/} true
North.}
\vskip -6pt
\tablenotetext{d}{Set to $+\infty$ if the source is saturated in all observations.}
\vskip -6pt
\tablenotetext{e}{Starting with CSC release 2.1: Detection displays flux variability (intra-observation variability index $\ge 6$) in one or more energy bands.}
\vskip -6pt
\tablenotetext{f}{In the CSC Release 2 Series, the source extent ellipse
  is approximated by a circle.}
\vskip -6pt
\tablenotetext{g}{The ``modified'' apertures exclude any overlapping detection source regions.}
\vskip -6pt
\tablenotetext{h}{This definition is changed from CSC release 1.}
\vskip -6pt
\tablenotetext{i}{Spectral model fits are only performed if the source has at
  least 150 net counts in the ACIS broad energy band.}
\vskip -6pt
\tablenotetext{j}{In the CSC Release 2 Series, the $\rhat$ values are not populated.}
\end{deluxetable*}
}

\begin{figure}[h!]
\epsscale{1.0}
\plotone{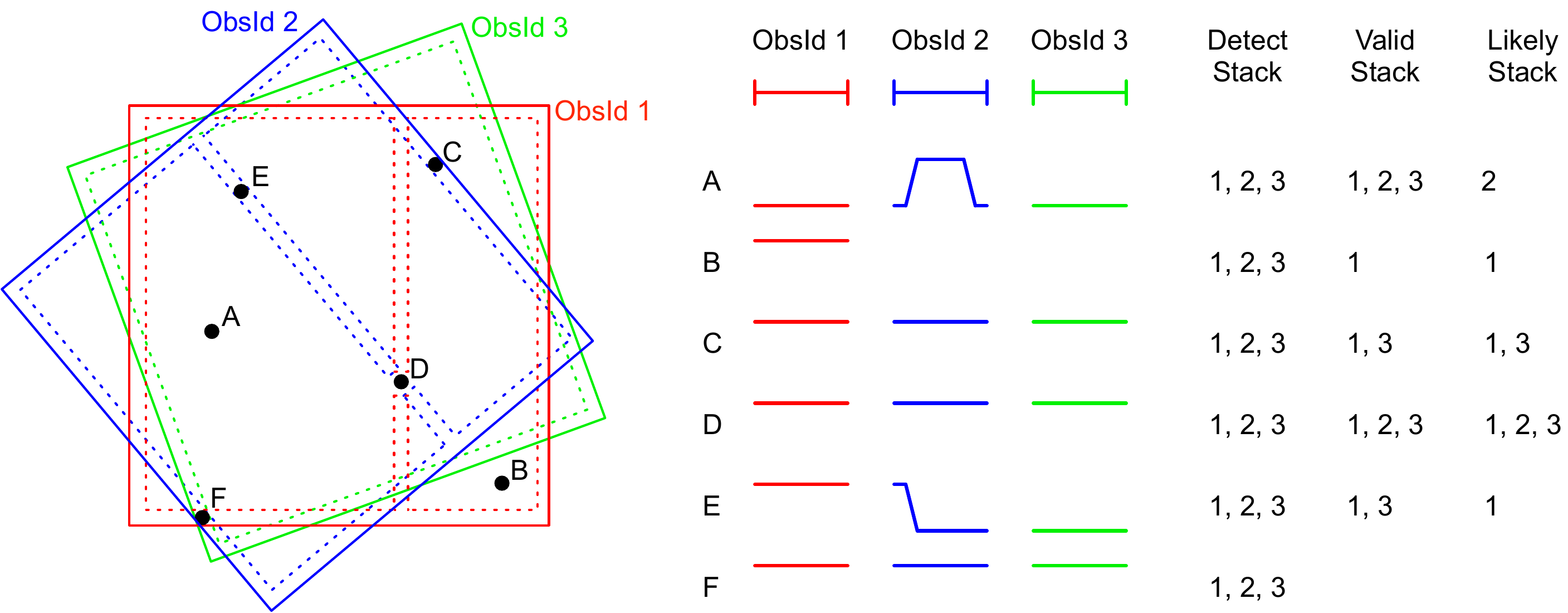}
\caption{\label{fig:stacks}
Example detect, valid, and likely stacks for 6 candidate detections (`A'--`F') identified in a stack of three observations (\obsid s 1--3) shown in red, blue, and green, respectively.  The field edges are shown by the solid lines while the dashed lines identify the pixel mask edges.  Pixels that fall outside of the pixel mask are considered invalid.  The light curves for each candidate detection are shown beneath the respective \obsid\ labels.  Candidate detection `A' is valid in all 3 observations, but because the source is bright only in \obsid~2 the likely stack for that detection only includes \obsid~2.  Detection `B' is only valid in \obsid~1 because it falls outside the field of view of the other observations in the stack.  Detection `C' falls within the field of view but off the pixel mask for \obsid~2, and so that observation is excluded from the valid stack.  Detection `D' is a bright source that is saturated in \obsid s~1 and~2; the readout streaks emanating from the source in these \obsid s are marked invalid by the pixel mask, although the pixel mask streak cutouts  are broken around the bright source itself.  Detection `E' falls on the pixel mask streak cutout for the bright source `D' in \obsid~2 and so that observation is excluded from the detection's valid stack; since the source is only bright in \obsid~1, only that observation is included in the likely stack. Finally, detection `F' falls outside the pixel masks (but within the fields of view) for all 3 observations, and so the corresponding valid stack is empty; this detection will not be present in the catalog.}
\end{figure}

The {\em Detect Stack\/} table (Table~\ref{tab:detectstack}) records the mappings between observation stacks and the component observations that comprise the stacks.  Although all the observations that are included in a stack are recorded in the Detect Stack table, observations that are part of the same stack may cover somewhat different sky regions and therefore overlap each other only partly.

{
\setlength{\descripwidth}{444pt}
\renewcommand\arraystretch{0.9}
\begin{deluxetable*}{lp{\descripwidth}}
\tabletypesize{\scriptsize}
\tablecolumns{2}
\tablecaption{Master Source/Stacked Observation Detection Associations Table Properties\label{tab:masterstackassoc}}
\tablehead{
\noalign{\vskip 2.5ex}
\colhead{Property} & \colhead{Description} \\
\noalign{\vskip -2.5ex}
}
\startdata
name							& Source name in the format ``2CXO J${\it hhmmss.s}\pm{\it ddmmss}$[X]'' \\
detect\_stack\_id					& Detect stack identifier (designation of observation stack used for source detection) in the format ``$\{\hbox{\rm acis}|\hbox{\rm hrc}\}$fJ${\it hhmmsss}\{\hbox{\rm p}|\hbox{\rm m}\}{\it ddmmss}\_{\it nnn}$'' \\
region\_id							& Detection region identifier (component number) \\
match\_type						& Type of match between master source and stacked observation detection; u$=$unique match, a$=$ambiguous match, n$=$non-detection \\
\enddata
\end{deluxetable*}
}

{
\setlength{\descripwidth}{444pt}
\renewcommand\arraystretch{0.9}
\begin{deluxetable*}{lp{\descripwidth}}
\tabletypesize{\scriptsize}
\tablecolumns{2}
\tablecaption{Stacked Observation Detection/Per-Observation Detections Associations Table Properties\label{tab:stackobiassoc}}
\tablehead{
\noalign{\vskip 2.5ex}
\colhead{Property} & \colhead{Description} \\
\noalign{\vskip -2.5ex}
}
\startdata
detect\_stack\_id					& Detect stack identifier (designation of observation stack used for source detection) in the format ``$\{\hbox{\rm acis}|\hbox{\rm hrc}\}$fJ${\it hhmmsss}\{\hbox{\rm p}|\hbox{\rm m}\}{\it ddmmss}\_{\it nnn}$'' \\
obsid							& Observation identifier \\
obi								& Observation interval number \\
cycle								& ACIS readout cycle for the observation, `P' (primary) or `S' (secondary) for alternating exposure (interleaved) mode observations, or `P' for other ACIS modes \\
region\_id							& Detection region identifier (component number) \\
\enddata
\end{deluxetable*}
}

{
\setlength{\descripwidth}{444pt}
\renewcommand\arraystretch{0.9}
\begin{deluxetable*}{lp{\descripwidth}}
\tabletypesize{\scriptsize}
\tablecolumns{2}
\tablecaption{Detect Stack Table Properties\label{tab:detectstack}}
\tablehead{
\noalign{\vskip 2.5ex}
\colhead{Property} & \colhead{Description} \\
\noalign{\vskip -2.5ex}
}
\startdata
detect\_stack\_id					& Detect stack identifier (designation of observation stack used for source detection) in the format ``$\{\hbox{\rm acis}|\hbox{\rm hrc}\}$fJ${\it hhmmsss}\{\hbox{\rm p}|\hbox{\rm m}\}{\it ddmmss}\_{\it nnn}$'' \\
obsid							& Observation identifier \\
obi								& Observation interval number \\
cycle								& ACIS readout cycle for the observation, `P' (primary) or `S' (secondary) for alternating exposure (interleaved) mode observations, or `P' for other ACIS modes \\
\enddata
\end{deluxetable*}
}

{
\setlength{\descripwidth}{444pt}
\renewcommand\arraystretch{0.9}
\begin{deluxetable*}{lp{\descripwidth}}
\tabletypesize{\scriptsize}
\tablecolumns{2}
\tablecaption{Valid Stack Table Properties\label{tab:validstack}}
\tablehead{
\noalign{\vskip 2.5ex}
\colhead{Property} & \colhead{Description} \\
\noalign{\vskip -2.5ex}
}
\startdata
detect\_stack\_id					& Detect stack identifier (designation of observation stack used for source detection) in the format ``$\{\hbox{\rm acis}|\hbox{\rm hrc}\}$fJ${\it hhmmsss}\{\hbox{\rm p}|\hbox{\rm m}\}{\it ddmmss}\_{\it nnn}$'' \\
region\_id							& Detection region identifier (component number) \\
obsid							& Observation identifier \\
obi								& Observation interval number \\
cycle								& ACIS readout cycle for the observation, `P' (primary) or `S' (secondary) for alternating exposure (interleaved) mode observations, or `P' for other ACIS modes \\
\enddata
\end{deluxetable*}
}

%This shouldn't be necessary but AASTeX leaves a huge gap if it's not present
\vskip -60pt

The {\em Valid Stack\/} table (Table~\ref{tab:validstack}) records for each stacked observation detection the subset of observations within the stack for which the detection falls within the observation pixel mask (see \S~\ref{sec:pixmask} for more detail) that defines the valid pixels within the observation field of view.

Similar to the Valid Stack table, the {\em Likely Stack\/} table (Table~\ref{tab:likelystack}) records for each stacked observation detection the subset of observations within the stack that were used by the maximum likelihood estimator analysis (\S~\ref{sec:mle}) to validate the candidate detection.  For CSC~2, source detection is performed both on the full valid stack and on each of the individual observations that comprise the valid stack, so the Likely Stack table record corresponding to a stacked observation detection will include either the entire set of observations included in the valid stack, or the single observation for which the likelihood was maximized.  The latter case corresponds primarily to detections for which the source intensity flared in a single observation.  Figure~\ref{fig:stacks} illustrates the difference between the detect, valid, and likely stacks for a set of conceptual stacked observation detections.

Finally, the {\em Limiting Sensitivity\/} table (Table~\ref{tab:limsens}) records the catalog limiting sensitivity, defined as the minimum flux in an energy band required for a point source to be detected and classified as either \marginalvalue\ or \truevalue\ by the source detection process in that energy band, as a function of position on the sky.
{
\setlength{\descripwidth}{444pt}
\renewcommand\arraystretch{0.9}
\begin{deluxetable*}{lp{\descripwidth}}
\tabletypesize{\scriptsize}
\tablecolumns{2}
\tablecaption{Likely Stack Table Properties\label{tab:likelystack}}
\tablehead{
\noalign{\vskip 2.5ex}
\colhead{Property} & \colhead{Description} \\
\noalign{\vskip -2.5ex}
}
\startdata
detect\_stack\_id					& Detect stack identifier (designation of observation stack used for source detection) in the format ``$\{\hbox{\rm acis}|\hbox{\rm hrc}\}$fJ${\it hhmmsss}\{\hbox{\rm p}|\hbox{\rm m}\}{\it ddmmss}\_{\it nnn}$'' \\
region\_id							& Detection region identifier (component number) \\
obsid							& Observation identifier \\
obi								& Observation interval number \\
cycle								& ACIS readout cycle for the observation, `P' (primary) or `S' (secondary) for alternating exposure (interleaved) mode observations, or `P' for other ACIS modes \\
\enddata
\end{deluxetable*}
}

{
\setlength{\descripwidth}{304pt}
\renewcommand\arraystretch{0.9}
\begin{deluxetable*}{lccp{\descripwidth}}
\tabletypesize{\scriptsize}
\tablecolumns{4}
\tablecaption{Limiting Sensitivity Table Properties\label{tab:limsens}}
\tablehead{
\noalign{\vskip 2.0ex} \colhead{Property} & \colhead{Multi-Band\tablenotemark{\scriptsize a}} & \colhead{Units} & \colhead{Description} \\
}
\startdata
healpix                                                        & No  &                                                               & HEALPIX number (NESTED Scheme)\tablenotemark{\scriptsize b} \\
flux\_sens							& Yes & $\rm erg\,cm^{-2}\,s^{-1}$			& Estimated aperture-corrected net energy flux from the PSF 90\% ECF region aperture required for a source to be detected and classified as \marginalvalue \\
flux\_sens	\_true					& Yes & $\rm erg\,cm^{-2}\,s^{-1}$			& Estimated aperture-corrected net energy flux from the PSF 90\% ECF region aperture required for a source to be detected and classified as \truevalue \\
photflux\_sens						& Yes & $\rm photon\,cm^{-2}\,s^{-1}$		& Estimated aperture-corrected net photon flux from the PSF 90\% ECF region aperture required for a source to be detected and classified as \marginalvalue \\
photflux\_sens\_true					& Yes & $\rm photon\,cm^{-2}\,s^{-1}$		& Estimated aperture-corrected net photon flux from the PSF 90\% ECF region aperture required for a source to be detected and classified as \truevalue \\
\enddata
\tablenotetext{a}{Indicates that tabulated properties include separate entries
  for each energy band.  The individual band entries are identified by the
  suffix ``\_$\canon{x}$'', where $\canon{x}$ is one of the energy band
  designations.}
\vskip -6pt
\tablenotetext{b}{CXC-provided user interfaces automatically translate HEALPIX number into ICRS Right Ascension and Declination, and vice-versa.}
\end{deluxetable*}
}

\section{CATALOG GENERATION}
This section describes the methods used to generate release 2 of the catalog and the algorithms for computing source and detection properties where they are new or differ from those described in \paperI.

\begin{figure}
\epsscale{1.175}
\plotone{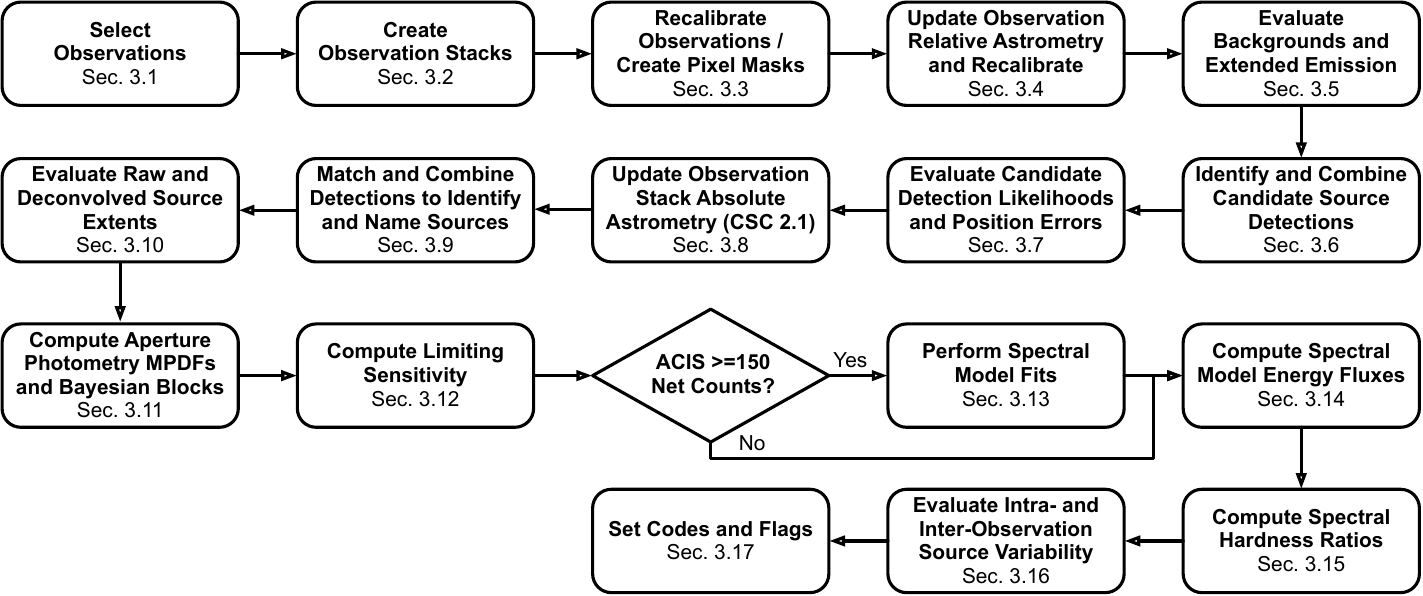}
\caption{\label{fig:flowchart}
High-level flow of the catalog processing steps.  The references identify the relevant sections of the text that describe the methods and algorithms used.}
\end{figure}

The overall flow of catalog processing consists of the steps that are shown in Figure~\ref{fig:flowchart}.  In the figure, each block references the section of the text that describes in detail the methods used.  First, \Chandra\ imaging observations that meet a defined set of criteria for inclusion in the catalog are identified and organized into observation stacks for co-adding.  The observations are recalibrated to ensure that the most current calibrations are applied, and then a preliminary source detection step is performed to identify bright X-ray sources present in each observation.  Detected sources are cross-matched between multiple observations that are part of the same stack, and those matches are used to compute the relative astrometric offsets necessary to align the individual observations prior to stacking them.  Background maps are then created for each observation \addedtext{and observation stack} using a Voronoi tessellation technique.  Regions of bright, extended ($\gtrsim\!30''$) X-ray emission are also identified. Next, candidate compact- ($\lesssim\!30''$) and point-sources are detected in each stacked observation using two different algorithms, and the detections are combined.  The likelihood of each candidate detection is evaluated against a set of thresholds to classify the detection as \truevalue, \marginalvalue, or \falsevalue\null.  Once the detections from all spatially overlapping observation stacks are evaluated, they are cross-matched to identify distinct X-ray sources on the sky.  A set of commonly useful source properties are then computed for each detection at both the individual observation and stacked observation levels, and best estimates of the properties are evaluated for each identified distinct X-ray source. These properties include spatial extents, aperture photometry, limiting sensitivity, spectral model fits, integrated fluxes using several canonical spectral models, hardness ratios, intra- and inter-observation temporal variability.  Finally, a set of codes and flags that identify interesting circumstances ({\em e.g.\/}, source is extended, or variable) or human-intervention, are added.

Data processing for CSC 2.0 was performed using \Chandra\ X-ray Center (CXC) data system \citep[CXCDS;][]{2006SPIE.6270E..1UE,2006SPIE.6270E..0NE} catalog processing system versions CAT\thinspace4.3--CAT\thinspace4.7 and calibration data extracted from \Chandra\ Calibration Database \citep[CalDB;][]{2006SPIE.6270E..1XG,2007ASPC..376..335E} version 4.6.7.  For CSC 2.1, catalog processing system version CAT\thinspace5.4 and CalDB 4.9.7 were used for data processed post-CSC~2.0.

\subsection{Observation Selection} \label{sec:obsel}
There are many filters that determine which data sets are included in, or excluded from, each release of the CSC.

Release 2.0 includes most imaging observations obtained using either the ACIS or HRC-I instruments that were released publicly prior to 2015 January 01, 00:00:00 TT\null.  Because most general observer programs from this period included a one year proprietary data period, this cutoff means that most observations included in the catalog were obtained prior to 2014 January 01.  However, some observations such as calibration observations or Director's Discretionary Time observations typically have no proprietary period and so the catalog does include some observations obtained during calendar year 2014.   For CSC 2.1, the public release cutoff date is 2022 January 01, 00:00:00 TT\null.

For the ACIS instrument, ``timed-exposure'' readout mode observations obtained using either the ``faint,'' ``very faint,'' ``faint with bias,'' or ``graded'' datamodes are included.  All ACIS ``continuous-clocking'' readout mode observations, and any observations obtained using a datamode other than those listed above, are excluded from CSC~2\null.  If an observation was obtained using ACIS ``alternating exposures'' (interleaved short and long frame-time exposures), then only the long frame-time exposures are considered, and the short frame-time exposures are ignored.  HRC-I imaging mode observations are included in CSC~2, but HRC-S observations continue to be excluded because of the presence of background features associated with the edges of ``T''-shaped energy-suppression filter regions that form part of the HRC-S UV/ion-shield.

Observations obtained using either of the insertable transmission gratings (high-energy transmission grating [HETG], or low-energy transmission grating [LETG]) are excluded from the catalog, as are ``moving target'' (solar system object) observations.  Furthermore, observations obtained with ${\rm defocus} < -0.1\,{\rm mm}$ or $> +0.2\,\rm{mm}$ are excluded, as are observations obtained with large science instrument module $Z$ offsets from the nominal instrument-specific locations (specifically, ${\rm sim\_z\_offset} < -20.0\,{\rm mm}$ or $> +20.0\,\rm{mm}$).  Experience obtained from release 1 of the CSC demonstrates that observations in the latter two categories have significantly poorer astrometric and other calibrations than is typical, and very few observations fall into these categories.

All observation datasets included in CSC~2 must have been processed through CXC data system (CXCDS) release DS~8.4.2 or later to ensure the availability of all input data products required by the catalog processing pipelines.  This cutoff corresponds to datasets that were processed through the 4th bulk reprocessing of \Chandra\  data (or later).  As a consequence, observations obtained prior to January 29, 2000 are not included in the catalog, with the exception of four HRC-I observations (\obsid s 00243, 00255, 00267, and 00279).  In particular this means that early observations obtained with ACIS focal plane temperatures higher than $-120\,^\circ{\rm C}$ are not present in release 2.  All processed datasets must have undergone successful ``Validation and Verification'' quality assurance checks as part of their latest processing.

The following observations that otherwise meet all of the other criteria are excluded from CSC~2: \obsid s 00984, 00985, 01996, 02000, and 02594.

Unlike release 1 of the CSC, there are no restrictions placed on observations that include bright, spatially-extended emission within the field of view.  The entire field of view of such observations is included in release 2.  In a very small number of cases, ``exclusion regions'' are created that reject all candidate detections from specific regions of a stacked observation field-of-view. These regions typically surround areas that have ACIS ``exclude'' windows in the individual observations comprising the stacked observation (so that no X-ray photons will be detected in these regions), or that are completely saturated by a region of extremely bright, extended emission. The stacked observation has no detection sensitivity within the exclusion region, and this is reflected in the limiting sensitivity maps.  The following observation stacks include one or more exclusions regions (see \S~\ref{sec:stackgen} for the observation stack naming scheme): {\tt acisfJ0534316p220052\_001}, {\tt acisfJ0534322p215837\_001}, {\tt acisfJ0955234p690445\_001}, and {\tt acisfJ2144487p382120\_001}.

\subsection{Observation Stack Creation} \label{sec:stackgen}
Separately for each instrument (ACIS or HRC-I) the set of observations selected in \S{\ref{sec:obsel}} for CSC 2.0 are grouped together to form observation stacks using the {\tt  treecluster} hierarchical clustering algorithm described by \citet{2004Bioinformatics.20.1453D} with pairwise complete linkage.  The metric distance between each pair of observations is computed as the great circle separation between their pointings, where the pointing of an observation is defined as the world coordinates $({\tt ra\_pnt}, {\tt dec\_pnt})$ of the computed optical axis of the telescope projected on the sky.  Each cluster of observations with a maximum metric separation of $60''$ is defined to be a separate observation stack.  The constraint on the maximum separation of the pointings ensures that the radial scale of the local PSFs are similar to each other for all observations as a function of off-axis angle across the entire field of view of the observation stack.  The roll angle of the telescope is {\em not\/} considered when constructing observation stacks, even though the detailed structure of the PSF depends on the azimuthal angle relative to the telescope structure for off-axis PSFs.

Observation stacks are named using the format ${\tt <\!\!inst\!\!>\!\!fJ}hhmmsss\{{\tt p|m}\}ddmmss{\_}nnn$, where ${\tt <\!\!inst\!\!>}$ is either {\tt acis} or {\tt hrc} depending on the instrument, $hhmmsss$ is the stack right ascension (truncated to the nearest $0.1''$), $\{{\tt p|m}\}$ is the character ${\tt p}$ or ${\tt m}$, depending on whether the stack declination is positive or negative, $ddmmss$ is the stack declination, and $nnn$ is the stack version number.  The latter is initialized to ``{\tt 001}'' and is incremented only if the list of observations included in the stack is modified to differentiate newly released stacked observation data from a previously published version.  The stack right ascension and declination, $({\tt ra\_stk}, {\tt dec\_stk})$ is the exposure-weighted average of the individual $({\tt ra\_pnt}, {\tt dec\_pnt})$ values for the observations included in the stack.

Observation stacks for release 2.1 are formed as described above, with one difference.  To minimize changes to stack definitions, each observation newly added in CSC 2.1 is first evaluated to determine whether there is an existing stack definition to which the new observation can be added while still meeting the $60^{\prime\prime}$  maximum metric separation criterion.  If so, that observation is added to the existing stack.  Any stack modified in this manner retains the previous stack instrument and position designation and has the stack version number incremented to ``{\tt 002}''\null.  After all observations that can be added to existing observation stacks are considered, new stacks are formed from the remaining observations using the {\tt treecluster} algorithm as previously described.  This approach avoids the possibility of splitting existing observation stacks into multiple new stacks when new observations are considered, which simplifies processing for the incremental release.  As a consequence of this approach, observation stacks in CSC 2.1 may be split into three categories: (a)~unmodified stacks, which are unchanged from CSC 2.0; (b)~modified stacks, which are stacks from CSC~2.0 with added observations; and (c)~new stacks, which are made up entirely of observations that were not included in CSC 2.0.

\begin{figure}
\epsscale{0.85}
\plotone{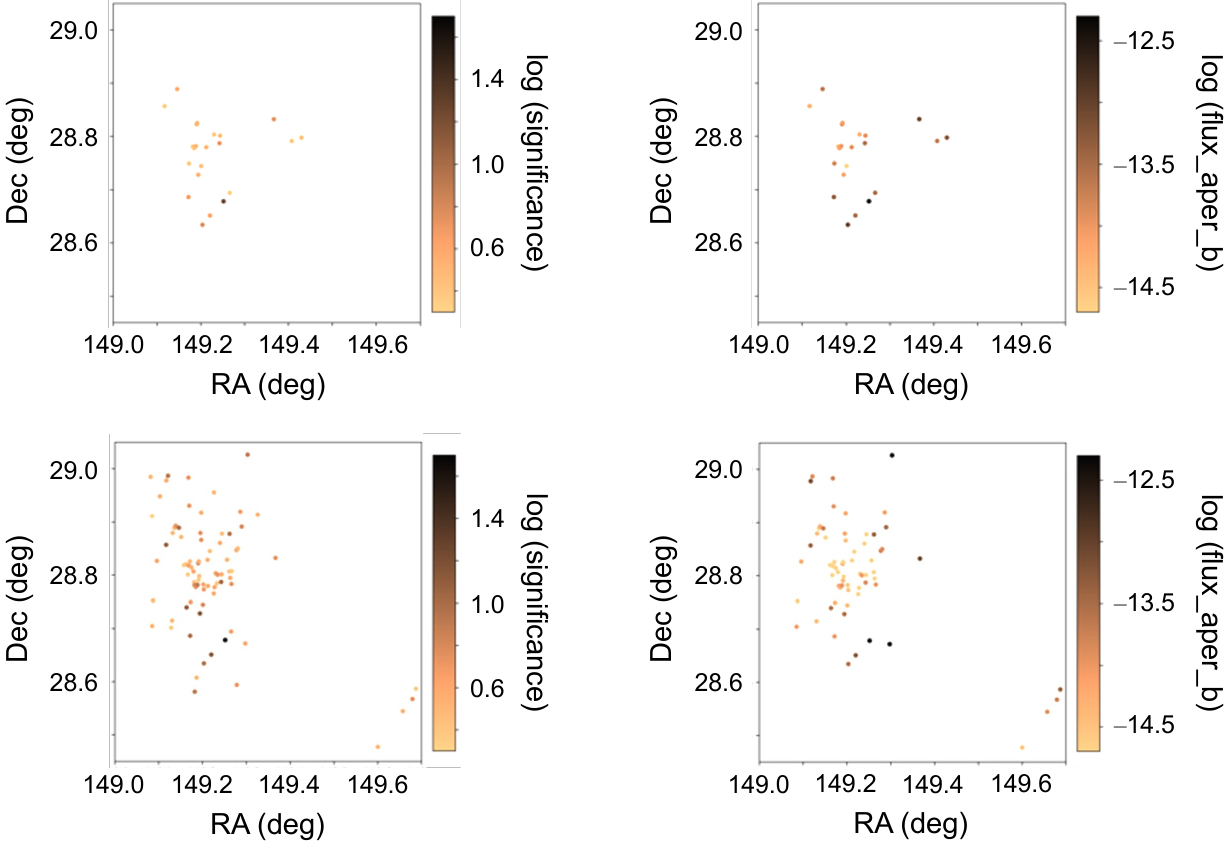}
\caption{\label{fig:modstack}
The upper panels show CSC 2.0 detections in stack {\tt acisfJ0956451p284947\_001}, which includes a single observation, color-coded by master source significance (left) and master source broad-band flux (right).  The corresponding plots for the modified stack in CSC 2.1, {\tt acisfJ0956451p284947\_002}, which includes five observations, are shown in the lower panels.  While the gain in the total number of detections is clearly evident, some detections present in CSC 2.0, for example at approximately $(\alpha, \delta) = (149.42, 28.79)$, are no longer detected in CSC 2.1.  This is due to detection thresholds \addedtext{(see \S~\ref{sec:lthresh})} changing as a result of the newly added observations.
}
\end{figure}

The addition of new observations in a modified stack may alter the region of the sky covered by the observation stack since the new observations may extend beyond the CSC 2.0 boundaries of the original stack.  Even in regions where the new observations fully overlap the original stack, the additional data can affect source detectability, photometric, spectral, and temporal variability properties.  While the additional observations will often result in the detection of sources that were previously undetectable in the original stack, in rare cases previously detected sources can become undetectable in the modified stack due to the increased background level arising from adding the new observations ({\em e.g.\/}, Fig~\ref{fig:modstack}).  In some cases the modified spatial distribution of counts can result in a single prior detection being split into more than one detection, or a close pair of detections being merged into a single detection.

\subsection{Observation Recalibration} \label{sec:recal}
For the reasons described in \paperI, all observations included in the CSC are reprocessed through the instrument-specific calibration steps of the CXCDS standard data processing (SDP) pipelines as the first step in catalog construction.  The recalibration steps performed for CSC~2 are similar to those described therein and use the same tools as the routine SDP pipelines used to process science data received from the satellite and included in the CIAO portable data analysis system \citep{2006SPIE.6270E..1VF}, principally \ape\ for ACIS data and \hpe\ for HRC data.  Calibration data are extracted from  \Chandra\  CalDB, which is held static for all observations processed in a given release to ensure that catalog detections have a uniform photometric calibration.  \addedtext{Unless otherwise noted in the text}, all  \addedtext{spatially binned} data products are created at single-pixel \addedtext{($0\farcs4920$)} resolution for ACIS observations and two-pixel \addedtext{($0\farcs2636$)} resolution for HRC-I observations. 

For ACIS observations, the main calibrations performed are the time-dependent gain calibration and the correction for CCD charge transfer inefficiency (CTI).  Observation-specific bad pixels, and ``streak'' events on CCD S4 (ACIS-8) are flagged with bad status for removal.  Hot pixels and pixel afterglow events are similarly flagged for removal using the CIAO {\tt acis\_find\_afterglow} tool.  This tool, which was developed after the previous release of the catalog, identifies afterglow events and hot-pixels by evaluating whether there is a statistically significant excess of events compared to the expected number of background events, and effectively identifies all afterglows that include four or more photon events.  The principal HRC-specific calibration step is application of the ``degapping'' correction to correct photon event positions for instrumental distortions caused by the readout electronics.  Average dead time corrections are calculated for HRC datasets.  Standard filtering is applied to the photon event data to include only ``Good Time Intervals'' (GTIs) that include scientifically valid data, and to exclude photon events with bad event grades (ACIS only) or that have been flagged with bad status.

Similar to \paperI, more aggressive background event screening is applied as part of the catalog observation recalibration step than is included in routine SDP, significantly reducing the total background and improving source detection efficacy for observations that include time intervals during which the background is flaring.  The method used to screen background events is similar to that used in CSC release~1.  Background regions are identified by constructing a histogram of pixel counts and rejecting all pixels that have a value greater than 3 standard deviations above the mean.  An optimally-binned light curve of the background region counts is created using the Gregory-Loredo algorithm \citep[][see \S~\ref{sec:intravar}]{1992ApJ...398..146G}, and time bins for which the ratio of the total background count rate, $\mathcal{B}_T$, to the quiescent background count rate (estimated as the minimum light curve value), $\mathcal{B}$, exceeds 15 are identified.  The corresponding intervals are considered to be background flares, and the GTIs are revised to exclude those periods.  This entire analysis is performed separately for each detector chip (ACIS CCD or HRC microchannel plate).  The CSC~2 background rate threshold $\mathcal{B}_T/\mathcal{B} > 15$ is optimized to improve the detectability of faint sources near $\sim\!5$ net counts while not signficantly impacting the S/N of brighter sources, and is somewhat more relaxed than the corresponding release~1 criterion ($\mathcal{B}_T/\mathcal{B} > 10$) where the detection threshold was roughly a factor of two higher.  For more details see \citet{nowak_bkgfilt}.

For each observation, the recalibrated photon event list and several other per-observation full-field and ancillary data products are  included in the set of available catalog FITS format data products (see Table~\ref{tab:dataprods}).

{
\setlength{\descripwidth}{289pt}
\renewcommand\arraystretch{0.9}
\startlongtable
\begin{deluxetable*}{llp{\descripwidth}}
\tabletypesize{\scriptsize}
\tablecolumns{3}
\tablecaption{Catalog FITS Format Data Products\label{tab:dataprods}}
\tablehead{
\noalign{\vskip 2.5ex}
\colhead{Data Product} & \colhead{Suffix} & \colhead{Description} \\
\noalign{\vskip -2.5ex}
}
\startdata
\cutinhead{Master Source Data Products}	
Bayesian Blocks							& {\tt blocks3}	& Aperture photometry-derived properties for each flux- and time-based Bayesian Block \\
Aperture Photometry							& {\tt phot3}	& Marginalized probability density functions for the combined master-source level photometry properties ({\em i.e.\/}, for the ``best'' Bayesian block) \\
Extended Source Region Polygons				& {\tt poly3}	& Polygons defining extended convex hull sources at the master-source level \\
\cutinhead{Stack Detection Data Products}
Source Region Event List						& {\tt regevt3}	& Photon event list, with associated GTIs recorded in consecutive FITS HDUs \\
Source Region Image						& {\tt regimg3}	& Per-energy-band background-subtracted, exposure corrected images ($\rm \hbox{photon}\,\hbox{cm}^{-2}\,\hbox{s}^{-1}$) \\
Source Region Exposure Map					& {\tt regexp3}	& Per-energy-band exposure map images ($\rm \hbox{cm}^2\,\hbox{s}\,\hbox{photon}^{-1}$) computed at the band monochromatic effective energy \\
Source Region								& {\tt reg3}	& Modified source region aperture and background region aperture region definitions \\
Source Region MLE Position Draws				& {\tt draws3}	& MCMC point source and compact extended source model position error draws from the maximum likelihood estimator detection position fit \\
Aperture Photometry							& {\tt phot3}	& Marginalized probability density functions for the combined photometry properties for the observation included in the valid stack \\
\cutinhead{Observation Detection Data Products}
Source Region Event List						& {\tt regevt3}	& Photon event list, with associated GTIs recorded in consecutive FITS HDUs \\
Source Region Image						& {\tt regimg3}	& Per-energy-band background-subtracted, exposure corrected images ($\rm \hbox{photon}\,\hbox{cm}^{-2}\,\hbox{s}^{-1}$) \\
Source Region Point Spread Function			& {\tt psf3}		& Per-energy-band local model PSF images computed at the band monochromatic effective energy \\
Source Region Exposure Map					& {\tt regexp3}	& Per-energy-band exposure map images ($\rm \hbox{cm}^2\,\hbox{s}\,\hbox{photon}^{-1}$) computed at the band monochromatic effective energy \\
Source Region PI Spectrum					& {\tt pha3}	& (ACIS-only) Per-energy-band pulse-invariant source region aperture and background region aperture spectra, with associated GTIs, in consecutive FITS HDUs \\
Source Region ARF							& {\tt arf3}		& Ancillary response file; table of telescope plus detector effective area ($\rm\hbox{cm}^2$) vs. energy bins \\
Source Region RMF							& {\tt rmf3}	& (ACIS-only) Detector redistribution matrix file \\
Source Region Light Curve					& {\tt lc3}		& Per-energy-band optimally binned light curve, computed using the Gregory-Loredo formalism \\
Source Region								& {\tt reg3}	& Modified PSF 90\% ECF region aperture region definitions \\
Source Region MLE Position Draws				& {\tt draws3}	& MCMC point source and compact extended source model position error draws from the maximum likelihood estimator detection position fit \\
Aperture Photometry							& {\tt phot3}	& Marginalized probability density functions for the photometry properties for the observation \\
\cutinhead{Stack Full Field Data Products}
Stack Event List							& {\tt evt3}		& Photon event list, with associated GTIs recorded in consecutive FITS HDUs \\
Stack Image								& {\tt img3}	& Per-energy-band background-subtracted, exposure corrected images ($\rm \hbox{photon}\,\hbox{cm}^{-2}\,\hbox{s}^{-1}$) \\
Stack Background Image						& {\tt bkgimg3}	& Per-energy-band background image generated by {\tt mkvtbkg} (counts) \\
Stack Exposure Map							& {\tt exp3}	& Per-energy-band exposure map images ($\rm \hbox{cm}^2\,\hbox{s}\,\hbox{photon}^{-1}$) computed at the band monochromatic effective energy \\
Stack Field of View							& {\tt fov3}		& Observation stack sky field-of-view region definition \\
Stack Limiting Sensitivity						& {\tt sens3}	& Observation stack limiting sensitivity map \\
Stack Merged Source Detection List				& {\tt mrgsrc3}	& Candidate-detection list including detection classifications and original detection algorithm properties \\
\cutinhead{Observation Full Field Data Products}
Observation Event List						& {\tt evt3}		& Photon event list, with associated GTIs recorded in consecutive FITS HDUs \\
Observation Image							& {\tt img3}	& Per-energy-band background-subtracted, exposure corrected images ($\rm \hbox{photon}\,\hbox{cm}^{-2}\,\hbox{s}^{-1}$) \\
Observation Background Image				& {\tt bkgimg3}	& Per-energy-band background image generated by {\tt mkvtbkg} (counts) \\
Observation Exposure Map					& {\tt exp3}	& Per-energy-band exposure map images ($\rm \hbox{cm}^2\,\hbox{s}\,\hbox{photon}^{-1}$) computed at the band monochromatic effective energy \\
Observation Aspect Solution					& {\tt asol3}	& Aspect solution (spacecraft dither position vs. time) including astrometric corrections \\
Observation Aspect Histogram					& {\tt ahst3}	& Table of $X$, $Y$ offsets (pixels) and roll-angle offsets (deg) vs. time due to spacecraft dither motion \\
Observation Bad Pixel Regions					& {\tt bpix3}	& Detector bad pixel region definitions, including observation-specific bad pixels \\
Observation Field of View						& {\tt fov3}		& Individual observation sky field-of-view definition \\
Observation Pixel Mask						& {\tt pixmask3}	& Individual observation pixel mask identifying pixels considered valid and included in catalog processing \\
Observation Extended Source Region Polygons	& {\tt poly3}	& Polygons defining extended sources at the individual observation level \\
\enddata
\end{deluxetable*}
}

% \vfil avoids the section heading being at the bottom of the page
\vfil

\subsubsection{Pixel Mask} \label{sec:pixmask}
To avoid an excessive false detection rate seen in release 1 of the catalog \citep{2011ApJS..194...37P} that is associated with either the edges of the field of view where the total exposure is changing rapidly on the scale of a few pixels, or with the inter-chip gaps between adjacent ACIS front- and back-illuminated CCDs for which the sensitivities and background rates differ significantly, in CSC 2 we introduce the concept of a {\em pixel mask\/} for each observation that defines the set of pixels in that observation's ``sky'' \citep[\hbox{[X, Y]};][]{mcdowell_coordsI} coordinate system that are considered in catalog processing.  Detections located on sky pixels that fall outside of an observation's pixel mask are not included in the catalog and consequently those sky pixels effectively have no detection sensitivity.  Pixel mask data products have pixel values set to 1 if the sky pixel has non-zero valid exposure; otherwise the pixel mask pixel value is nominally set to 0 (in some cases the IEEE 754 NaN [Not a Number] value is used \addedtext{instead of 0, with identical meaning}).

The pixel mask for an observation is constructed initially by comparing that observation's {\em normalized\/} exposure map to the normalized {\em ideal\/} exposure map for the observation.  An exposure map is normalized by dividing the map by its peak value.  The ideal exposure map is based on a detector model with uniform 100\% quantum efficiency and no contamination, although the effects of X-ray vignetting and spacecraft dither {\em are\/} included.  Sky pixels for which the normalized exposure map value is $\geq 90\%$ of the normalized ideal exposure map value are marked as valid (set to 1) in the pixel mask while sky pixels with lower exposure map values are marked as invalid.

Pixel masks computed in this matter are subsequently adjusted for some common circumstances.  For ACIS, pixels in the inter-chip gaps between adjacent ACIS-I CCDs (ACIS-0 through ACIS-3) are marked as valid in the pixel mask, since the default spacecraft dither permits some source detectability across the inter-chip gaps.  This is not done for the inter-chip gaps between adjacent ACIS-S CCDs (ACIS-4 through ACIS-9) because 4 of the 5 chip gaps occur between front-illuminated and back-illuminated CCDs and the background rates on the two types of CCDs are significantly different.  ACIS pixel masks are also adjusted to exclude the regions surrounding readout streaks (see \S~\ref{sec:rostreak}) generated by bright detections.  The readout streak regions are broken around those detections so that the detections primarily responsible for the streaks are not themselves excluded from consideration by the pixel mask.

Finally, any exclusion regions (see \S~\ref{sec:obsel}) are also excluded from the pixel mask since no useful X-ray events are present.  The per-observation pixel mask is recorded in the {\tt pixmask3} data product.

\subsection{Observation Relative Astrometry} \label{sec:ofa}
For each observation stack that includes more than one observation interval, additional steps are performed to ensure that all of the observations are registered to a common astrometric frame prior to stacking.

Once the individual observations included in the stack are recalibrated, a preliminary source detection step is performed separately on each observation using the CIAO \wavdetect\ wavelet-based source detection algorithm \citep{2002ApJS..138..185F} to identify bright point sources within the field of view.  The \wavdetect\ tool is run as described in \S~\ref{sec:wavdet}, except that the limiting significance level $S_0$ is set to $1.0\times10^{-7}$ (corresponding to $\sim\!0.5$ false sources due to random fluctuations per $2048\times2048$ pixel image), and the local background level is estimated internally by the algorithm.

Since all of the individual observations included in a single stack have telescope pointings that are co-located within a circle with diameter $60''$, the relative astrometric corrections are computed on a common tangent plane.  All of the observations are first reprojected onto a single  tangent plane that is referenced off the median of the individual observations' tangent plane reference positions.  

The relative astrometric corrections to align the individual observation datasets are computed on a pairwise basis.  The observation with the highest number of detected sources that have  ${\rm S/N} \geq 10$ is designated the ``reference'' observation, and all other observations are separately aligned with the reference observation.  The CIAO tool {\tt wcs\_match} is used to perform the matching.  This tool works by minimizing the tangent plane position differences between matched detections in a least-squares sense ({\em i.e.}, the sum in quadrature of the residuals is minimized.)

A detection in one of the observations is only considered if that detection is uniquely matched to a single detection in the other observation, and the separation between the two measured positions is not greater than $2''$.  Since the rotation angle and plate scale of the \Chandra\  detectors are well determined,\footnote{\url{https://cxc.cfa.harvard.edu/cal/ASPECT/celmon/}} only translations are allowed.  An iterative process is used to determine the optimal translation required to map the second observation to the same astrometric frame as the reference observation.  At each step, the algorithm computes the two-axis translation that minimizes the sum in quadrature of the residuals between the sets of matched detections.  If the maximum residual between any matched pair of detections is $\leq0\farcs5$, then the algorithm has converged and the optimal translation has been found.  Otherwise, the matched pair of detections with the largest residual is discarded and another iteration is performed.

To ensure that the computed astrometric correction is satisfactory, any pair of observations for which there are $\leq3$ matched pairs of detections remaining after the algorithm converges are sent for quality assurance review by a human.  Human quality assurance review is also triggered if the remaining RMS residual error after convergence is $>0\farcs3$ for ACIS observations ($>0\farcs2$ for HRC observations), or if the computed translation required to align the observations is $>0\farcs5$ for ACIS observations ($>0\farcs3$ for HRC observations).  The review process allows a trained individual to execute the {\tt wcs\_match} tool interactively, manually select pairs of detections to be included in/excluded from the solution, and decide when convergence is reached.

Once the pairwise astrometric corrections are computed for all of the observations included in the stack, the individual observation translations are updated to conform to an astrometric zero-point position that is computed from the exposure-weighted mean of the computed corrections, in order to remove any position bias introduced by aligning all observations to the reference observation.  The stack tangent plane reference position is also updated accordingly.

The computed astrometric correction for each observation is then applied to that observation's aspect solution.  The updated aspect solution is recorded in the CSC~2 aspect solution ({\tt asol3}) data product for the observation, and a transformation matrix representation of the applied astrometric correction is stored in the {\tt XFM} (``transform'') Hierarchical Data Unit (HDU) included in the {\tt asol3} file.

Finally, each observation is recalibrated as described in \S~\ref{sec:recal}, with the updated aspect solution applied in the \ape\ or \hpe\ step (as appropriate) to correct the astrometry. 

\subsection{Background Map} \label{sec:bmc}
The background maps used for automated source detection in the first release of the CSC were created using a modified Poisson mean to construct a low spatial frequency representation of the smooth background separately for each observation, following the precepts of \citet{2008ASPC..394..559M}.

For observations obtained using the ACIS instrument, the impact of ``readout streaks'' was addressed by combining the low spatial frequency background with an additional high spatial frequency background using a technique developed by \citet{2005ASPC..347..478M}.  Readout streaks arise from source photons that are detected during the ACIS CCD readout frame transfer interval ($\sim\!40\,\mu$s per row) following each exposure ($\sim\!3.2\,$s per exposure for a typical observation).  These ``out-of-time'' events effectively expose all pixels along a given readout column to all points on the sky that lie along that column during the frame transfer interval.  As a result, columns including bright X-ray sources have enhanced count rates along their length.

The release~1 approach creates high-quality backgrounds in fields where the background intensity is approximately uniform spatially.  However,  the low spatial frequency background can introduce artifacts in localized areas surrounding bright compact sources or in regions where the background intensity varies spatially.  This typically results in an overestimate of the local background level surrounding bright sources and a poor representation of the background structure in areas with spatially varying intensity {\em e.g.\/}, Figure~11 of \paperI\null.  While these backgrounds were adequate for release 1 of the catalog, the inclusion of extended emission regions and the  fainter limiting sensitivity of CSC~2 require a more robust approach for background determination.

\subsubsection{Voronoi Tessellation Background}
Release 2 backgrounds are constructed \addedtext{for each observation} using a newly-developed tool, \mkvtbkg\ \citep{houck_mkvtbkg}, that applies a Voronoi tessellation approach based on an extension of the ideas described by \citet{1993PhRvE..47..704E}.  The distribution of background photon events is modeled as a homogeneous Poisson point process that deposits events randomly over the detector area with a uniform spatial distribution.  The Voronoi tessellation finds a unique set of non-overlapping polygons that cover the entire detector region, each of which contains a single photon event.  

\mkvtbkg\ uses the {\tt Triangle} code developed by \citet{shewchuk1996triangle} to construct a Voronoi tessellation in sky $(X, Y)$ coordinates separately for each detector chip (ACIS CCD or HRC-I microchannel plate).  Exposure variations across the field of view are corrected by scaling the geometric area of each polygon $k$ by the local normalized effective area \citep{2001ApJ...548.1010D} determined from the exposure map:
\begin{displaymath}
a_k=a_{{\rm geom},k}A_k,
\end{displaymath}
where $a_{{\rm geom},k}$ is the geometric area of polygon $k$, $A_k\in[0, 1]$ is the normalized effective area, and $a_k$ is the resulting exposure-corrected polygon area.  Rather than work directly with the exposure-corrected polygon areas $a_k$, it is convenient to define a dimensionless reduced area, $\alpha_k = a_k/\bar{a}$, where $\bar{a}$ is a suitably chosen value described below.  

For the largest polygons that are assumed to be associated with photon events drawn from the  background distribution, the statistical distribution of Poisson Voronoi polygon reduced areas, $\alpha$, is well characterized by a generalized Gamma distribution of the form \citep{2003Forma...18..221T}
\begin{equation}
\label{eqn:VParea}
f(\alpha)=\frac{ab^{c/a}}{c(c/a)}\alpha^{c-1}\exp{(-b\alpha^a)},
\end{equation}
where the constants $\{a, b, c\}$ have been determined empirically to be $\{1.07950, 3.03226, 3.31122\}$ (ibid).

An appropriate value for $\bar{a}$ can be determined by iteratively fitting the large area tail of the observed histogram of exposure-corrected polygon areas with a function of the form
\begin{displaymath}
F(\alpha;\mathcal{N}, \bar{a})=\mathcal{N}\!\int\! d\alpha\,f(\alpha),
\end{displaymath}
where $f$ is the distribution of Poisson Voronoi polygon reduced areas, equation~(\ref{eqn:VParea}), and the overall normalization $\mathcal{N}$ depends on the total number of detected photon events.

\begin{figure}
\epsscale{0.55}
\plotone{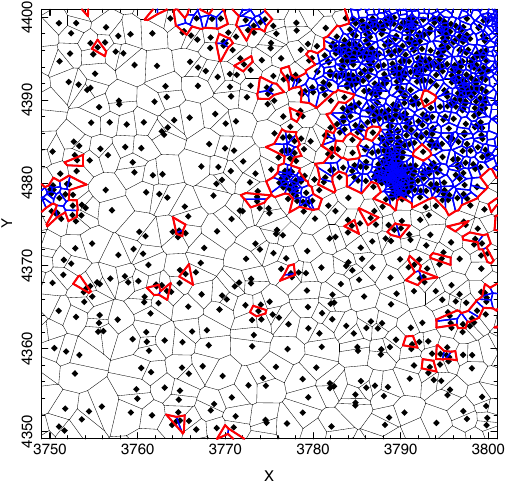}
\caption{\label{fig:srcpolygons}
Voronoi tessellation of a region containing both source and background events.  Voronoi polygons with areas $\alpha < \alpha_{\rm thresh}$ (blue) are identified with sources. Merging Voronoi source polygons yields non-convex source polygons (red).}
\end{figure}

X-ray photon events associated with a detected source are localized spatially, and the resulting polygons are therefore more compact in the vicinity of a detection.  These polygons must be excluded when fitting the background.  Voronoi polygons whose reduced area falls below a certain threshold
\begin{equation}
\label{eqn:alphaThresh}
\alpha < \alpha_{\rm thresh},
\end{equation}
where $\alpha_{\rm thresh}$ is an adjustable threshold parameter, typically $\sim\!0.25$, are associated with candidate source detections, and adjacent polygons are merged to form a source polygon.  Note that the merged source polygons may not be convex and may contain included holes (Fig.~\ref{fig:srcpolygons}).

\subsubsection{ACIS Readout Streak Identification} \label{sec:rostreak}
Regions surrounding ACIS readout streaks are identified from the set of source  polygons using the following procedure.  The polygons are first rotated into a $(x^\prime, y^\prime)$ coordinate system in which the readout direction is parallel to $y^\prime$ axis.  Next, a large number of spatially uniformly-distributed points in $(x^\prime, y^\prime)$ space is generated randomly, and for each $x^\prime$ column (binned at ACIS single pixel resolution) a count of the number of points that fall inside candidate source polygons is recorded.  Columns that are mostly ($\gtrsim\!80\%$) covered by candidate source  polygons are labeled as readout streaks and the associated polygons are flagged as readout streak polygons.

\subsubsection{Creating the Background Image} \label{sec:bkgimg}
Prior to constructing the background image \addedtext{for an observation}, events present in source polygons must be filtered and replaced by a smaller randomly-generated replacement set of uniformly distributed background events.  The set of source polygons is further partitioned into three categories: (1)~{\em small\/} polygons ($\alpha \le\alpha_{\rm large}$), associated with detections of compact sources; (2)~{\em large\/} polygons ($\alpha >\alpha_{\rm large}$), associated with detections of extended sources; and (3)~{\em streak\/} polygons, identified in \S~\ref{sec:rostreak}.  The threshold $\alpha_{\rm large}$ is chosen to distinguish point and compact sources from extended sources such as galaxies.

Events falling in small polygons are replaced with a set of uniformly distributed events that match the spatial density of background events.  \addedtext{The latter is computed as the density of uniformly distributed events such that the mean of their corresponding Voronoi polygon exposure-normalized geometric areas, $a_k$, matches the mean of the exposure-normalized geometric areas for polygons that populate the large-area tail of the observed geometric area distribution in a given observation. This is because the polygons associated with background photons will have the largest geometric areas.  The mean area is determined by fitting the large-area tail of the area distribution with a Gamma function that mimics the exponential decay shape of the tail.  Specifically, fitting  the Gamma function with argument $(a_k/\bar{a})$ is performed iteratively, using the mean area, $\bar{a}$, as a free parameter and updating the range of considered areas in each iteration until convergence is reached.}

For events in large polygons, the spatial density of replacement events is chosen to be large enough to mask the diffuse emission.  This choice allows \wavdetect\ to detect bright embedded point sources sitting on top of the diffuse emission, while simultaneously reducing the number of spurious point source detections associated with the diffuse emission from extended sources.  Low discrepancy Halton sequences are used to generate the replacement events to minimize the possibility of creating local spatial groupings of randomly generated events that could be detected as a faint source.

All of the streak events associated with streak polygons should be included in the background image to reduce the number of spurious detections associated with readout streaks.   However, the streak polygons {\em must\/} include one or more embedded point sources that are responsible for generating the readout streak, so some additional event filtering is necessary.  This is achieved by constructing an image of the events, selecting those pixels that exceed a defined threshold related to the average intensity along the streak, and filtering out enough events to bring the bright pixels below that threshold.

\begin{figure}
\epsscale{1.0}
\plotone{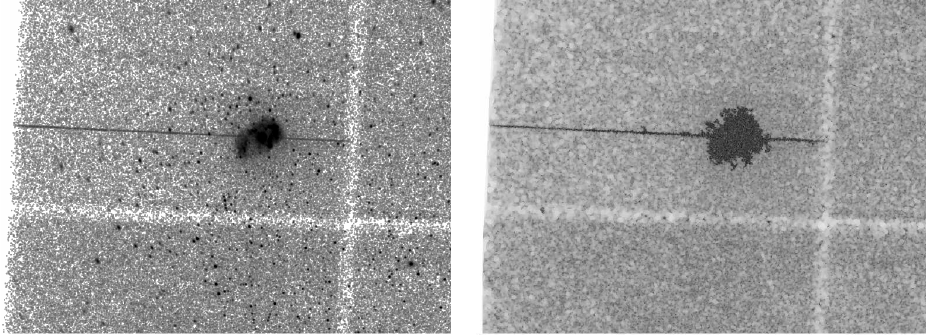}
\caption{\label{fig:bkgimg}
{\em Left:\/} \Chandra\ level 2 event file for \obsid\ 6204, smoothed with a Gaussian kernel with radius 3 pixels. {\em Right:\/} DTFE background image for the same region constructed using the approach described.  The background image contains the bright readout streak and  the polygon containing the galaxy has been filled in with a relatively high uniform surface brightness.  Both images are grey scale, showing pixel values in the range 0.001 (white) to 5 counts (black) on a logarithmic scale.}
\end{figure}

Once the small and large source polygons and the streak polygons have been filtered, the background image is constructed from the Delaunay triangulation of the remaining events using the Delaunay Tessellation Field Estimator \citep[DTFE; ][]{2007PhDT.......433S}.  The DTFE estimates the count density at any position based on a two-dimensional linear interpolation between the count density values defined at each vertex in the Delaunay triangulation.  The count density, $\sigma_k$, at each vertex $k$ in the Delaunay triangulation is given by
\begin{displaymath}
\sigma_k=\frac{3}{\sum_{j=1}^{M_k}\hat{A}_j(k)},
\end{displaymath}
where $M_k$ is the number of triangles that share vertex $k$ and $\hat{A}_k(j)$ is the area of triangle $j$ that shares vertex $k$.  Integrating over the linearly interpolated count density function yields the number of counts in every background image pixel.  This approach preserves exactly the total counts in the filtered data and linearly interpolates the spatial structure into every pixel while preserving small scale features associated with bad columns, chip edges, and readout streaks (see Figure~\ref{fig:bkgimg}).

\addedtext{Background images for observation stacks are created by summing the background images created above for the set of individual observations that comprise the stack.  This is done to preserve any discontinuities in the observation stack background intensity that arise at the edges of the fields-of-view of overlapping individual stacked observations.  The observation stack background images are used when performing source detection on the stack using the wavelet algorithm (\S~\ref{sec:wavdet}).}

% \vfil avoids the section heading being at the bottom of the page
\vfil

\subsection{Source Detection} \label{sec:srcdet}
\addedtext{Source detection is performed at the observation stack level.}  For each observation stack, the photon events lists of the observations \addedtext{that} comprise the stack and that were aligned astrometrically in \S~\ref{sec:ofa} are summed to create the stacked observation photon event list that is input to source detection process.  The stacked observation photon event list, as well as other similar stacked observation FITS data products, independently enable multiple science use cases that are not source-centric.

Candidate  detections of compact X-ray sources with observed spatial scales $\lesssim\!30\arcsec$ are identified using the wavelet-based source detection algorithm \citep[CIAO \wavdetect;][]{2002ApJS..138..185F} that was used for release~1 of the catalog, together with the \mkvtbkg\ Voronoi tessellation source detection method described in \S~\ref{sec:vtsrcdet}.  The \wavdetect- and \mkvtbkg-identified candidate detections are merged as described below, and the resulting detections are further evaluated using a maximum likelihood estimator to eliminate likely spurious detections and refine the measured sky positions.   Similar to CSC~1, detection of candidate X-ray sources is performed in each energy band except for the ACIS ultra-soft band, which is impacted heavily both by increased background and by decreased effective area because of ACIS focal plane contamination.

Extended X-ray sources with observed spatial scales significantly larger than  $\sim\!30\arcsec$ are identified using only the \mkvtbkg\ Voronoi tessellation source detection method.

\begin{figure}
\epsscale{1.15}
\plotone{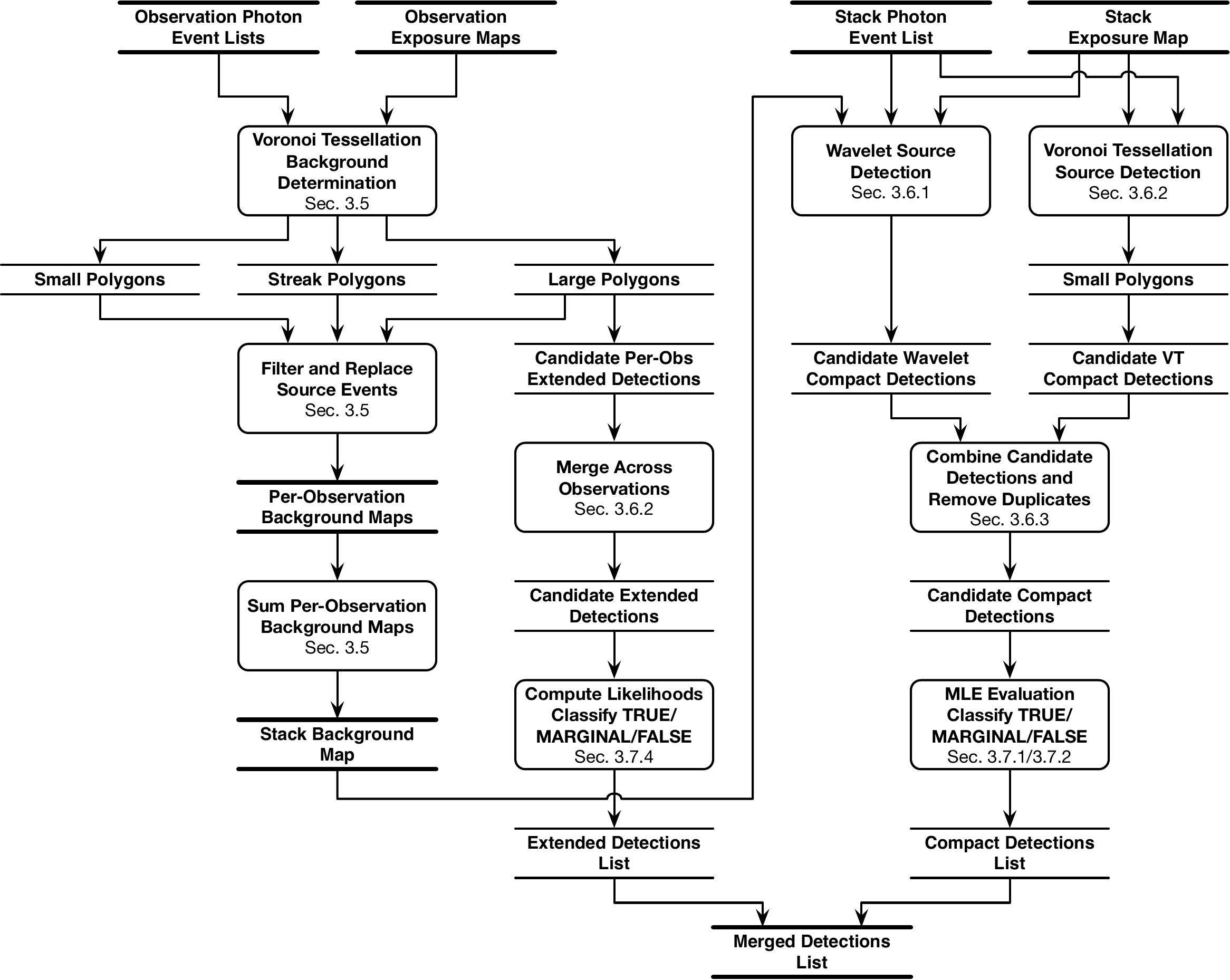}
\caption{\label{fig:detectflow}
\addedtext{High-level flow of the background determination and source detection steps.  Functional steps are shown in boxes with rounded corners.  Archival data products are identified between thick horizontal bars, while temporary data products are shown between thin horizontal bars.  For more information on the individual steps and data products, consult the text.  The references identify the relevant sections of the text that describe the methods and algorithms used.}}
\end{figure}

\addedtext{Figure~\ref{fig:detectflow} presents the overall background determination and source detection flow, with references to individual sections where the various steps are described in detail.}

\subsubsection{Wavelet Source Detection} \label{sec:wavdet}
Use of the wavelet algorithm to detect candidate sources in CSC~2 is similar to release~1 of the catalog, and will not be repeated in detail  here.  Suffice to say that \wavdetect\ searches for local maxima of the two-dimensional correlation integral of the product of the image data and  a set of Marr (``Mexican Hat'') wavelets with pre-specified scale sizes.  Twelve different wavelet scale sizes are used, ranging from $\sqrt{2}$ to 64 pixels, increasing by a factor of $\sqrt{2}$ between steps.  These correspond to spatial scales from $\sim\!0.7\arcsec$ to $\sim\!31\arcsec$ for ACIS and $\sim\!0.4\arcsec$--$17\arcsec$ for HRC-I (recall from \S~\ref{sec:recal} that HRC-I data products are binned at two-pixel resolution), and provide good detection sensitivity for sources with observed angular extents $\lesssim\!30\arcsec$.  Detection sensitivity may be decreased in some energy bands for point sources with extreme off-axis angles, typically $\theta \gtrsim 20\arcmin$, where the size of the local PSF exceeds the largest wavelet scale size.

An image pixel $(i, j)$ is identified as a source pixel if the detection significance
\begin{displaymath}
S_{i, j} = \int_{C_{i, j}}^\infty dC\,p(C|n_{B, i, j}) \leq S_0,
\end{displaymath}
where $C$ is the two-dimensional correlation integral \citep[equation 2 of ][]{2010ApJS..189...37E}, $p(C|n_{B, i, j})$ is the probability of $C$ given the background $B$, and $S_0$ is the defined limiting significance threshold.  \addedtext{The observation stack Voronoi tessellation background map constructed in \S~\ref{sec:bmc} is used as a robust estimate of the spatially varying background, $B$, by \wavdetect\null.  This choice of background is especially appropriate for detection of compact sources since in addition to the intrinsic background underlying small source polygons, the map includes contributions from both large source polygons ({\em i.e.\/}, extended emission) and streak polygons ({\em i.e.\/}, readout streaks) that may underlie them.}

To improve the detection sensitivity of CSC~2 over that of release~1, the significance threshold for identifying a pixel as belonging to a detected source is set to $S_0=5\times10^{-6}$.  This significance threshold formally corresponding to $\sim\!80$ false candidate detections in a $4096\times4096$ pixel image due to random fluctuations; however, the subsequent application of the maximum likelihood estimator to filter the candidate detections (see \S~\ref{sec:mle}) eliminates almost all of the false detections.

\subsubsection{Voronoi Tessellation Source Detection} \label{sec:vtsrcdet}
As a side effect of background map creation (see \S\ref{sec:bkgimg}), the \mkvtbkg\ tool identifies {\em small\/} polygons associated with compact source detections and {\em large\/} polygons associated with extended source detections.  These polygons are computed for multiple surface brightness contour levels by varying $\alpha_{\rm thresh}$ in equation~(\ref{eqn:alphaThresh}), with
\begin{equation}
\label{eqn:alphathreshvals}
\alpha_{\rm thresh}=[0.9, 2^{-(n/2)}],
\end{equation}
for $n=1$, $2$, $\ldots$.

\addedtext{To maximize detection sensitivity, \mkvtbkg\ is rerun on the observation stack and} candidate compact detections are identified from the combined set of \addedtext{observation stack} small polygons extracted from multiple surface brightness contour levels.  Starting at the highest surface brightness contour in which a small source polygon is detected, the set of enclosing small source polygons (one for each successive fainter contour level) are evaluated by comparing the derivatives of the polygon areas and their enclosed counts (as a function of contour level) to the values determined from the local PSF at the location of the candidate detection.  Candidate detections that are smaller than the local PSF are discarded.  An elliptical source region is generated for each accepted candidate compact detection by identifying the convex hull that surrounds the outermost source polygon, and then computing the minimally-sized ellipse that encloses that hull.

{\em Large\/} source polygons are used to identify candidate extended detections.   \addedtext{Unlike candidate compact detections, candidate extended detections are extracted directly from large source polygons from the individual observation \mkvtbkg\ runs rather than from the observation stack run.  This is done because discontinuities in the apparent surface brightness of an extended detection may occur at the edges of the fields-of-view of overlapping individual observations that comprise the stack, potentially impacting the reliability of the detections.} Polygons with contour levels $n=3$ and above in equation~(\ref{eqn:alphathreshvals}) form the initial list of candidate extended detections.  This threshold value was chosen empirically to maximize sensitivity while simultaneously minimizing detection of random background fluctuations.  These candidate detections are subject to several tests to help establish their reality (see \S~\ref{sec:combidet})\null.  For example, the raw flux enclosed by a polygon is compared to the flux from the sum of the compact detections located within that polygon to determine whether the candidate detection is really a close cluster of compact detections.  The polygons for extended detections that meet all of the criteria are recorded in the {\tt poly3} data product for the observation.

Because source polygons are constructed by combining adjacent Voronoi polygons, the number of vertices that define a large source polygon can be very large, potentially numbering in the thousands.  To minimize the complexity of the catalog, these polygons are converted to convex hull representations containing a limited number of vertices.  While convex hull polygons may not be an ideal approximation of some extended emission geometries, this representation is adequate in many cases such as galaxy halos and supernova remnants and provides a simple representation that is easily manipulated. However, the full polygon representations can be recovered from the {\tt poly3} file if desired.  \addedtext{Convex hull polygons from multiple energy bands and multiple observations that comprise an observation stack are merged by taking the ``most encompassing'' polygon vertices of the overlapping polygons to form a new convex hull polygon at the observation stack level.} 

\subsubsection{Combining Candidate Detections} \label{sec:combidet}
Detections from \wavdetect\ and \mkvtbkg\ are compared and combined to eliminate duplicates and form a single list of candidates for further processing.  Any detections (either compact or extended) that are identified by only one of the algorithms are automatically included in the candidate detection list.

In cases where one or more compact \wavdetect\ detections overlap an extended ({\em i.e.\/}, large source polygon) \mkvtbkg\ detection, establishing the validity of the extended detection requires consideration of behaviors commonly exhibited by the two detection algorithms.  In regions of bright and spatially variable extended X-ray emission, \wavdetect\ is prone to identify significant numbers of spurious compact detections rather than a single, large detection because the wavelet scale sizes are optimized to identify compact detections.  Additionally, the background map that is computed by \mkvtbkg\ and used by \wavdetect\ is never a perfect representation of the actual background and local variations in intensity between them can result in invalid compact detections.  Conversely, in crowded fields containing multiple spatially adjacent compact sources, \mkvtbkg\ tends to identify a single extended detection in preference to multiple compact detections, because the overlapping wings of the adjacent PSFs are interpreted as extended emission by the algorithm.

The validity of the extended \mkvtbkg\ detection is determined using a set of criteria that consider the detection areas and counts.  We define the ratio of the summed areas of all overlapping \wavdetect\ compact source regions to the area of the overlapped \mkvtbkg\ source polygon, $R_A$, as
\begin{displaymath}
R_A = \sum_i{A_{{\rm wav,}i}} / {A_{\rm vt}},
\end{displaymath}
where $A_{\rm vt}$ is the area of the \mkvtbkg\ region and $A_{{\rm wav},i}$ is the area of the $i$th overlapping \wavdetect\ region. $R_A$ can be used to divide phase space into two regions $R_A \le R_{A,{\rm thresh}}$ and $R > R_{A,{\rm thresh}}$, where the threshold value, $R_{A,{\rm thresh}} = 0.45$, is empirically determined.

For $R_A \le R_{A,{\rm thresh}}$, the \mkvtbkg\ detection is considered to be valid if
\begin{displaymath}
R_n={{\sum_i{n_{{\rm wav,}i}}} \over {n_{\rm vt}}} < R_{n,{\rm thresh}},
\end{displaymath}
where $n_{\rm vt}$ is the number of counts in the \mkvtbkg\ region, $n_{{\rm wav},i}$ is the number of counts in the $i$th overlapping \wavdetect\ region, and the empirically determined threshold value $R_{n,{\rm thresh}}=0.65$.  This combination of criteria (on $R_A$ and $R_n$) corresponds to cases where the extended (\mkvtbkg) detection includes both significantly more area and significantly more counts than the sum of the embedded compact (\wavdetect) detections.

\begin{figure}
\epsscale{0.8}
\plotone{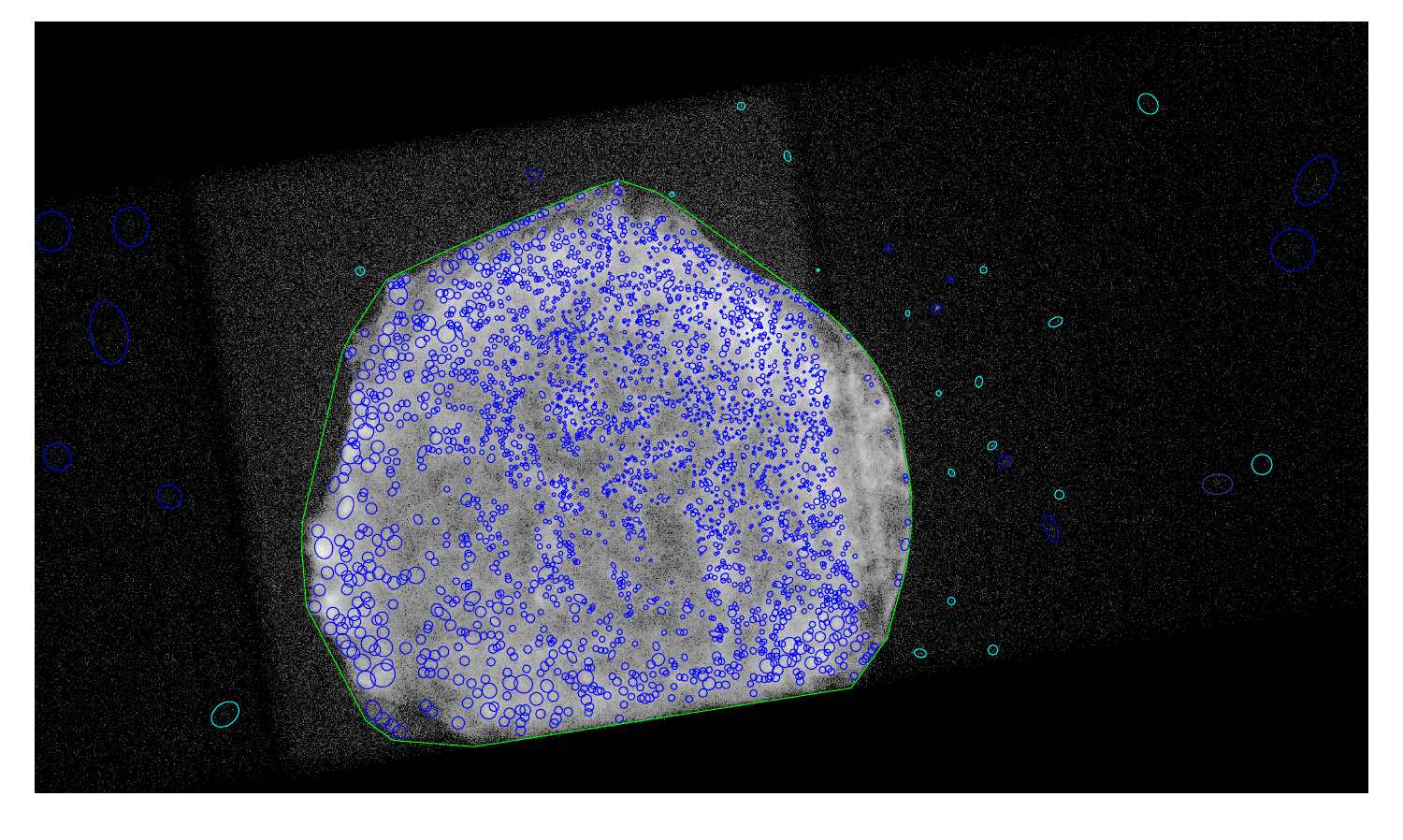}
\caption{\label{fig:extended_detect}
Example candidate detections overlaid on a single observation of the Tycho supernova remnant.  Compact detections from \wavdetect\ and \mkvtbkg\ are shown in blue and cyan respectively, and the \mkvtbkg\ extended detection is shown in green.   This plot was created for human quality assurance review of the field.  As described in the text, \wavdetect\ identifies numerous spurious compact detections on the extended emission of the SNR\null.}
\end{figure}

If $R_A > R_{A,{\rm thresh}}$, a more complex empirically determined validity criterion on the counts applies:
\begin{equation}
\label{eqn:hi_area_cutoff}
R_n < R_{n,{\rm thresh}} + (R_A - R_{A,{\rm thresh}})
\end{equation}
where the additional factor effectively increases the counts threshold in cases where the summed area of the embedded compact detections occupies most of the area of the extended detection region.  Equation~(\ref{eqn:hi_area_cutoff}) is satisfied typically in fields containing resolved supernova remnants where \wavdetect\ will sometimes identify very large numbers of (often spurious) compact detections superimposed on the extended emission, whereas \mkvtbkg\ will identify a single extended source region surrounding the bulk emission ({\em e.g.\/}, Fig.~\ref{fig:extended_detect}).

If a \mkvtbkg\ extended detection region is determined to be valid using the above criteria, then that detection is added to the candidate detection list {\em and any embedded \wavdetect\ compact detections are discarded\/}.  Overlapping compact ({\em i.e.\/}, small source polygon) detections that are embedded within the extended detection region and that were identified simultaneously by the \mkvtbkg\ algorithm are retained, and added to the candidate detection list.

Outside of a \mkvtbkg\ extended detection region, if one or more \wavdetect\ compact detections and a \mkvtbkg\ compact ({\em i.e.\/}, small source polygon) detection overlap each other, then a different test is used to determine whether the \wavdetect\ or \mkvtbkg\ compact detections should be retained.  Define the {\em difference region\/} as the \mkvtbkg\ detection region with any overlapping \wavdetect\ detection regions excluded. If
\begin{displaymath}
n_{\rm dif} / A_{\rm dif} > 0.15\sum_i {n_{{\rm wav},i}/A_{{\rm wav},i}},
\end{displaymath}
and
\begin{displaymath}
A_{\rm dif}>0.3A_{\rm vt}, 
\end{displaymath}
where $n_{\rm dif}$ is the number of counts in the difference region, $A_{\rm dif}$ is the area of the difference region, and the remaining symbols have the same definitions as above, then the \mkvtbkg\ compact detection is retained and the \wavdetect\ compact detections are discarded.  Otherwise the \wavdetect\ compact detections are retained and the \mkvtbkg\ detection is discarded.

The final step in this process is to merge detections that are present in more than one energy band by selecting the candidate detection with the highest detection significance.  

Once automated candidate source detection for an observation stack is completed, an automated assessment is performed to determine whether a human quality assurance review is required.  The evaluation includes criteria for individual candidate detections, such as the presence of saturated ACIS compact detections, possibly extended detections that do not satisfy all extended detection criteria, as well as observation-level criteria such as an excessive number of detections, too many observations with the same spacecraft roll angle (which can result in amplifying the intensity of summed readout streaks in the observation stack), or a forced manual review for complex fields.  If human review is required, then that review is performed by a trained individual.  The reviewer can add, delete, or manually modify any of the candidate detection regions as needed.  All changes to detection regions are flagged in the catalog, as described in \S~\ref{sec:flags}\null.  Note that any such changes to candidate extended detections do {\em not\/} result in re-evaluation of the overlapping compact/extended detection selection criteria described above; if any changes are required they will be applied as part of the quality assurance review process.

\subsubsection{Source Apertures} \label{sec:srcaper}
As in release 1 of the catalog, numerous detection-specific catalog properties are evaluated within defined apertures, and these apertures are defined similarly to CSC~1, with minimal changes to account for the combination of candidate compact detections from \wavdetect\ and \mkvtbkg.  

The ``PSF 90\% ECF aperture'' is defined identically to CSC~1 as the ellipse that encloses 90\% of the total counts in a model PSF located at the position of the candidate detection. Because the size and shape of the \Chandra\  PSF is energy dependent, the dimensions of the PSF 90\% ECF aperture vary with energy band and are computed at the effective monochromatic energy of the band.

The lengths of the semi-axes of the energy-independent, elliptical ``source region aperture'' for each detection are initially defined to be equal to the $3\sigma$ orthogonal deviations of the distribution of the counts in the source cell for candidate detections identified by the \wavdetect\ algorithm (for more information, see \S~3.4 of \paperI), and equal to the minimally-sized elliptical source region for candidate compact detections derived from small source polygons identified by the \mkvtbkg\ algorithm (see \S~\ref{sec:vtsrcdet}, above).  This definition of the source region aperture is smaller than the definition used in CSC~1 by a factor of $2/3$, which significantly reduces the occurrence of overlapping source region apertures in crowded fields.

To ensure that the source region aperture is not significantly smaller than the size of the local PSF, the lengths of the ellipse semi-axes are compared to the radius of an energy-independent circular approximation to the size of the PSF 90\% ECF aperture, $R_{90}$, defined as
\begin{equation}
\label{eqn:R90}
R_{90} = 1.1136 - 0.15636\times\theta + 0.14133\times\theta^2 - 0.0019723\times\theta^3 \hbox{\rm\ arcseconds},
\end{equation}
where $\theta$ is the off-axis angle of the candidate compact detection in arcminutes and the coefficients have been determined empirically.  If either source region aperture semi-axis is smaller than $R_{90}$ then that semi-axis is re-sized to $R_{90}$.  If both semi-axes of the initial source region aperture are smaller than $R_{90}$ then the final source region aperture will be circular.  While the use of the circular approximation for the PSF size may seem somewhat simplistic, source region aperture expansion is needed infrequently and in all cases any PSF fraction aperture corrections are evaluated using the {\em actual\/} local PSF (see \S~\ref{sec:mle}) evaluated through the appropriate source aperture.

CSC~2 follows release 1 by defining an elliptical ``background region aperture'' as an annulus surrounding and co-located with the source region aperture.  The inner radii of the annulus are enlarged by 10\% compared to the CSC~1 definition, and are set equal to $1.1\times$ the radii of the corresponding source aperture, with the same position angle.  The radii of the outer edge of the annulus are set equal to $5\times$ the radii of the source region aperture, as in CSC~1\null.  The background region aperture defined in this manner typically includes $\sim\!5$--10\% of the X-rays from the detection, and this contamination is accounted for explicitly when computing aperture photometry fluxes.

The source, PSF 90\% ECF, and background region apertures are defined to be simple elliptical regions and the corresponding elliptical annuli.  In practice arbitrary areas from either aperture may need to be excluded to avoid contamination from nearby (overlapping) sources, or because of missing data for a variety of reasons such as detector edges.  Apertures with the exclusion regions applied are designated as ``modified'' source, PSF 90\% ECF, and background region apertures, as appropriate.  The modified source and background region apertures are recorded in the CSC 2 region ({\tt reg3}) data product for the stack since they are observation-independent, while the PSF 90\% ECF regions for each energy band, which are observation-dependent, are recorded in the region ({\tt reg3}) data product for the observation.  For CSC 2.0, if the modified background region aperture includes $\le10$ counts, then the outer elliptical annulus radii are iteratively scaled up until the region includes at least 30 counts or the background region area exceeds $100\times$ the source region area.  This background expansion helps to ensure that there are sufficient counts in the background region for the aperture photometry algorithms to converge reliably.  A more robust method is used when evaluating such cases in CSC 2.1, so background expansion is not performed for that release.

\begin{figure}
\epsscale{0.65}
\plotone{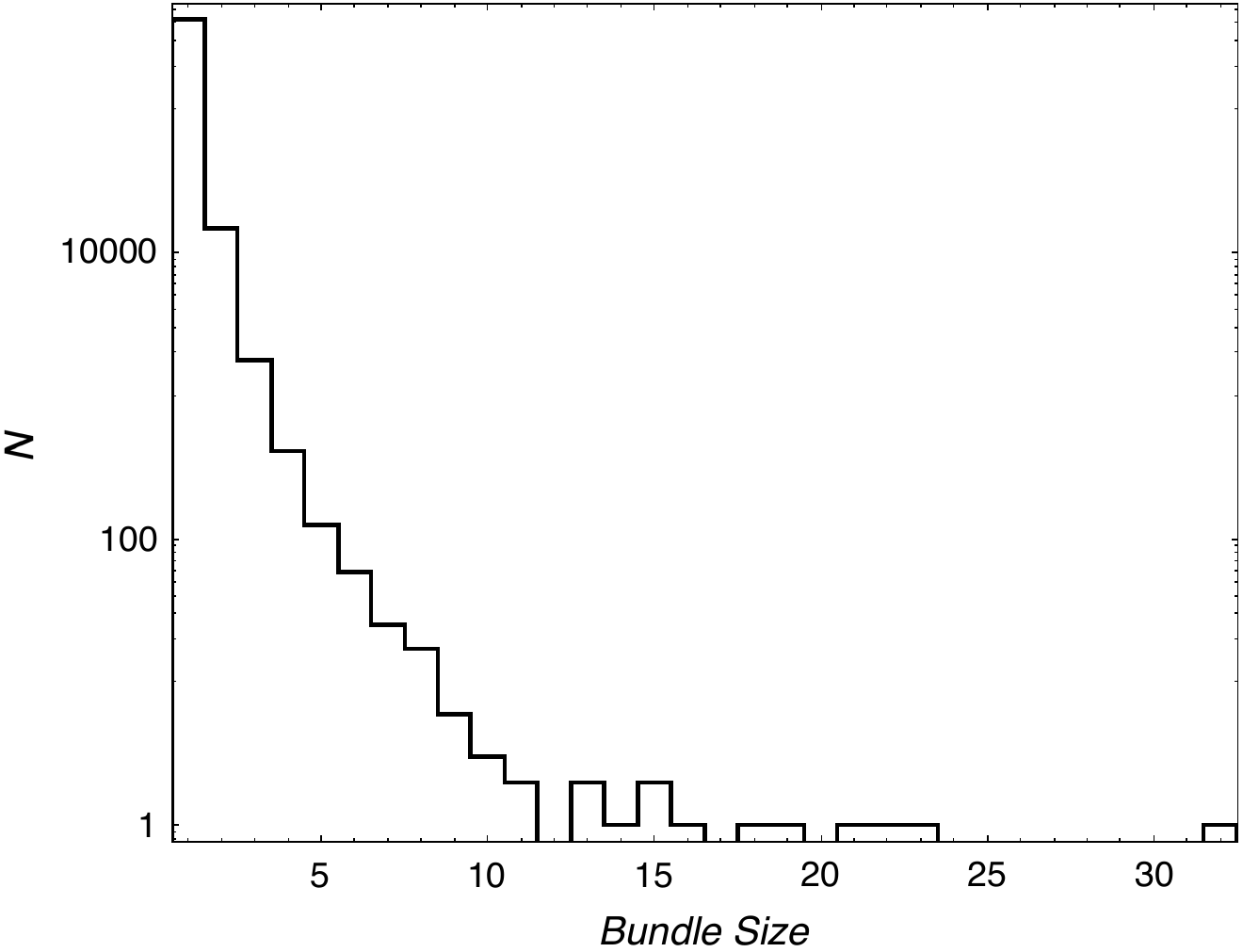}
\caption{\label{fig:bndlhist}
Number of candidate detections per detection bundle for CSC 2.0 observation stacks.  Approximately $\sim\!96.1\%$ of detection bundles include only a single candidate detection, and $\sim\!99.97\%$ of detection bundles include 5 or fewer candidate detections.}
\end{figure}

For later use, we define a ``detection bundle'' as a set of detections that have overlapping source regions.  Most source regions do not overlap another source region, and these cases can be considered to constitute single-detection detection bundles.  However, as shown in Figure~\ref{fig:bndlhist}, a significant fraction ($\sim\!18\%$) of detection bundles include two or more detections, with $\sim\!2\%$ including five or more.

\subsection{Candidate Compact Detection Validation and Detection Position Determination} \label{sec:detvalidation}
\subsubsection{Maximum Likelihood Estimator Analysis} \label{sec:mle}
Each candidate compact detection is evaluated using a Bayesian maximum likelihood estimator (MLE) analysis.  The algorithm, which is based loosely on the XMM-Newton Scientific Analysis System\footnote{\url{https://www.cosmos.esa.int/web/xmm-newton/sas/}} tool {\tt emldetect} (\addedtext{\citealt{2009A&A...493..339W};} see also \citealt{1988ESOC...28..177C}), uses the CIAO modeling and fitting package \sherpa\ \citep{2001SPIE.4477...76F} to evaluate the likelihood that the observed distribution of counts is consistent with either a point-source model or a compact-source model superimposed on the local background.

Both the point- and compact source models are constructed by convolving the local PSF with a two-dimensional rotated elliptical Gaussian.  The \sherpa\ model, $\mathcal{M}$, is defined as
\begin{equation}
\label{eqn:mlemodel}
\mathcal{M}=\mathcal{B}\times\mathcal{E} + \left[\mathcal{P}\ast\hbox{\tt sigmagauss2d}\left(A, x, y, \sigma_a, \sigma_b, \phi\right)\right]\times\mathcal{E},
\end{equation}
where $\mathcal{B}$ is the local background model, $\mathcal{E}$ is the exposure map, $\mathcal{P}$ is the local PSF, and $\hbox{\tt sigmagauss2d}$ is the two-dimensional rotated elliptical Gaussian model with parameters $A, x, y, \sigma_a, \sigma_b$, and $\phi$ being the  peak amplitude, $(x, y)$ position [which maps directly to $(\alpha, \delta)$ through the stack's World Coordinate System transform], standard deviations of the semi-major and semi-minor axes, and rotation angle of the semi-major axis, respectively.  The amplitude $A$ and spatial position $(x, y)$ parameters of $\mathcal{M}$ are optimized to maximize the detection likelihood for both the point- and compact-source models, and for the compact-source model, the parameters $(\sigma_a, \sigma_b, \phi)$ are also fitted.  However, for the point-source model $\sigma_a$ and $\sigma_b$ are both frozen at one image pixel (and $\phi$ is frozen at zero).

The point- and compact-source models defined by equation~(\ref{eqn:mlemodel}) are evaluated independently in each energy band.  For observation stacks that include multiple observations, the models are evaluated for the stacked observation and also separately for each individual observation.  The stacked observation models will yield a higher likelihood for the majority of detections because of the higher S/N of the summed observations.  However, if the underlying source is temporally variable, the model fits to a single observation where the source is bright may yield a higher likelihood.  The highest likelihood across all evaluated models is used when evaluating the candidate detection quality based on the detection thresholds (\S~\ref{sec:lthresh}).

Similar to CSC release 1, the local PSF model $\mathcal{P}$ is evaluated at the detected position of each candidate compact detection using version 2.0 of the SAOTrace simulation code \citep{1995ASPC...77..357J, 2004SPIE.5165..402J} to compute the ray-trace at the effective monochromatic energy of each energy band.  In each case, the ray-density for the simulation is selected to produce a high-quality model with $\sim\!50,000$ counts in the PSF\null.  This choice of ray-density is a good compromise between PSF fidelity and computation time, as only $\sim\!0.004\%$ of non-saturated compact detections included in the catalog have more than $50,000$ net counts in any energy band.

The simulated rays are further processed by MARX \citep{2012SPIE.8443E..1AD} to model the projection of the ray-trace onto the detector focal plane.  For ACIS observations, MARX simulates the event grade distribution of the X-ray event charge clouds in the CCD silicon so the resulting ACIS PSF model event lists can have the Energy-Dependent Subpixel Event Repositioning (EDSER) algorithm \citep{2004ApJ...610.1204L} applied.  This is necessary because the EDSER algorithm improves the quality of reconstructed images of sources with arcsecond-scale features and was enabled by default in \Chandra\ pipeline processing starting in June, 2011.  Based on point-source simulations, \citet{2004ApJ...610.1204L} conclude that the EDSER algorithm reduces the FWHM of ACIS PSFs by approximately 30--50\% for bright ($\sim\!1200$--1500 X-ray counts) point sources, depending on off-axis angle and whether the target falls on a front-illuminated or back-illuminated CCD\null.  MARX adds a Gaussian aspect blur of $0\farcs07$ when projecting the PSF onto the detector focal plane, to account for blurring of the PSF of the X-ray PSF due to aspect reconstruction.\footnote{\url{https://cxc.harvard.edu/cal/ASPECT/img\_recon/report.html}}

Prior to performing the optimization, all of the input data (the detected photon events, background map, local PSF, exposure map, and pixel mask) are sub-pixel binned if necessary so that there is a minimum of 3--6 image pixels across the PSF FWHM, to better estimate detection extent when the PSF is small.  In addition, the image pixel binning is adjusted by up to half a pixel in each axis so that the PSF is positioned at the center of an image pixel.  This has the effect of minimizing any artificial widening of the PSF due to image pixel quantization.  The model PSFs are recorded in the {\tt psf3} data product for the observation detection.

For both source models, the overall normalization of the exposure-normalized background map ($\mathcal{B}\times\mathcal{E}$) is optimized in the modified background region associated with the candidate detection ({\em i.e.\/},with any overlapping source regions masked out), and then subsequently frozen while evaluating the source models.  The normalization should be very close to unity since the \mkvtbkg\ background map is a good representation of the true background, but an accurate estimate of the background level in the source region is fundamental to determining the source likelihood robustly.

The source models (with frozen background) are each evaluated using a two-step process.  First, the models are evaluated in an elliptical region that has twice the radius of the PSF 90\% ECF aperture, centered on the position of the candidate detection.  This step optimizes the starting position for the second fit, because the candidate detection regions from the detection algorithms are not necessarily well-centered.  Subsequently, the source models are re-evaluated in the PSF 90\% ECF aperture recentered on the best-fit position from the first optimization.  Restricting the source model fits to the PSF 90\% ECF aperture typically yields better results than fitting in a larger aperture in crowded fields since most candidate detections are consistent with point sources.

In most cases this two-step fitting process yields good results, but human quality assurance review is triggered under any of the following circumstances: (a)~the two candidate detections have fitted positions that are too close to each other (within 110\% radius of PSF 90\% ECF aperture for either detection), (b)~the best-fit position from the first fit falls outside of the expanded region, (c)~the best-fit position from the second fit falls outside 70\% of the radius of the PSF 90\% ECF aperture, or (d)~for the best-fit compact source model the eccentricity exceeds 0.9.  In any of these cases, the review process allows the reviewer to modify the regions and/or initial positions for each fit, and if necessary manually force a final fit position.  Once the review is completed, the fit is rerun with the new parameters (and can subsequently be sent for review again if any of the conditions is triggered), and the appropriate manual review flags are set for the candidate detections.

If the source regions of more than one candidate compact detection overlap ({\em i.e.\/}, the detection bundle includes multiple candidate detections), in CSC~2.0 for each fit the candidate detections are evaluated in order from the highest number of counts to the lowest number of counts, with each optimized detection being included with frozen parameters as part of the background model for subsequent detections.  In all cases, overlapping source regions are masked out while performing the source model optimizations.

Maximization of the posterior probability distribution is performed using the \sherpa\ \citep{2001SPIE.4477...76F} Monte Carlo optimizer, which is an implementation of the differential evolution algorithm \citep{1997JGlOpt.11..341S}.  Since counts of individual photons are recorded by each detector pixel, these models are fitted by minimizing the Poisson log-likelihood $C$-statistic \citep{1979ApJ...228..939C}
\begin{equation}
\label{eqn:cstat}
C = 2\sum_i \left(\mathcal{M}_i - \mathcal{D}_i + \mathcal{D}_i\left[\log\mathcal{D}_i - \log\mathcal{M}_i\right]\right),
\end{equation}
where $\mathcal{M}_i$ is the sum of the source and background model amplitudes in pixel $i$ and $\mathcal{D}_i$ is the number of observed counts in pixel $i$.

To at least partially address the impacts of possible source temporal variability, the source models are fitted to the data both as individual-observation fits for all observations included in a stack, and also as a simultaneous fit to all observations included in the stack.  The latter fit will generally yield better results for non- or weakly-variable sources, but the former may be better for flaring X-ray sources since the total background will be lower.   The impact of spectral shape on the detection likelihood is minimized by performing all of the fits separately in each science energy band.  After the fits are performed, the optimization that yields the highest likelihood is used to establish both the detection likelihood (for comparison with the detection thresholds) and the optimal model parameters, regardless of whether this fit is from a single observation or the entire stack.

For a compact detection, the detection log-likelihood is calculated in a manner similar to \citet{2009A&A...493..339W} as
\begin{equation}
\label{eqn:likelihood}
\likelihood = -\ln P,
\end{equation}
where the probability $P$ that a Poisson fluctuation in the background would yield at least the detected number of counts in the source region aperture given the expected background is given by
\begin{equation}
\label{eqn:likelyprob}
P = 1-\Gammafunc{(\nu / 2, \Delta C / 2)},
\end{equation}
where $\Gammafunc{(a, x)}$ is the upper incomplete Gamma function, $\nu$ is the number of free (non-frozen) model parameters in the source model, and
\begin{equation}
\label{eqn:deltaC}
\Delta C = C_\mathcal{B\times E} - C_\mathcal{M}
\end{equation}
is the difference between the $C$-statistic (equation~[\ref{eqn:cstat}]) of the ``background-only'' null hypothesis $\mathcal{B\times E}$ and the best-fitting \sherpa\ model $\mathcal{M}$, evaluated in the source region.  

The null hypothesis is rejected (or the alternative hypothesis of a non-zero flux detection is accepted) for small values of $P$, (or equivalently, large values of $\likelihood$)\null.  Large values of $\likelihood$ therefore imply that the distribution of counts observed within the detection region are extremely unlikely to have been generated by a random fluctuation of the background in absence of a real detection.

The likelihood of a saturated ACIS compact detection cannot be determined in this manner because the detection is so bright that photon pile-up has eroded the core of the observed detection profile, which takes on a ``cratered'' appearance.  Such detections are identified during human quality assurance review and the Bayesian MLE analysis is subsequently skipped.  They are assigned infinite likelihood.

To reduce the number of cases sent for human quality assurance review when running the MLE analysis on candidate compact detections, the approach used to perform the fits was modified \citep{martinezgalarza_mleupdate} in CSC 2.1\null.  For any detection bundle, the release 2.0 sequential fit approach is attempted first as described above.  However, if the fit would have triggered human quality assurance review through the criteria identified earlier in this section, an alternative approach that depends on the clustering properties in the vicinity of the candidate detection is attempted.  If the detection bundle includes only a single candidate detection then the \sherpa\ fit is performed as before, except that the model is evaluated in the source region aperture rather than in the PSF 90\% ECF aperture.  This significantly reduces fit failures due to mis-centering of the aperture on the counts distribution.  If the detection bundle includes 2--5 candidate detections, a simultaneous fit over the detections is performed in an aperture that is the union of the individual source region apertures.  The background region used to optimize the background normalization is similarly the union of the individual background region apertures, with any overlapping source regions excluded.  Performing a simultaneous fit for all of the candidate detections produces more reliable fit parameters, especially in cases where candidate detections have similar amplitudes where the position of the first fit when using the sequential fit approach may be dragged towards the second candidate detection. The simultaneous fit is potentially much more computationally expensive than the sequential fit method, especially for observation stacks that include large numbers of individual observations, since the number of parameters scales as the product of the number of observations in the stack and the number of candidate detections in the bundle.  If the detection bundle includes more than 5 candidate detections, then the original sequential fit is retained and human quality assurance review will be triggered.  As shown in Figure~\ref{fig:bndlhist}, only about 0.03\% of detection bundles include $>5$ candidate detections.  In the case of simultaneous fitting, when computing the likelihood of a candidate detection, the $C$-statistic for the ``background-only'' null hypothesis is evaluated with the best-fit amplitude of the background and best-fit parameters for all of the candidate detections other than the detection of interest frozen at their optimum fit values.  The results of the fit are evaluated using the same quality assurance criteria identified previously, and human review will be triggered if any of the criteria are violated.

\subsubsection{Maximum Likelihood Detection Thresholds} \label{sec:lthresh}
The highest model likelihood for each candidate compact detection is evaluated against threshold values to classify the detection as either \truevalue, \marginalvalue, or \falsevalue\null.  \truevalue\ and \marginalvalue\ detections are subject to further analysis to evaluate their properties, and are included in the final catalog.  The best fitting parameter values for all fits performed as part of the MLE analysis, including for detections classified as \falsevalue, are recorded in the {\tt mrgsrc3} data product for the stack, but \falsevalue\ detections are otherwise discarded with no further analysis and are not included in the final source catalog.  The {\tt mrgsrc3} file includes the computed likelihood values for each candidate detection.

The \marginalvalue\ and \truevalue\ likelihood thresholds were established by processing a set of blank sky simulations through the catalog pipelines and evaluating the false detection rate as a function of likelihood in $3\arcmin$ width annuli centered on the telescope optical axis.  The likelihood thresholds were chosen conservatively to achieve a false detection rate of approximately one false detection per observation-stack for point-source detections brighter than the \marginalvalue\ threshold, and 0.1 false detections per observation-stack for point-source detections brighter than the \truevalue\ threshold.

For the ACIS instrument, the likelihood thresholds follow the empirical form \citep{nowak_threshacis}
\begin{equation}
\label{eqn:Lthreshacis}
\likelihood_x = N^d_x\left(A^d \theta^2 + B^d\right) + \max\left[S^d_x \log_{10}\left(T_{\rm stack}\over {10^4}\right), L^d_x - N^d_x B^d\right],
\end{equation}
where $x$ refers to the threshold ($m$ for the \marginalvalue\ threshold or $t$ for the \truevalue\ threshold), $d$ is the detector at the observation stack tangent point ($I$ for ACIS-I or $S$ for ACIS-S; if the stack comprises both ACIS-I and ACIS-S observations, then the slightly more conservative ACIS-S thresholds are used), $\theta$ is the mean off-axis angle of the detection in arcminutes, and $T_{\rm stack}$ is the effective stacked observation exposure time in seconds.

For ACIS-I, the remaining constants in equation~(\ref{eqn:Lthreshacis}) are set as follows:
\begin{displaymath}
N^I_m = 1, N^I_t = 2, A^I = 0.032, B^I = 6.0, S^I_m = S^I_t = 4.6, L^I_m = 1, L^I_t = 2,
\end{displaymath}
while for ACIS-S the corresponding values are
\begin{displaymath}
N^S_m = 1, N^S_t = 2, A^S = 0.08, B^S = 3.6, S^S_m = S^S_t = 4.6, L^S_m = 1, L^S_t = 2.
\end{displaymath}

For the HRC-I instrument, a different form based on the logistic function is used \citep{nowak_threshhrc}:
\begin{equation}
\label{eqn:Lthreshhrc}
\log_{10}\likelihood_x = A_x + {{B_x-A_x}\over{1+\exp\left[-\left(\theta-C_x\right)/2\right]}},
\end{equation}
where $x$ is defined as for equation~(\ref{eqn:Lthreshacis}), and $A_x$, $B_x$, and $C_x$ are defined as
\begin{gather}
A_x = A_{x1} + A_{x2} l_b, \label{eqn:LthreshhrcA} \\
B_x =A_x+{{B_{x1}}\over{1+\exp\left[-B_{x2} \left(l_b-B_{x3}\right)\right]}}, \label{eqn:LthreshhrcB} \\
C_x =C_{x1} + C_{x2} l_b. \label{eqn:LthreshhrcC}
\end{gather}
Here, $l_b\equiv\log_{10}\sum_i\mathcal{B}_iT_i$, where $T_i$ is the effective exposure time for each observation, $i$, included in the stack, and $\mathcal{B}_i$ is the background count rate of the $i$th observation per square arcminute per second computed within the central $18\arcmin$ diameter region of the field-of-view.  The remaining empirically determined constants in equations (\ref{eqn:LthreshhrcA})--(\ref{eqn:LthreshhrcC}) are
\begin{gather*}
A_{m1} = 0.6825, A_{t1} = 0.8364, A_{m2} = 0.0711, A_{t2} = 0.0561, \\
B_{m1} = 1.5219, B_{t1} = 1.5197, B_{m2} = 5.2369, B_{t2} = 7.5655, B_{m3} = 3.1628, B_{t3} = 2.9419, \\
C_{m1} = 17.0781, C_{t1} = 21.8523, C_{m2} = -1.8660, C_{t2} = -3.2626.
\end{gather*}

\subsubsection{Detection Position Uncertainty}
For compact detections classified as \truevalue\ or \marginalvalue, the 95\% confidence detection position error ellipse is evaluated using a Markov chain Monte Carlo (MCMC) analysis.  The semi-axes and position angle of the ellipse are determined by sampling the computed posteriors with the implementation of the Bayesian Low-Count X-ray Spectral \citep[pyBLoCXS;][]{2001ApJ...548..224V, 2011ASPC..442..439S} routine included in \sherpa\ (function {\tt get\_draws}).  Additionally, independent lower and upper 68\% confidence bounds are calculated for all other fitted parameters.  The confidence intervals are recorded, together with the corresponding best fitting ({\em i.e.\/}, maximum likelihood) model parameter values, in the stack {\tt mrgsrc3} data product.

Note that the detection position error ellipse includes only the model fit errors and does not consider uncertainties in the absolute position of the \Chandra\  data relative to the ICRS astrometric reference frame.  The detection position error ellipses for all stacked observations of an X-ray source are combined together with the systemic astrometric uncertainty relative to the ICRS reference frame to determine the final source position uncertainty (see \S~\ref{sec:masterpos} below).

The best fitting model parameter values are used as the initial parameter settings for the MCMC run.  The assumed prior probabilities for the model parameters $x, y, \sigma_a, \sigma_b$, and $\phi$ are flat within the boundary of the fitted regions, while the prior for the amplitude is assumed to be uniform and non-negative.  A total of 5,000 draws from the posterior probability distribution are computed for the point-source model, while 15,000 draws are computed for the compact-source model. In both cases the first 100 samples are discarded as burn-in.  The MCMC draws for a detection are recorded in the detection {\tt draws3} data product.

The convergence and correlations within the draws for each parameter are calculated using the $\rhat$ diagnostic parameter \citep[see Chapter~11 of][]{2013bda3.book.....G}.  When calculating $\rhat$ the full MCMC chain is split into 10 sub-sections and the variances within each section and between sections are used as input to the calculation.  The MCMC chain is accepted provided $\rhat < 1.2$ {\em and\/} the Metropolis-Hastings sampler draw acceptance rate is greater than 0.2.  For the compact-source model an additional condition, that the offset between the compact-source model position and the point-source model position must not be greater than $R_{90}$ (equation~\ref{eqn:R90}), is imposed.

If the MCMC chain is accepted, a minimally sized ellipse on the $(x, y)$-plane that includes $100\pm1\%$ of the draws that have a fit statistic value that falls below the 95th percentile confidence interval of the minimum fit statistic is computed, and this ellipse is the 95\% confidence position error ellipse for the detection.

If the MCMC chain does not satisfy the convergence criteria described above, a circular position error estimate is computed in a similar manner to  CSC release~1.  The radius of the 95\% uncertainty position error circle, $R^{err}_{95}$, is given by
\begin{equation}
\label{eqn:circerr}
\log R^{err}_{95} = C_0 + C_1\times\theta + C_2\times\log n + C_3\times\theta\times\log n,
\end{equation}
where $n$ is the net number of counts included in the detection, and $C_0$ ... $C_3$ are energy-band dependent coefficients from Table~{\ref{tab:circerr}}.  The radius computed using this equations is capped to fall in the range $0\farcs1 \le R^{err}_{95} \le 42\farcs2$.

A circular position error approximation is also used for saturated ACIS detections.  We assume that the count rate in the brightest $3\times3$ ACIS pixel is 0.2 counts per ACIS exposure (typically $3.2\,\rm s$), corresponding to $\sim\!10\%$ pile-up fraction, which is the value at which the pile-up warning flag is set.  Saturated detections should have at least this many counts.  The total number of counts in the brightest $3\times3$ ACIS pixel is then $C_{33} = 0.2\times N_{{\rm exp}}$, where $N_{{\rm exp}}$ is the total number of ACIS exposures during the observation.  The fraction of the PSF 90\% ECF detection flux in the  brightest $3\times3$ ACIS pixel can be estimated empirically as
\begin{displaymath}
f_{33} = A_1 \times \exp(-(\theta / S_1) ^ 2 / 2) + A_2 \times \exp(-(\theta / S_2) ^ 2 / 2) + A_3 \times \exp(-(\theta / S_3) ^ 2 / 2),
\end{displaymath}
where $A_1 = 0.132$, $A_2 = 0.875$, $A_3 = -0.0267$, $S_1 = 7.097$, $S_2 = 2.88$, $S_3 = 0.68$ and $\theta$ is the off-axis angle in arcminutes.  This yields an estimate of the actual number of counts in the PSF 90\% ECF aperture $n = C_{33}/f_{33}$ that can be used to estimate the radius of the 95\% uncertainty position error circle, $R^{err}_{95}$, using equation~\ref{eqn:circerr}.  The total circular position error radius for a saturated ACIS detection is estimated as
\begin{displaymath}
R^{err}_{tot} = \sqrt{{(R^{err}_{95})}^2 + {(R^{err}_{man})}^2},
\end{displaymath}
where $R^{err}_{man}$ is an estimate of the uncertainty introduced in assigning positions to saturated detections during human quality assurance review.  In release 1, one could usually assign such positions with an accuracy of $\lesssim1\,\hbox{\rm pixel}$ for a single observation; however stacking observations in release 2 increases this uncertainty, especially for observation stacks with multiple spacecraft roll angles.  We therefore conservatively set $R^{err}_{man} = 1.0\,\hbox{\rm  arcsec}$ for release 2.

\begin{deluxetable}{ccccc}
\tabletypesize{\small}
\tablecolumns{5}
\tablecaption{Circular Position Error Coefficients\label{tab:circerr}}
\tablehead{
\colhead{Band} & \colhead{$C_0$} & \colhead{$C_1$} & \colhead{$C_2$} & \colhead{$C_3$}
}
\startdata
Ultra-soft & $\phn 0.242$ & $0.134 $ & $-0.714$ & $\phn 0.001$ \\
Soft & $\phn 0.088$ & $0.153$ & $-0.582$ & $-0.018$ \\
Medium & $\phn 0.056$ & $0.154$ & $-0.578$ & $-0.018$ \\
Hard & $\phn 0.054$ & $0.176$ & $-0.607$ & $-0.021$ \\
Broad & $-0.031$ & $0.173$ & $-0.526$ & $-0.023$ \\
Wide & $\phn 0.075$ & $0.206$ & $-0.544$ & $-0.026$ \\
\enddata
\end{deluxetable}

\subsubsection{Candidate Extended Detection Validation and Detection Position Determination}
Since we expect that the spatial distribution of the extended emission within an extended detection region will be complex with no single spatial model appropriate to all anticipated source shapes and profiles, we evaluate the positions and likelihoods of candidate extended detection regions using a simplified approach.

We choose to use simple flux-weighted position centroids and variances to define the location of the extended detection, which in CSC~2 is represented by a convex hull polygon as described in \S~\ref{sec:vtsrcdet}.  The extended detection position is defined as
\begin{displaymath}
(\bar x_{\rm ext}, \bar y_{\rm ext}) = \left(\frac{\sum f_i x_i}{\sum f_i}, \frac{\sum f_i y_i}{\sum f_i}\right),
\end{displaymath}
where $x_i$ and $y_i$ are the $x$ and $y$ positions of pixel $i$, $f_i = n_i/e_i$ is a simple exposure-normalized net flux for pixel $i$ with net counts $n_i$ and exposure $e_i$, and the sum is performed over the set of pixels included in the extended detection region {\em excluding\/} any embedded compact detection regions.  Weighting by exposure-normalized net flux rather than  counts addresses the rather common situation where the exposure is spatially variable over the extended detection, for example because the region overlaps the edge of the field of one or more observations that comprise the observation stack.

The extended detection position variance in $x$ is 
\begin{displaymath}
\sigma^2_{\bar x_{\rm ext}} = \sum\sigma^2_{x_i}\left({{\partial \bar x_{\rm ext}}\over{\partial x_i}}\right)^2
= \frac{\sum f^2_i}{{\left(\sum f_i\right)}^2}\times\left( \frac{\sum f_i x^2_i}{\sum f_i} - \bar x^2_{\rm ext}\right),
\end{displaymath}
where we further make the assumption that the individual pixel variances $\sigma^2_{x_i} \sim \sigma$, the weighted sample variance for all $x_i$.  The $y$ position variance is defined similarly.

To compute net fluxes and likelihoods, we define a circular background region centered at $(\bar x_{\rm ext}, \bar y_{\rm ext})$ with 
radius
\begin{displaymath}
R_{\rm bkg} = \sqrt{\frac{f \times A_{\rm ext}}{\pi}},
\end{displaymath}
where $A_{\rm ext}$ is the area of the extended detection region with any embedded compact detection regions excluded, and the scale factor $f\sim2$.  The final background region excludes both the original extended detection region and also any embedded compact detection regions.  With these definitions, the background area should be comparable to the extended detection region area; however, depending on the shape of the extended detection region, the background region may not wholly enclose the former, so the actual background area may be greater than $A_{\rm ext}$.

For simplicity, within the extended detection region a constant 2-D model is adopted with intensity $b_i=b^0\times e_i$ in pixel $i$\null.  For the background-only null hypothesis, $\mathcal{B}\times\mathcal{E}$, $b^0$ is set to
\begin{displaymath}
b^0_{\mathcal{B}\times\mathcal{E}} = \frac{1}{N}\sum\frac{n_i}{e_i},
\end{displaymath}
where the sum is over the $N$ pixels included in the background region, and the background-only $C$-statistic \citep{1979ApJ...228..939C}, $C_{\mathcal{B}\times\mathcal{E}}$, is evaluated.  The normalization $b^0$ is then varied to minimize the best-fitting model $C$-statistic, $C_\mathcal{M}$, and the candidate extended detection log likelihood is computed using equations (\ref{eqn:likelihood})--(\ref{eqn:deltaC}), similar to the compact detection case.

Only candidate extended detections classified as \truevalue, which corresponds to a log likelihood threshold $\likelihood=350$, are included in the catalog.  However, in CSC 2 all candidate extended detections are evaluated by human quality assurance review and detections with log likelihoods below this value may be manually included in the catalog during this step based on visual inspection.

\subsection{Observation Stack Absolute Astrometry (CSC 2.1)}  \label{sec:absastrom}
Starting with CSC 2.1, the absolute astrometry of the entire catalog is referenced to the Gaia DR3 realization of the International Celestial Reference Frame \citep[Gaia-CRF3;][]{2022A&A...667A.148G}.  Astrometric corrections to the Gaia reference frame are computed for each observation stack by matching CSC stack detections to source positions in either the Gaia Early Data Release 3 main source catalog \citep{2021A&A...649A...1G} directly, or via an intermediate matching step to the AllWISE source catalog \citep{2014yCat.2328....0C}. 

A weighted least squares approach that minimizes the Euclidean distance between the adjusted CSC detection positions and the matching reference catalog positions is employed to calculate the astrometric corrections.  As discussed in \S~\ref{sec:ofa}, only translations are allowed since the rotation angle and plate scale of the \Chandra\ detectors are well determined.   The implementation of the algorithm is described in detail by \citet{martinezgalarza_stackfaspec}.  Suffice to say here that the method identifies the astrometric translation $(\Delta\xi, \Delta\eta)$ that minimizes
\begin{displaymath}
S = \sum_i w_i \left[(\xi_i^{\rm ref} - \xi_i^{\rm trans})^2 + (\eta_i^{\rm ref} - \eta_i^{\rm trans})^2\right],
\end{displaymath}
where $(\xi_i^{\rm ref}, \eta_i^{\rm ref})$ are the standard coordinates for the position of the $i$th matching source in the reference catalog, and $(\xi_i^{\rm trans}, \eta_i^{\rm trans}) = (\xi_i^{\rm CSC}+\Delta\xi, \eta_i^{\rm CSC}+\Delta\eta)$ are the standard coordinates for the translated position of the $i$th matching CSC detection in the observation stack.  The values $w_i$ are the weights, which are proportional to the inverse of the area of the CSC stack detection position error ellipse for direct Gaia matches, and the quadrature sum of that value and the equivalent AllWISE $1\sigma$ uncertainties in $\alpha$ and $\delta$ for two-step matches using AllWISE\null.  To limit potential degradation of the solution due to the large off-axis \Chandra\ PSF, only compact detections located within $10^\prime$ of the telescope optical axis are matched to the reference catalog.

\begin{figure}
\epsscale{0.75}
\plotone{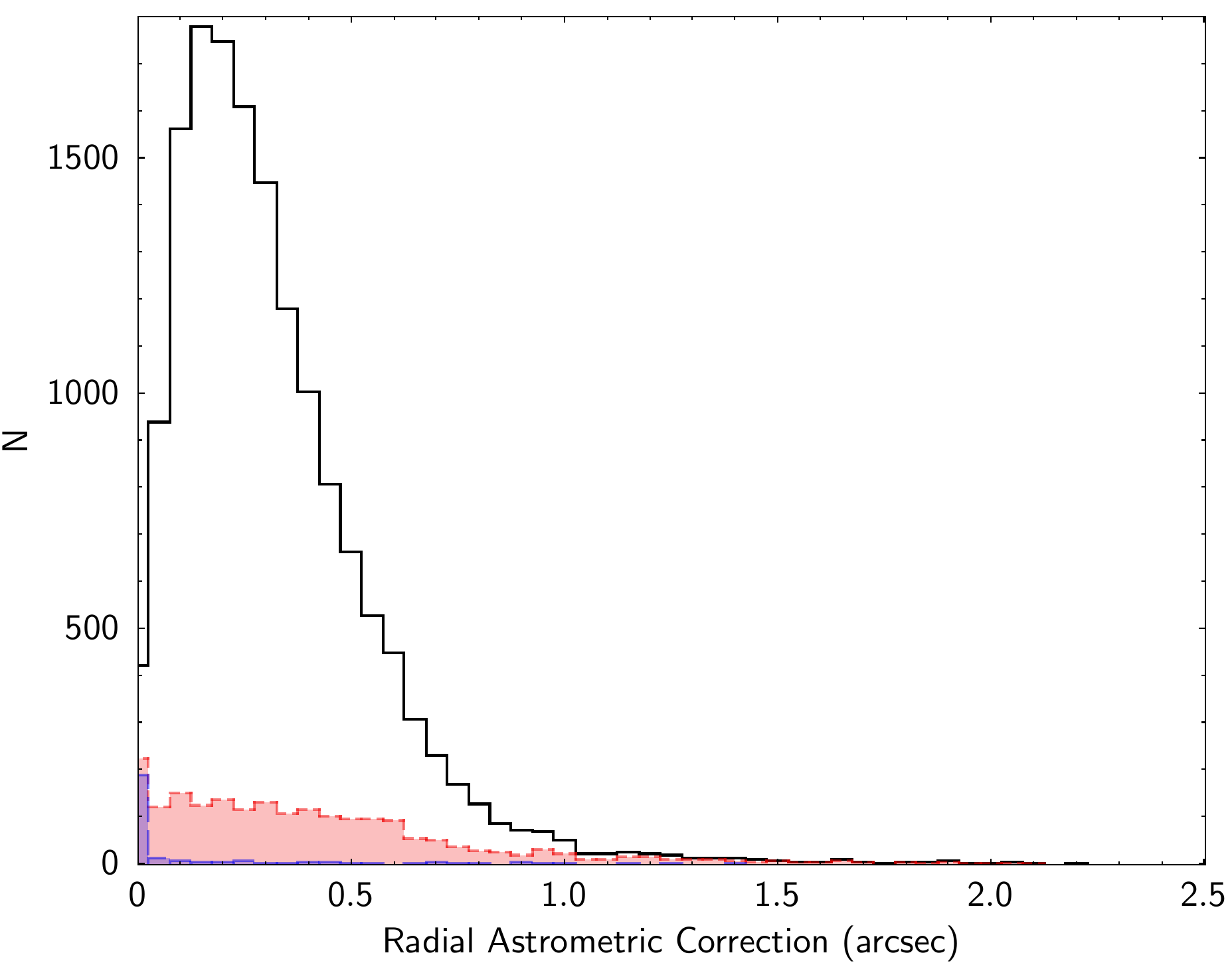}
\caption{\label{fig:gaiaastrom}
Histogram of the total radial astrometric correction relative to the Gaia-CRF3 reference frame for all observations included in CSC 2.1, computed by combining the individual observation relative astrometry corrections with the stack absolute astrometry corrections.  \addedtext{The short-dashed red histogram reflects the distribution for which the observation stack absolute astrometry corrections were computed from matches with the Gaia reference catalog that were modified by human review ({\em i.e.\/}, $\hbox{\tt man\_astrom\_flag} = \hbox{\tt TRUE}$).  The long-dashed blue histogram identifies cases where no observation stack absolute astrometric correction could be computed due to a lack of reliable X-ray source--Gaia optical source matches.}}
\end{figure}

Human quality assurance review is triggered if there are fewer than 3 matches, if the computed radial translation exceeds $1\farcs0$, or if the reference catalog source density exceeds 100 sources per square arcminute and there are less than 10 matches.  \addedtext{For the latter two cases, human review typically confirms the matches identified by the automated algorithm, and where necessary eliminates incorrect matches when multiple close candidates are present in the reference catalog.  The applied absolute astrometric correction should be reliable in these cases.  When there are fewer than 3 matches, human judgement is required to confirm the matches.  In many cases we find that additional faint X-ray detections are present that visually match reference catalog sources, but are not considered by the automated algorithm ({\em e.g.\/}, because of low S/N, or with off-axis angle $\theta > 10'$).  Nevertheless, these additional detections can be used as a visual pattern match to help confirm the matches selected by the algorithm.  Although the astrometric correction should still be reliable in this case, the accuracy may be somewhat degraded because of the limited number of matches included in the solution.  If the matches are modified by human review then the {\tt man\_astrom\_flag} is set to true for the observation stack.  If human review is unable to confirm good matches between the X-ray detections and the reference catalog then no absolute astrometric correction is applied.  In such cases, the observation stack astrometric translations {\tt deltax} and {\tt deltay} are set to zero.  Approximately $2\%$ of observation stacks in CSC~2.1 do not have absolute astrometric corrections.}

Figure~\ref{fig:gaiaastrom} presents a histogram the total radial astrometric correction to the Gaia reference frame for all observations included in CSC 2.1, computed as the magnitude of the vector sum of  the relative astrometric correction for the individual observation (\S~\ref{sec:ofa}) and the associated stack absolute astrometric correction.  This is a measure of the absolute astrometric error for individual \Chandra\ observations relative to Gaia-CRF3\null. \addedtext{Human review cases and cases where no correction is applied are identified by the red and blue histograms, respectively.}

A transformation matrix representation of the applied astrometric correction is recorded in the {\tt STKXFM} (``transform'') Hierarchical Data Unit (HDU) included in the CSC~2.1 aspect solution ({\tt asol3}) file for each observation included in the stack.  The stack astrometric corrections and a flag indicating whether the matches were manually adjusted as part of the review are recorded in the Stack Detections Table (Table~\ref{tab:stackobsproperties}) starting with release~2.1\null.

The computed astrometric updates to the world coordinate systems for each observation stack are applied to all stack detections, observation-stack level FITS data products, individual observation FITS data products for all observations that comprise the stack, and master source level FITS data products derived from the stack.  This ensures that astrometric information for all catalog products is corrected to the Gaia-CRF3 reference frame.  {\em This is done even for properties and FITS data products that are otherwise unchanged from CSC 2.0\null}.  Note that properties measured in defined apertures, such as aperture photometry, are unaffected by the astrometric update since the positions of the apertures are defined in terms of the \Chandra\ ``sky'' coordinate system \citep{mcdowell_coordsI}, which does not depend on the mapping to world coordinates.

\subsection{Combining Stacked Observation Detections to Identify Master Sources} \label{sec:matchdets}
\subsubsection{Matching Stacked Observation Detections and Combining Detection Positions} \label{sec:masterpos}
Matching compact detections from multiple overlapping observation stacks to identify master sources on the sky is performed using the algorithm described in Appendix~A of \paperI, which should be consulted for more detail.  The method is based on evaluation of the overlap fractions between the detections' 90\% ECF PSF apertures, and enables the strong dependence of PSF extent on off-axis angle to be considered explicitly.  This approach handles cases where a far off-axis detection in one observation stack is resolved into multiple detections close to the optical axis in another observation stack more robustly than a simple positional cross-match.

Since the individual 90\% ECF PSF apertures will vary from observation to observation within the stack, a composite aperture is determined for each stacked observation detection for input to the algorithm.  This is done by summing the individual observation PSFs weighted by their MLE model amplitudes and relative exposure, $A\times\mathcal{E}$, determined in \S~\ref{sec:mle}\null.  The composite 90\% ECF PSF aperture for a stacked observation detection which is used as input to the detection matching algorithm is defined to be the best fitting ellipse that includes 90\% of the composite PSF encircled counts.

Once the matching step is completed, each stacked observation detection is associated with either a single master source using unique linkage, or with multiple master sources using ambiguous linkage. Conversely, each master source is uniquely associated with one or more stacked observation detections (which must all be from different observation stacks), and in addition may be ambiguously associated with one or more stacked observation detections that each have additional ambiguous links to other master sources.

Compact source positions in the Master Sources table, and the {\em internal\/} component of the associated position uncertainties, are computed by combining the fitted positions and model-fit position uncertainties of the {\em contributing\/} stacked observation detections identified above, using the method described in Appendix~B of \paperI\null.  Contributing detections are defined as those that are uniquely linked to the master source, with the exception that for observation stacks constructed solely from ACIS observations, the stacked observation detections are not used if the estimated photon pile-up fraction \citep{Davis_pileup} exceeds $\sim\!10\%$ in all of the individual per-observation detections that comprise the stacked observation detection, {\em unless\/} this threshold is exceeded for all stacked observation detections that are uniquely linked to the master source (in which case, no other {\em un\/}-piled-up stacked observation detections exist for the master source).  Unless otherwise noted, master source properties are  evaluated using only contributing stacked- or per-observation detections, as appropriate.

The {\em total\/} position uncertainty of a source is estimated by adding the internal component of the position uncertainty of the source in quadrature to a global estimate of the catalog systemic astrometric error.  The latter value is determined using the approach of \citet{2011ApJS..192....8R} by comparing the distribution of total position uncertainties to a Raleigh distribution.  

For CSC~2.0 the catalog systemic astrometric error computed in this manner is $0\farcs71$ {\em in each axis\/} for 95\% confidence, based on comparison of the positions of 17,703 X-ray sources that are unambiguously matched with SDSS DR13 sources.  \addedtext{This value} is roughly 80\% larger than the catalog systemic astrometric error estimate estimated by \citet{2011ApJS..192....8R} for CSC 1.0, which was $0\farcs16$ $1\sigma$ in each axis, equivalent to $0\farcs39$ for 95\% confidence assuming Gaussian statistics.  Since CSC 1.0 only includes observations released publicly prior to 2009, this increase is likely due to a slow degradation of the \Chandra\ absolute pointing accuracy with epoch.  Temporal variation of the catalog systemic astrometric error is {\em not\/} modeled in CSC 2.

\addedtext{Since the absolute astrometric correction described in \S~\ref{sec:absastrom} is not applied to individual observation stacks in CSC~2.0, we use the distribution of separations between the detected sky positions for the same X-ray source across multiple overlapping observations to evaluate the reliability of the reported position uncertainties in CSC~2.0\null.}  Based on \addedtext{this analysis}, we conclude that the actual position uncertainty for a source \addedtext{in CSC~2.0} should be consistent with the reported position uncertainty for roughly 95\% of the sources included in the catalog.  For sources in a small subset of single-observation stacks that do not overlap any other observations \addedtext{(and which therefore will not have an observation relative astrometry correction applied in \S~\ref{sec:ofa}), we conclude that} the absolute position uncertainty may be significantly larger, up to $\sim\!2\farcs0$\addedtext{, based on the worst-case observation-to-observation detection position separations observed.}

\addedtext{The 95\% confidence catalog systemic astrometric error for CSC~2.1 is $0\farcs29$ in each axis, based on 44,573 X-ray source position matches with Gaia DR3\null.  This decrease in the catalog systemic astrometric error estimate compared to CSC~2.0 results from the application of the absolute astrometric correction for individual observation stacks.}

In these releases of the catalog, all stacked observation extended detections are combined manually to form master sources.  This is done because the extended detection regions identified automatically by \mkvtbkg\ may not be adequate in a few circumstances.  For example, differentiation between large source polygons and readout streak polygons may not be optimal for bright, extended sources such as supernova remnants where there may be many adjacent columns that include a mix of source photon events and out-of-time readout streak events.  As a guide for the human reviewer, a candidate combined master source convex hull is created by re-gridding the convex hulls from each observation stack to a common pixel grid, and then including in the candidate master convex hull each pixel that is included in the convex hulls for at least 20\% of the number of contributing observation stacks.  The master convex hull created in this way is either accepted, which is usually the case, or modified manually to correct for any defects in one or more of the contributing stacked observation convex hulls.

Unlike compact detections, once a master source convex hull is created in this way, that master source convex hull is used to re-evaluate the stacked observation and per-observation detection properties for the source, so that the same extended source region definition is used for all levels. \addedtext{Tabulated source properties for these extended detections are extracted from the region included within the master source convex hull.}

\subsubsection{Source Naming} \label{sec:naming}
Following release~1 of the catalog, CSC~2 sources are named in compliance with the International Astronomical Union (IAU) Recommendations for Nomenclature.\footnote{\url{http://cdsweb.u-strasbg.fr/Dic/iau-spec.html}} Specifically, CSC~2 sources are assigned designations of the form ``2CXO J{\em HHMMSS.s\/}$\pm${\em DDMMSS\/}[Z],'' where {\em HHMMSS.s\/} and $\pm${\em DDMMSS\/} are, respectively, the sexagesimal representations of the source ICRS right ascension and declination (computed as described in \S~\ref{sec:masterpos}) {\em truncated\/} to the indicated precision.  The suffix [Z] is normally absent, but may be present for compact sources where necessary to differentiate between nearby sources that would otherwise have the same designations.  In such cases, the name of the source with the numerically smallest right ascension will not include a suffix and the other sources' names will include a suffix that is one of the letters ``A,'' ``B,'' and so on, in that order.  The names of all extended sources include the letter ``X'' as a suffix to simplify their identification.

To comply with the IAU Recommendations for Nomenclature, once a source is included in a released version of the catalog that source designation will remain unchanged in future catalog releases that use the same name prefix ({\em e.g.\/}, ``2CXO'').  Therefore, master sources identified in unmodified or modified stacks in CSC 2.1 that were present in CSC 2.0 will retain the source name previously published in CSC 2.0 {\em even if the source positions are updated\/}.  As a result, small discrepancies can arise between the source position and the source designation, so the latter should never be used as a proxy for the former when full accuracy is required.  If new observations included in CSC 2.1 resolve an apparently single source included in CSC 2.0 into multiple distinct sources, or conversely if multiple close sources present in CSC 2.0 are merged to a single source when detected in CSC 2.1 (which can happen because the increased background due to the additional observations can change source detectability in some circumstances), then the previous source designation is retired and the CSC 2.1 sources will be assigned new names based on their positions, {\em unless\/} (in the case of a split) one of the components is clearly and unambiguously associated with the pre-existing source.

\subsection{Compact Source Extent Estimates}
Observed and intrinsic (deconvolved) spatial extent estimates for compact detections are computed using a similar approach to release~1 of the catalog.  At the individual observation level, the observed spatial extents of the detection and the local PSF in each energy band are estimated using a rotated elliptical Gaussian parameterization, where the $1\sigma$ radii of the ellipse semi-axes and the position angle of the semi-major axis are determined using a wavelet-based approach (for details, see \S~3.6 of \paperI)\null.  Since this method and the \wavdetect\ source detection algorithm both utilize Marr (``Mexican Hat'') wavelet functions, in this release we utilize \wavdetect\ to provide initial parameter guesses for the detection extent and centroid position.  The former is set to the \wavdetect\ wavelet scale, $a_i$, that maximizes the correlation integral (equation~\hbox{[2]} of \paperI), while the centroid position is determined by maximizing a circularly symmetric wavelet with scale $a_i$\null.  Therefore, the position of the pixel where the maximum occurs is found first, and then the orientation and size of the ellipse are optimized for subsequently.

As in the earlier release, for individual observation detections included in the Per-Observation Detections table, the catalog provides a circularly-symmetric root-sum-square intrinsic source extent estimate ($a_{\rm rss}$, equation~\hbox{[10]} of \paperI) as a proxy for a true deconvolved source size.  The telescope dither motion effectively increases the raw instrumental PSF extent during an ACIS exposure (typically $\sim\!3.2\,{\rm s}$), and so an additional Gaussian blur is added in quadrature to the model PSF $\sigma$ to increase the total aspect blur to $0\farcs2$ prior to computing the effective PSF width.  

For stacked observation detections and master sources, the $a_{\rm rss}$ values for the observations included in the valid stack (Stacked Observation Detections table) or included in the valid stacks of observation stacks with unique linkage to the master source (Master Sources table) are combined using the approach described in \paperI.

\subsection{Aperture Photometry} \label{sec:aperphot}
Aperture photometry in CSC~2 is performed at the individual observation detection, stacked observation detection, and master source levels.  For the first of these, aperture photometry is computed using data from a single observation.  At the stacked observation detection level, data from all of the observations that comprise the valid stack for the detection are combined to produce average aperture photometry estimates.  Finally, at the master source level data are combined from observations of detections uniquely associated with a source.  This is done both as averages across all observations and also as averages across observations segregated by a Bayesian Blocks temporal variability analysis (see \S~\ref{sec:bblocks}) that ensures the individual observation aperture photometry is consistent across all of the observations in all energy bands for each Bayesian Block.

Similar to release~1, aperture photometry quantities (including net counts, count rates, and photon and energy fluxes) for compact detections are computed from counts and exposure information in independent source and background region apertures defined in \S~\ref{sec:srcaper}.  The Bayesian formalism of \citet{2014ApJ...796...24P} is used to determine the marginalized posterior probability distribution (MPDF) for photon flux for each detection and energy band.

When computing aperture photometry properties, if the spillover counts from the PSF wings of one detection will affect the flux determination of a spatially adjacent detection, then the detections are analyzed together.  In such cases, the MPDFs for the adjacent detections are determined simultaneously, rather than separately as they were in release~1.  Sets of adjacent detections are termed ``photometric bundles,'' and for CSC~2 they are constructed by modifying the original detection bundles defined in \S~\ref{sec:srcaper}.  The details of photometric bundle construction are discussed in a technical memorandum by \citet{primini_aperphot} and the salient points are repeated below for convenience.  We designate the detections included in the original detection bundle as ``primary detections.''  The photometric bundle background region is defined as the union of the background regions of the primary detections.  

Adjacent detections with source regions $\{R_i\}$ that overlap the background region are promoted to photometric bundle membership if the spillover counts from their source regions are comparable to the statistical error in the background, even if their source regions do not overlap the source regions of any primary detections.  This condition is approximated as
\begin{displaymath}
(1 - f_i)\,C_i\gtrsim n\sqrt{B},
\end{displaymath}
where $f_i$ is the enclosed counts fraction of the local PSF in source region $R_i$, $C_i$ is the total number of counts in source region $R_i$, $B$ is the total number of counts in the background region, and $n$ is a factor of order unity.  If this inequality is satisfied then the contaminating detection is included in the photometric bundle as a ``secondary detection''; otherwise the contaminating detection source region is excluded from the background region.

If a photometric bundle includes two or more primary detections, $i$, $j$, and $C_j/C_i>5000$ for any $i$, $j$, then detections $i$ are demoted from the current photometric bundle and form a new bundle.  This ensures that very bright detections adjacent to much fainter detections are included in the background when computing the joint posterior probability distribution (JPDF) for the expected net counts of the faint detections, which experiment shows is more computationally robust than including all of the detections in a single photometric bundle.

Once the set of detections that comprise a photometric bundle are identified, and assuming statistical independence of the individual source and background region counts, the JPDF for obtaining expected net counts $\{\mu_{s_i}\}$ and expected background counts $\mu_b$, given total source region counts $\{C_i\}$ and total background region counts $B$, can be written as \citep[equation B1 of][]{2014ApJ...796...24P}
\begin{equation}
\label{eqn:jpdf}
P(\mu_{s_1} ... \mu_{s_n}, \mu_b | C_1 ... C_n, B) = K P(\mu_b) {\rm Pois}(B | \mu_b) \prod_{i=1}^n P(\mu_{s_i}) {\rm Pois}(C_i | \mu_{s_i}),
\end{equation}
where $K$ is a normalization constant, ${\rm Pois}(N | \mu_x)$  is the Poisson probability for obtaining $N$ total counts given $\mu_x$ expected total counts, and $P(\mu_{s_i})$, $P(\mu_b)$ are the prior probability distributions for the expected counts of the detections and  background, respectively.

The expected total counts $\{\mu_{s_i}\}$ and $\mu_b$ in the source and background region apertures are then \citep[{\em cf.} equation 3 of][]{2014ApJ...796...24P}
\begin{gather*}
\mu_{s_i} = E_i \times \left[ \sum_{j=1}^n f_{ij} s_j + \Omega_i b \right]; \quad \mu_b = E_b \times \left[\sum_{i=1}^n g_i s_i + \Omega_b b\right],
\end{gather*}
where $f_{ij}$ is the fraction of the PSF for detection $j$ enclosed in source region aperture $R_i$, $g_i$ is the fraction of the PSF for detection $i$ enclosed in the background region aperture, $\Omega_i$ is the area of source region aperture $R_i$, $\Omega_b$ is the area of the background region aperture, $E_i$ is the exposure of source region aperture $R_i$, and $E_b$ is the exposure of the background region aperture.

The MPDF for a single detection is obtained by marginalizing the JPDF over all other detections and the background.  Assuming non-informative gamma-prior distributions for $\{s_i\}$ and $b$, and setting the normalization constant to unity, the MPDF for detection $i$ can be written as
\begin{equation}
\label{eqn:mpdf}
P(s_i | C_1 ... C_n, B)= \idotsint\limits_{b,s_j\ne s_i} db\,\left({\prod_{j\ne i}ds_j}\right)P(\mu_{s_1} ... \mu_{s_n}, b | C_1 ... C_n, B),
\end{equation}
where we have combined equation~(17) of \citet{2014ApJ...796...24P} with equation~(\ref{eqn:jpdf}) above.

Evaluating the MPDF for a particular detection or background by direct numerical integration of the JPDF hyper-cube over all other dimensions is feasible but computationally expensive.  Instead, the maximum-likelihood values for the detection and background net counts, $\{\hat{s_i}\}$, and the associated Gaussian statistical errors, $\{\sigma_{\hat{s_i}}\}$ \citep[equations 8 and 9 of][]{2014ApJ...796...24P}, are evaluated using \sherpa\ optimization, and the MPDFs are subsequently constructed \citep{primini_aperphotcomb} using the \sherpa's PyBLoCXS Markov chain Monte Carlo routines.

In the context of optimization in \sherpa, the raw aperture counts can be considered as a 1-dimensional dataset,
\begin{equation}
\label{eqn:aperphotdata}
\mathcal{D}=(\{C_i\}, B),
\end{equation}
of length $(n+1)$ where there are $n$ detections in the photometric bundle.  The model, 
\begin{equation}
\label{eqn:aperphotmodel}
\mathcal{M}=(\{\mu_{s_i}\}, \mu_b),
\end{equation}
with parameters 
\begin{displaymath}
\mathcal{P}=(\{s_i\}, b),
\end{displaymath}
has the same dimensionality.  The best fitting \sherpa\ model evaluated using the $C$-statistic \citep{1979ApJ...228..939C} yields essentially the maximum-likelihood net counts values, while the associated Gaussian statistical errors can be determined by computing the covariance matrix.  This approach has the added advantage that the same technique can be used when combining data from multiple observations.  

The detection MPDFs are generated from MCMC samples of the JPDF\null.  Each MCMC draw consists of a set of $\mathcal{P}$ parameter values determined by the MCMC algorithm to accurately sample the JPDF\null.  Five thousand draws are generated, of which the first 100 are discarded as burn-in.  Modes and confidence bounds for each parameter are estimated from uniformly-sampled MPDF arrays derived from the distribution of draws for that parameter.  The samples range from $\hat{s} - 5\sigma_{\hat{s}}$ to $\hat{s} + 5\sigma_{\hat{s}}$, where $\hat{s}$ and $\sigma_{\hat{s}}$ are defined above.  The MPDF is explicitly renormalized to unity at the end of the calculation.  The functional form of the MPDFs is assumed to be a $\gamma$ distribution,
\begin{equation}
\label{eqn:mpdf_gammadist}
\hbox{\rm MPDF}={{\beta^\alpha s^{\alpha-1}e^{-\beta s}}\over{\Gamma(\alpha)}},
\end{equation}
where parameters $\alpha$ and $\beta$ for the posterior $\gamma$ distribution can be estimated from the draws $\{s_i\}$ using the relations
\begin{gather*}
\alpha=\Mean(\{s_i\})^2/\Var(\{s_i\}); \quad \beta=\Mean(\{s_i\})/\Var(\{s_i\}).
\end{gather*}
Confidence bounds are derived by integrating the MPDF symmetrically from the mode of the distribution until the desired confidence level is reached.

Combining data from multiple observations to produce average aperture photometry estimates for a detection is accomplished by treating aperture data from each observation as a separate 1-dimensional dataset, given by equation~(\ref{eqn:aperphotdata}), with models defined by equation~(\ref{eqn:aperphotmodel}) and using \sherpa\ to perform a simultaneous fit across all datasets \citep{primini_aperphotcomb}.  The simultaneous fit forces each of the parameters $\{s_i\}$ to be the same across all observations, while allowing the parameter $b$ to vary from observation to observation.

Although the MCMC draws computed by the \sherpa\ pyBLoCXS algorithm are generally satisfactory, for roughly 1\% of the (mostly faint) detections and sources the algorithm fails to converge.  In CSC 2.1, if the original algorithm does not converge then a more robust and computationally expensive MCMC algorithm, the No-U-Turn Sampler (NUTS) \citep{2014JMLR...15.1593H} implemented in {\tt PyMC3} \citep{2016PJCS....2..E55S}, is used.  This improves the reliability of aperture photometry flux computations, particularly in the extreme low-count regime.  The details of the approach used are provided by \citet{primini_pymc3}, but the essential differences when using the NUTS are (a)~when computing the MPDF (equation~\ref{eqn:mpdf}), uniform priors (with zero lower bound and predetermined upper bounds that span the expected maximum values for the property being computed) are used instead of $\gamma$-priors for $\{s_i\}$ and $b$,  and (b)~multiple MCMC chains are computed with 1000 draws (after 1000 burn-in samples) and the distributions of the properties being computed are determined from the combined set of draws smoothed with an Epanechnikov Kernel Density Estimator, scaled to the RMS about the mean.

A word of caution is necessary when interpreting ACIS ultrasoft and soft energy band aperture photometry photon fluxes from observations obtained after $\sim\!2013$, {\em i.e.\/}, primarily new observations in CSC 2.1\null.  As discussed in \S~2.5.2 of \paperI, the instrument responses used to compute catalog photometric properties are computed at the monochromatic effective energies of the individual energy bands, rather than by integrating the (typically not available {\em a priori\/}) source spectrum over the energy band.  For the ultrasoft and soft energy bands, the ACIS effective area has continued to decrease over time because of buildup of a spatially varying contaminant, and the rate of decrease accelerated after $\sim\!2013$.  Combined with the significant slope of the intrinsic detector quantum efficiency as a function of energy in the ultrasoft and soft energy bands, for a fixed monochromatic energy this variation of the detector effective area with time will cause the detection photon fluxes to be overestimated for recent observations.\footnote{See \url{https://cxc.cfa.harvard.edu/ciao/why/low_energy_mono.html}.}  Prior to $\sim\!2013$, the effect in the soft energy band was limited to $\sim\!10\%$, but by the end of 2021 the overestimate could be as large as $\sim\!70\%$, depending on the source spectrum.

\subsubsection{Determining Detection Significance}
The likelihood, equation~(\ref{eqn:likelihood}), is the fundamental metric used to decide whether a detection is included in CSC~2 through the application of the thresholds described in \S~\ref{sec:lthresh}.

The likelihood of a detection is closely related to the probability, $P_{\rm Pois}$, that a Poisson distribution with a mean background in the source region aperture would produce at least the number of counts observed in the aperture.  This quantity, termed the {\em detection significance\/}, is also reported in the catalog.  Smoothed background maps are used to estimate the mean background, and the detection significance is expressed in terms of the quantities $\sigma$, $z$, in a zero-mean, unit standard deviation Gaussian distribution that would yield an upper integral probability $P_{\rm Gaus}$, from $z$ to $\infty$, equivalent to $P_{\rm Pois}$. That is, $P_{\rm Pois}=P_{\rm Gaus}$, where
\begin{displaymath}
P_{\rm Gaus} = \int_z^\infty\! dx\, {{e^{-x^2/2}} \over {\sqrt{2\pi}}}.
\end{displaymath}

The {\em flux significance\/} of a detection is the ratio of the mode of the photon flux MPDF, equation~(\ref{eqn:mpdf}), to its average uncertainty, $\sigma_e=(\sigma_{\rm hi} - \sigma_{\rm lo})/2$, where $\sigma_{\rm lo}$ and $\sigma_{\rm hi}$ are the lower and upper $1\sigma$ confidence limits computed from the MPDF:
\begin{displaymath}
\int_{-\infty}^{\sigma_{\rm lo}}\!ds\,P(s|C, B)=\int_{-\infty}^{\sigma}\!dx\,  {{e^{-x^2/2}} \over {\sqrt{2\pi}}}; \quad
\int_{\sigma_{\rm hi}}^\infty\!ds\,P(s|C, B)=\int_{\sigma}^\infty\!dx\,  {{e^{-x^2/2}} \over {\sqrt{2\pi}}}.
\end{displaymath}  
This is a simple statistic that is robust to calculate, easily interpretable by non-expert users, and consistent with the classical S/N definition for high count sources.

%Added next line for arXiv to stop subsubsection heading being at bottom of page
\vfil

\subsubsection{Bayesian Blocks Analysis} \label{sec:bblocks}
Simultaneous fitting of single intensities to data from multiple observations works best when the intensities in the contributing observations share the same parent distribution.  At the master source level, when more than one observation contributes aperture photometry information for the source, a Bayesian Blocks analysis \citep{2013ApJ...764..167S} is performed to group observations into blocks that have consistent average observation fluxes in each energy band across all observations in the block.

Given a set of $n$ observations with flux measurements, $\{O_j(P(s_j|\dots))\}$, the analysis groups them into blocks $\{B_i\}$ such that a constant flux is consistent with the flux measurements in each block.  The probability of obtaining a set of blocks $\{B_i\}$ given a set of observations $\{O_j\}$ may be written as 
\begin{displaymath}
P(\{B_i\}|O_j) = P(N_{\rm Blocks}) \prod_{i=1}^{N_{\rm Blocks}} F(B_i | O_j \in B_i),
\end{displaymath}
where the function $P(N_{\rm Blocks})$ is the prior probability distribution for the number of blocks $N_{\rm Blocks}$.  Since {\em a priori\/} $N_{\rm Blocks}\ll n$ is assumed to be much more likely than $N_{\rm Blocks}\approx n$, the prior probability distribution is assumed to have the form \citep{2013ApJ...764..167S}
\begin{displaymath}
P(N_{\rm Blocks}) \sim \gamma^{N_{\rm Blocks}},\quad 0<\gamma<1,
\end{displaymath}
which assigns a smaller probability to a larger number of blocks

The fitness function, $F(B_i | O_j \in B_i)$, is a measure of how well the data in the observations included in each block may be represented by a constant source flux.  The fitness function is defined as the product of the MPDFs of the individual observation flux measurements \citep{primini_aperphot},
\begin{displaymath}
F(B_i | O_j \in B_i) = \int_0^\infty\!ds\, \left[ \prod_{j|O_j \in B_i} P(s_j|\dots)\right],
\end{displaymath}
as this maximizes the fitness when the flux measurements for all of the observations included in a block are consistent with a single flux.

Individual MPDFs are re-gridded to a common grid using an Akima interpolation \citep{1970JACM...17..589A} of $\log P(s_j|\dots)$ with 1000 bins, and data from different energy bands are summed when computing $\log[F(B_i)]$ so that a single set of blocks will describe data from all bands.  Separate sets of blocks are determined for ACIS and HRC observations as their energy bands are defined independently.

The actual blocking is determined by selecting the set of blocks, $\{B_i\}$, that maximizes $\log[P(\{B_i\}|O_j)]$, where
\begin{eqnarray*}
\log[P(\{B_i\}|O_j)] &=& N_{\rm Blocks}\log\gamma + \sum_{i=1}^{N_{\rm Blocks}}\log[F(B_i)] \\
&=& \sum_{i=1}^{N_{\rm Blocks}}\left[\log[F(B_i)]+{\log\gamma}\right], \\
\end{eqnarray*}
where the parameter ${\log\gamma}$ is set to $-4.75$ based on simulations.  The set of Bayesian Blocks for each energy band is computed independently, after which the final set of multi-energy-band Bayesian Blocks is determined by intersecting the individual energy band sets.

The Bayesian Blocks analysis is performed twice, once with the observations sorted in order of observing epoch, and once with the observations sorted in order of increasing flux (specifically, the mode of the MPDF).  The former segregates observations of a variable source by time, so that if the source flux changes, prior and subsequent observations will be separated into different blocks {\em even if the source flux and spectral shape return to their previous values\/}.  For the flux-ordered observation analysis, all observations that have the same multi-band source fluxes will be grouped together in the same block even if they are separated temporally by other observations that have different source fluxes.  The latter enables multiple observations of a periodic variable source obtained at the same phase, or the multiple observations of a particular state of a flaring source, to be combined for improved S/N\null.

\begin{figure}
\epsscale{0.65}
\plotone{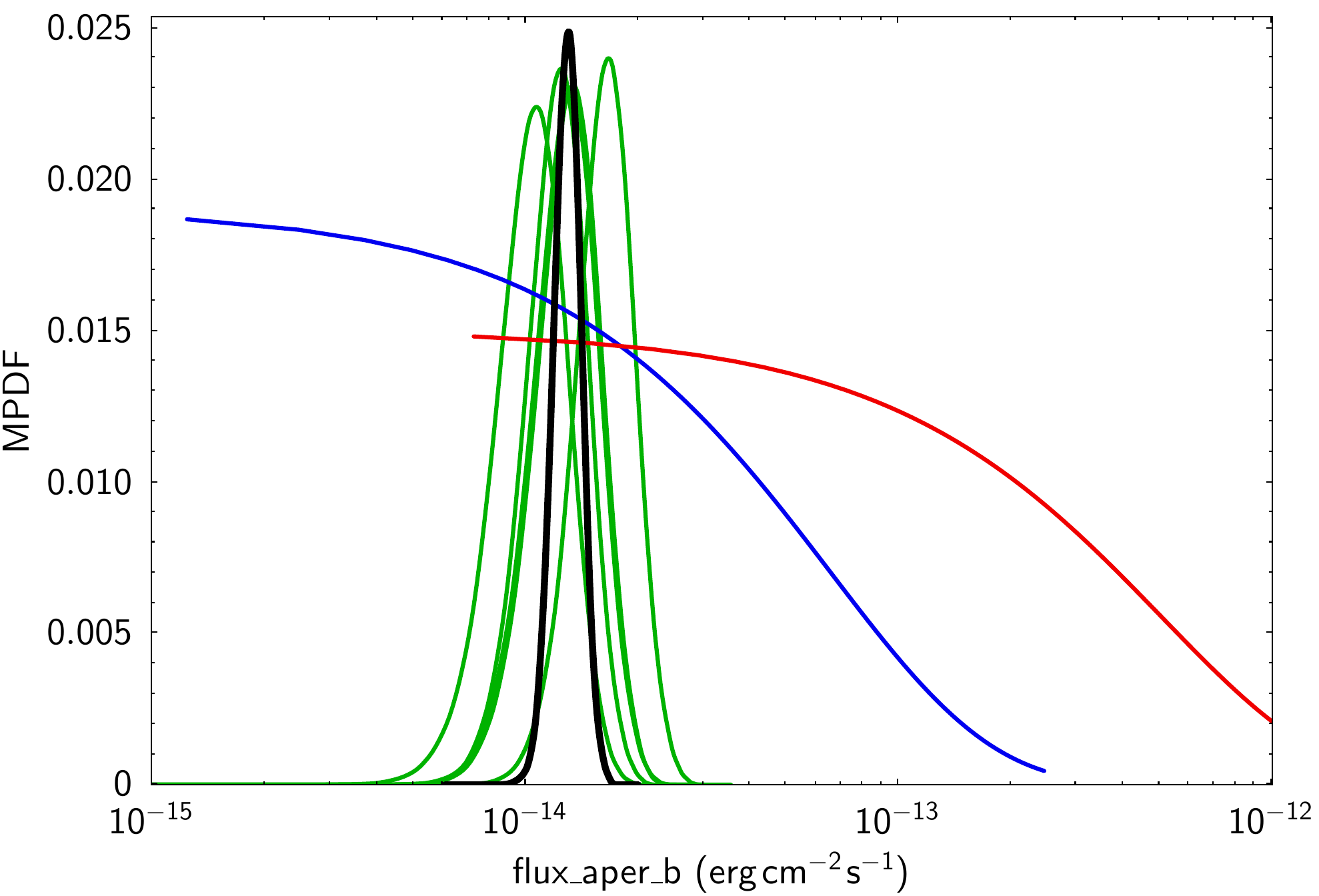}
\caption{\label{fig:bayesblks}
Marginalized probability density functions (MPDFs) for ACIS broad band energy flux in 7 observations contributing to master source $\rm 2CXO\,J004152.6-092213$\null.  The green, blue, and red curves represent the MPDFs for observations included in the three identified flux-ordered Bayesian Blocks.  The flux-ordered block with the longest exposure time includes the observations in green.  The black curve is the master source ``best-estimate'' MPDF, which combines data from all observations included in that block.  The observations included in the remaining blocks shown in blue and red are photometric upper limits and so have exponential MPDFs.}
\end{figure}

Because the physical state of a variable source may differ between observations included in different Bayesian Blocks, aperture photometry properties and properties derived from aperture photometry measurements (such as hardness ratios, model energy fluxes, and spectral model fits) are computed for each block separately.  The properties for all blocks are recorded in the {\tt blocks3} data product for the master source.  ``Best estimate'' properties included in the master sources table are those from the flux-ordered block with the largest total exposure time (Figure~\ref{fig:bayesblks}).  The master sources table also includes ``average'' values of the basic aperture photometry properties, computed from all contributing observations, to facilitate comparison with other catalogs.

\subsubsection{Non-Detections} \label{sec:nondet}
Once a \addedtext{master} source is identified by the catalog processing pipelines, all observation stacks that overlap the location of the source on the sky and do not include a detection at the location of the source are identified.  For each such observation stack an artificial detection, termed a ``non-detection,'' is then created at the position of the source.  The full set of detection properties is not evaluated for non-detections.  Rather, they are used only to compute upper limits for the detection's aperture photometry fluxes at the epochs of the observations that comprise the observation stack, and these photometric upper limits are considered when evaluating inter-observation source variability (see \S~\ref{sec:intervar}).  Aperture photometry fluxes for non-detections are computed in a canonical source region defined to be the largest PSF 90\% ECF aperture ellipse across the various energy bands for the detector, as well as in the individual PSF 90\% ECF apertures for each energy band.

The MPDFs for non-detections are computed as described above, except that the functional form of the MPDF is usually forced to have an exponential distribution, {\em i.e.\/},
\begin{displaymath}
\hbox{\rm MPDF} \sim e^{-\alpha s},
\end{displaymath}
instead of a $\gamma$ distribution (equation~\ref{eqn:mpdf_gammadist}).  This ensures that the mode of the MPDF is zero, consistent with a photometric upper limit.  A $\gamma$ distribution is retained instead of an exponential distribution if
\begin{displaymath}
\Mean(\{s_i\}) - 3\times\sqrt{\Var(\{s_i\})} > 0,
\end{displaymath}
which corresponds to the case where a source was not detected in the observation stack, but is bright enough that a photometric upper limit is not appropriate.

\subsection{Limiting Sensitivity}
Limiting sensitivity data are computed for each observation stack independently and report the estimated fluxes required for a point source to be detected in each energy band at the \marginalvalue\  and \truevalue\ likelihood thresholds.  Catalog limiting sensitivity is subsequently determined by combining the individual observation stack limiting sensitivity data on a {\em Hierarchical Equal Area iso-Latitude Pixelization\/} \citep[HEALPIX;][]{2005ApJ...622..759G} grid of order $k=16$ ($N_{\rm sides}=2^{16}$), corresponding to a pixel angular resolution $\theta_{\rm pix}\simeq3\farcs 22$, taking the faintest sensitivity value from all observation stacks overlapping each HEALPIX pixel.  The ``NESTED'' HEALPIX pixel ordering scheme is used.

\begin{figure}
\epsscale{0.5}
\plotone{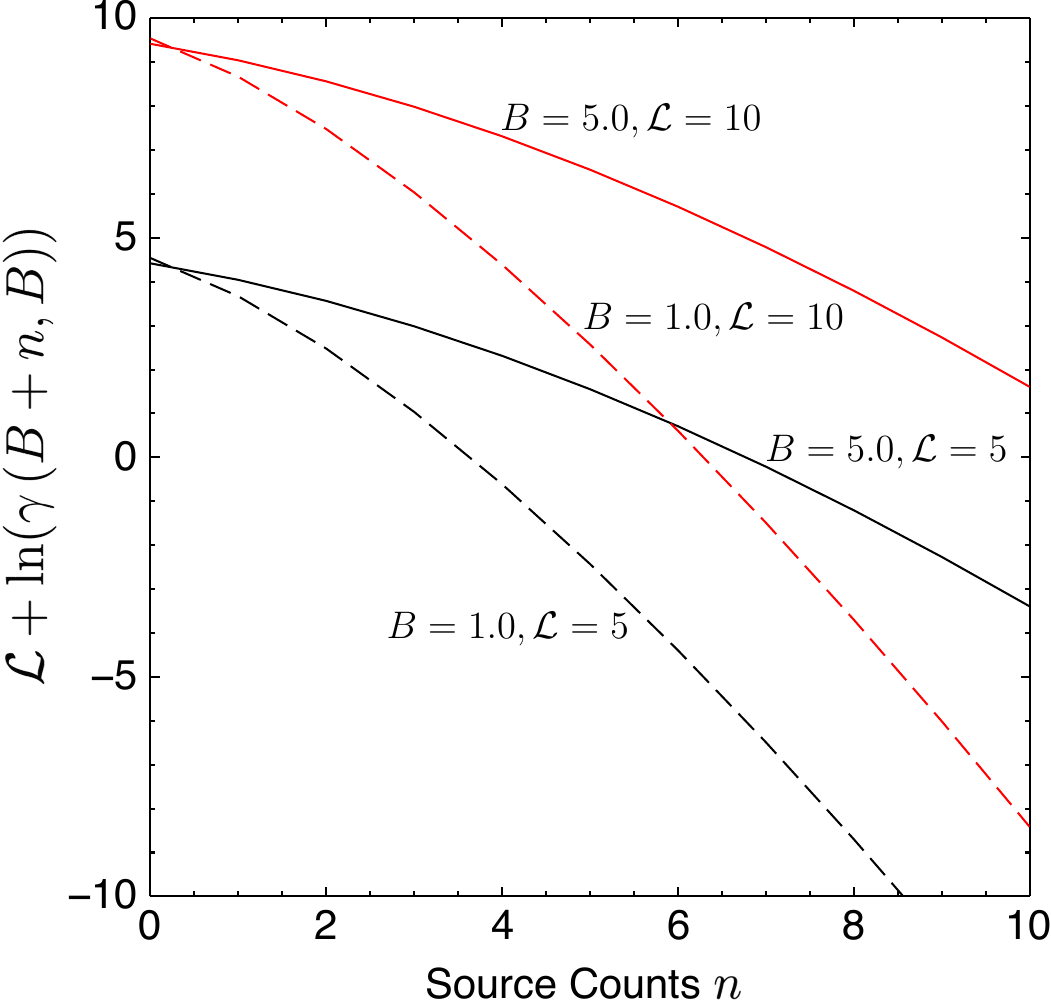}
\caption{\label{fig:limsens}
The function $\likelihood + \ln(\gammafunc{(B+n, B)})$ varies smoothly with $n$ and has a single root that can be easily determined numerically.  Example plots of $\likelihood + \ln(\gammafunc{(B+n, B)})$ as a function of source counts $n$ are shown for two representative values of $B$ and $\likelihood$.}
\end{figure}

Since the limiting sensitivity of an observation stack is defined by the likelihood thresholds discussed in \S~\ref{sec:lthresh}, we follow the approach used by the {\em 2XMM\/} catalog \citep{2009A&A...493..339W} to determine a linear empirical relationship between the log-likelihood, $\likelihood$, and net source counts, and use that relationship to estimate the source photon flux corresponding to the likelihood threshold.  The algorithm is described in detail in Appendix~A of \citet{2007A&A...469...27C}.  Recasting their equation~(A.1), \citep[see][]{primini_limsens},
\begin{displaymath}
\likelihood = -\ln(P(N\geq B + \hbox{\em crpoisim\/}\times E | B)),
\end{displaymath}
where $P(N\geq  M|B)$ is the cumulative Poisson probability of obtaining $M$ or more counts in a source aperture due to Poisson fluctuations if the average background in the aperture (obtained from the corresponding background map) is $B$, $E$ is the average exposure map value in the aperture, and the quantity $\hbox{\em crpoisim\/}\times E = \hbox{\em ctpoisim\/}$ has units of counts.\footnote{$E$ has units of $\hbox{\rm cm}^2\,\hbox{\rm s}\,\hbox{\rm count}\,\hbox{\rm photon}^{-1}$ and {\em crpoisim\/} has units of $\hbox{\rm photon}\,\hbox{\rm cm}^{-2}\,\hbox{\rm s}^{-1}$.} The cumulative Poisson probability is given by
\begin{displaymath}
P(N\geq B + \hbox{\em ctpoisim\/}|B) = \gammafunc{(B+\hbox{\em ctpoisim\/}, B)},
\end{displaymath}
where $\gammafunc{(a, x)}$ is the lower incomplete gamma function.  The quantity {\em ctpoisim\/} is then determined by finding the root of the equation
\begin{equation}
\label{eqn:ctpoisim}
\likelihood + \ln(\gammafunc{(B+\hbox{\em ctpoisim\/}, B)})=0.
\end{equation}
An example plot of $\likelihood + \ln(\gammafunc{(B+n, B)})$ vs.~$n$ is shown in Figure~\ref{fig:limsens} for two representative values of $\likelihood$ and $B$.  The function varies smoothly with $n$ and so its root is easily determined numerically.

Since source detection is performed at the stacked observation level, the procedure for determining limiting sensitivity in an energy band for any location within the field of view of an observation stack requires determining the background, $B$, and the average exposure map value, $E$, in the energy band in a source region aperture appropriate to the location.  The appropriate likelihood threshold at the location, $\likelihood$, is computed using equations~(\ref{eqn:Lthreshacis}) and (\ref{eqn:Lthreshhrc}), after which {\em ctpoisim\/}, and therefore {\em crpoisim\/}, are computed by finding the root of equation~(\ref{eqn:ctpoisim})\null.  The individual stacked observation limiting sensitivity point source photon fluxes, {\em crpoisim\/}, corresponding to the \marginalvalue\ and \truevalue\ likelihood thresholds are recorded in the {\tt sens3} file.

As described in \S~\ref{sec:detvalidation}, the CSC~2 detection thresholds are not based on likelihoods computed from Poisson fluctuations, but on likelihoods from fitting a point source model to the observed counts data.  We chose not to use the latter for computing sensitivity maps since that would be computationally expensive, requiring construction of PSFs for each sensitivity map pixel.  Instead, we assume that the flux associated with a likelihood derived from aperture quantities is related to the actual flux of a source detected at the source detection likelihood threshold, {\em i.e.\/}, $F_{\rm thresh}\propto\hbox{\em crpoisim\/}$, where $F_{\rm thresh}$ is the photon flux corresponding to the catalog likelihood threshold.

\begin{figure}
\epsscale{0.9}
\plotone{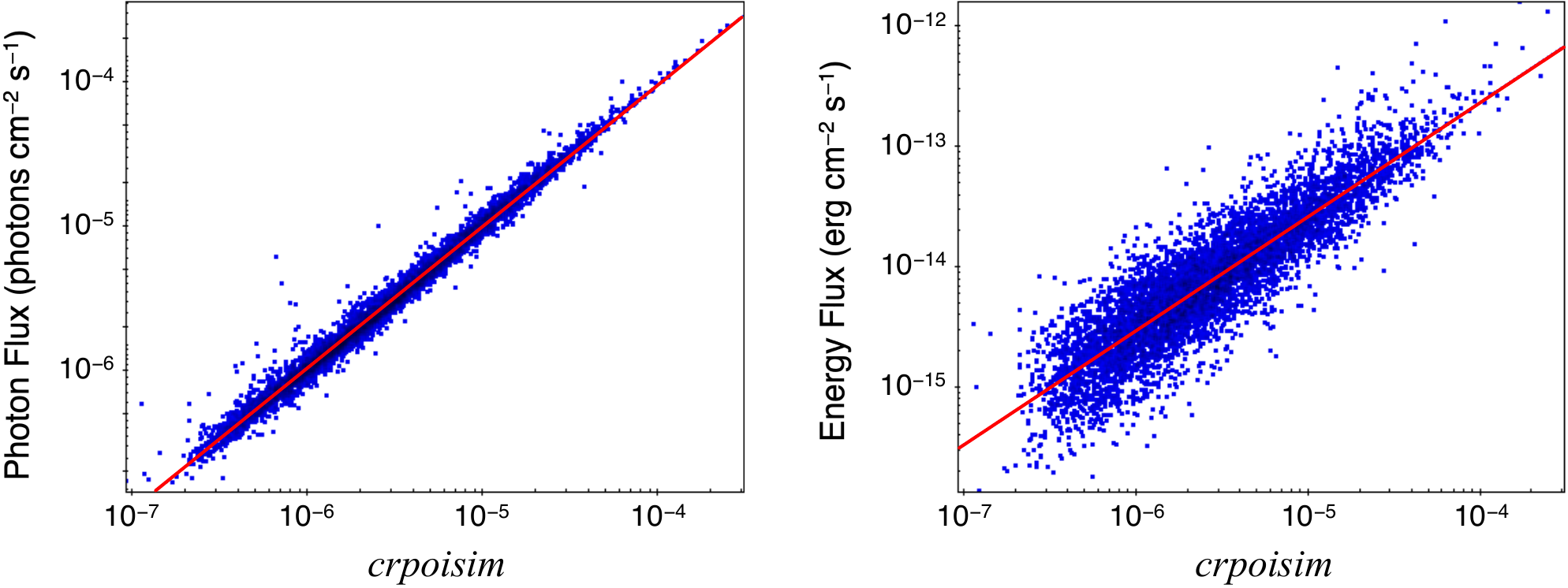}
\caption{\label{fig:limsensb}
{\em Left:\/} Comparison of ACIS broad energy band photon flux in the ECF 90\% PSF aperture, {\tt photflux\_aper90\_b}, with {\em crpoisim\/} determined using detection likelihoods, for a sample of isolated point sources; the best linear fit computed from equation~(\ref{eqn:limsenscoeffs}) using the broad band scale factors from Table~\ref{tab:limsenscoeffs} is shown in red.  {\em Right:\/} Same, except energy flux, {\tt flux\_aper90\_b}, is plotted.}
\end{figure}

To calibrate this relation, we selected a sample of isolated point sources and calculated {\em crpoisim\/} from the available aperture quantities, using the actual detection likelihoods, and compared these to actual photon fluxes and energy fluxes, as reported in the corresponding {\tt photflux\_aper90\_<band>} or {\tt flux\_aper90\_<band>} quantities for each energy band.  For all energy bands, the data are well-fit with relations of the form
\begin{equation}
\label{eqn:limsenscoeffs}
\log_{10} \left( F \right) = m \log_{10} \left( \hbox{\em crpoisim\/} \right) + c,
\end{equation}
where $F$ is the detection photon flux (and a similar form for energy flux).  See Figure~\ref{fig:limsensb} for example fits for the broad energy band.  The values of best-fitting scale factors $m$ and $c$ used in equation~\ref{eqn:limsenscoeffs} for each energy band\footnote{Source detection is not performed in the ACIS ultra-soft energy band, so limiting sensitivity is not computed for this  band.} are recorded in Table~\ref{tab:limsenscoeffs}.

After rescaling {\em crpoisim\/}, the observation stack limiting sensitivity maps are resampled onto the HEALPIX grid described above, and if multiple observation stacks overlap the same location on the sky the catalog limiting sensitivity is determined as the {\em minimum\/} value from the contributing observation stacks.

\begin{deluxetable}{ccccc}
\tabletypesize{\small}
\tablecolumns{5}
\tablecaption{{\em crpoisim\/} Scale Factors\label{tab:limsenscoeffs}}
\tablehead{
\colhead{Band} & \multicolumn{2}{c}{Photon Flux} &  \multicolumn{2}{c}{Energy Flux} \\
\colhead{} & \colhead{$m$} & \colhead{$c$} & \colhead{$m$} & \colhead{$c$}
}
\startdata
Soft & $0.988$ & $-0.049$ & $1.028$ & $-8.595$ \\
Medium & $0.988$ & $-0.053$ & $0.983$ & $-8.701$ \\
Hard & $0.990$ & $-0.057$ & $0.993$ & $-8.222$ \\
Broad & $0.993$ & $-0.034$ & $0.960$ & $-8.781$ \\
Wide & $0.952$ & $-0.264$ & $0.950$ & $-8.896$ \\
\enddata
\end{deluxetable}

\subsection{Spectral Model Fits}
Similar to CSC release~1, a set of spectral model fits is performed for each individual observation detection that was obtained using the ACIS instrument and that has at least 150 net counts in the broad ($0.5$--$7.0\,\rm keV$) energy band.  Unlike the earlier catalog release, spectral model fits are also performed at the master source level for each Bayesian Block that includes one or more uniquely linked observation detections and that meets the 150 net counts threshold.  For these fits, the required net counts may be distributed across multiple observation detections, and the individual observation detections are fitted simultaneously using the same spectral model and model parameters.

As in release~1, forward fitting is used to compute the predicted counts in detector channel space arising from the spectral model, and optimization of the model parameters to match the observed counts distribution is performed using \sherpa\ \citep{2001SPIE.4477...76F}.  For more details regarding the implementation of the models in \sherpa\ for CSC~2, see \citet{mccollough_specfitspec}.

As described in \paperI, and following the simplifying assumptions therein, the model $M(E')$ that describes the expected distribution of counts in the energy bin $E'$ arriving at the detector can be written as
\begin{equation}
\label{eqn:specmodel}
M(E')=\int_{E'} dE\,R(E';E)A(E)S(E),
\end{equation}
where $S(E)$ is the physical model that defines the energy spectrum of the source, and the redistribution matrix, $R(E'; E)$, and effective area, $A(E)$, are the instrumental response functions that define the mapping between physical (source) space and detector space \citep{2001ApJ...548.1010D}.

For all sources detected in ACIS observations, the observed pulse-invariant spectra of the photons included in the source and background regions are recorded in the {\tt pha3} file in a standard format \citep[PHA file;][]{arnaud_pha} for each individual observation detection.  The appropriate associated redistribution matrix and effective area are computed by weighting the instrumental responses based on the history of how the source and background regions move over the surface of the detector due to spacecraft dither motion.  They are recorded in the {\tt rmf3} and {\tt arf3} files for the observation detection, respectively, following the standard RMF and ARF calibration file formats defined by \citet{george_rmfarf}.

Again following release~1, each PHA spectrum is grouped to a minimum of 16 counts per channel bin to fit the background-subtracted data, and the model parameters are varied to minimize the $\chi^2$ statistic with data variance, $\sigma_i^2 = N_{i,S} + (A_S/A_B)^2 N_{i,B}$, where $N_{i,S}$ and $N_{i,B}$ are the source and background counts \addedtext{extracted from the modified source region aperture and modified background region aperture (defined in \S~\ref{sec:srcaper}), respectively,} in the $i$th channel bin, and $A_S$ and $A_B$ are the geometric areas for the \addedtext{modified} source and background region \addedtext{apertures, respectively}.  Independent two-sided 68\% ($1\sigma$) confidence limits are calculated for each model parameter using a similar method to \paperI.

CSC~2 attempts to fit several distinct spectral models to the data to cover a range of X-ray emitting plasmas.  Two of the models follow from the earlier version of the catalog: an absorbed power-law model, $f(E)=e^{-N_{\rm H}\sigma_E}\,AE^{-\Gamma}$, and an absorbed black body model,  $f(E)=e^{-N_{\rm H}\sigma_E}\,A(E^2/(e^{E/{kT}}-1))$,  where $N_{\rm H}$ is the equivalent Hydrogen column density, $\sigma_E$ is the photoelectric cross-section \citep{1992ApJ...400..699B}, $A$ is the model normalization at 1\thinspace keV, $\Gamma$ is the power-law photon index, and $kT$ is the black body temperature.  The metal abundances from \citet{1989GeCoA..53..197A} are assumed for all models, unless otherwise noted.

Release 2 adds an absorbed bremsstrahlung model (XSPEC [\citealp{1996ASPC..101...17A}] spectral model {\tt xsbremms}\footnote{\url{https://heasarc.gsfc.nasa.gov/xanadu/xspec/manual/XSmodelBremss.html}}),
\begin{displaymath}
f(E)=e^{-N_{\rm H}\sigma_E}\,{A(kT)^{-1/2}} \left[g_{\rm H}(E, kT)+4g_{\rm He}(E, kT)n_{\rm He}/n_{\rm H}\right]e^{-E/{kT}},
\end{displaymath}
where $g_{\rm H}(E, kT)$ and $g_{\rm He}(E, kT)$ are the energy-dependent Gaunt factors for Hydrogen-like and Helium-like plasmas, respectively, $n_{\rm He}/n_{\rm H}$ is the Helium to Hydrogen abundance ratio (which is fixed at $0.15$ by the model), and $kT$ is the plasma temperature. The factor $(kT)^{-1/2}$ is included to make the normalization independent of temperature.  The Gaunt factors are based on the polynomial fits of \citet{1975ApJ...199..299K} to numerical values from \citet{1961ApJS....6..167K}.

Finally, CSC 2.1 adds an absorbed thermal plasma \citep[APEC;\footnote{Astrophysical Plasma Emission Code}][]{2001ApJ...556L..91S} spectral model (XSPEC model {\tt xsapec}\footnote{\url{https://heasarc.gsfc.nasa.gov/xanadu/xspec/manual/XSmodelApec.html}}).  For these fits, all model parameters ($N_{\rm H}$, $kT$, $Z$ [abundance], $z$ [redshift]) are allowed to vary during the optimization, as this was found to provide better fits than a two-parameter fit with abundance and redshift frozen at $Z=1.0$ and $z=0.0$, respectively \citep{mccollough_apecspecfitspec}.  For model fits with $\chi^2 < 2.0$, the fit parameters are well behaved.  There are a small fraction of detections and sources for which the best fit total neutral hydrogen column density converges to artificially low values, $N_{\rm H}\sim\!10^{15}\,\rm cm^{-2}$, which are unphysical as discussed in \S~\ref{sec:modelfluxes}.  A significant fraction of the best-fit absorbed APEC models with $\chi^2 < 2.0$ have negative redshifts; for the majority of these either $Z\lesssim0.05$ or $kT\gtrsim7.0\,\rm keV$\null.  In these cases, emission lines in the X-ray spectrum are either very weak or not present, and so the redshift fits are unreliable.  We nevertheless chose not to impose a parameter limit to force redshift to be positive because doing so appears to create a bias in the best fit $kT$ values. Instead, we chose to exclude from the catalog all APEC model fits for which the upper redshift confidence limit is negative ({\em i.e.\/}, where a zero redshift is excluded for a negative fitted value of $z$).

\subsection{Spectral Model Energy Fluxes} \label{sec:modelfluxes}
Again like CSC release~1, energy fluxes are estimated for all \addedtext{individual} observation detections, \addedtext{as well as for each} master source \addedtext{Bayesian Block}, for a set of  canonical spectral models.  

For a canonical source model, $S(E)$, whose integral over energy band $E'$ is $S'$, the predicted total model counts, $M'$, is given by equation~(\ref{eqn:specmodel}), where we now integrate over the entire energy band.  The total energy flux over the band, $F'$, can then be determined by scaling $S'$ by the ratio $C/M'$, {\em i.e.\/} $F'=S'\times C/M$, where $C$ represents the detection or source net counts in the energy band computed in \S~\ref{sec:aperphot}.  For HRC observations a diagonal RMF is assumed, so equation~(\ref{eqn:specmodel}) simplifies to $M(E')=\int_{E'} dE\,A(E)S(E)$\null.  Since all of the model parameters are frozen except for the normalization, spectral model energy fluxes can be calculated even when there are too few net counts to compute reliable spectral model fits.

For CSC~2, four canonical spectral models are used: absorbed power-law, absorbed black body, absorbed bremsstrahlung, and absorbed thermal plasma (APEC)\null.  The first two models were also used in release~1 of the catalog, but with different parameters.

The parameters for the canonical models used in the current catalog release were evaluated by fitting models to 3,891 CSC~1.1 PHA spectra and responses (ARF and RMF) and evaluating the parameter distributions \addedtext{for model fits with reduced $\chi^2\le 1.25$} \citep{mccollough_specfitpars}.  \addedtext{If there are multiple observations, and therefore multiple spectral model fits, of the same X-ray source then all model fits with reduced $\chi^2\le 1.25$ are considered when evaluating the model parameter distributions to define the canonical parameter values.}  For the absorbed power-law and absorbed black body model fits with reduced $\chi^2\le 1.25$ \addedtext{($N=2349$)}, the distributions of best-fit photon index (for the power-law) and temperature (for the black body) are approximately Gaussian distributed, with median values of $\Gamma=2.02$ and $kT=0.72\,{\rm keV}$, respectively.  We therefore {\em define\/} canonical parameter values $\Gamma=2.0$ and $kT=0.75\,{\rm keV}$, respectively, for these two spectral models.

The median value of the best-fit temperature for the absorbed bremsstrahlung model is $kT=4.03\,{\rm keV}$ \addedtext{($N=2076$)}.  While the distribution of the best-fit temperature is also approximately Gaussian distributed, there is an excess of model fits with $kT=100\,{\rm keV}$ (even for fits with reduced $\chi^2\le 1.25$), which is the maximum allowed model parameter value.  This excess skews the median value to a higher temperature.  We therefore define a slightly lower canonical bremsstrahlung model temperature, $kT=3.5\,{\rm keV}$, that is consistent with the mode of the distribution {\em excluding\/} the $kT=100\,{\rm keV}$ excess.

The canonical APEC model has abundances fixed to be solar and redshift fixed to be zero.  With these constraints, the distribution of the best-fit model plasma temperature is broadly peaked around a median value $kT=6.07\,{\rm keV}$ \addedtext{($N=1303$)}, with both a pronounced lower-temperature tail and an excess at the maximum allowed model parameter value.  The mode of the distribution is located at a somewhat higher temperature than the median, and so we define a slightly higher canonical APEC model plasma temperature, $kT=6.5\,{\rm keV}$.

For all of the canonical spectral models, we define the absorption model component total neutral hydrogen column density to be fixed with $N_{\rm H} = N_{\rm H}(\hbox{\rm Gal})$, the measured Galactic absorption column density from \citet{1990ARAA..28..215D}.  

For the black body canonical spectral model this is a change from CSC release~1, where a lower fixed absorption, $N_{\rm H}=3\times10^{20}\,\rm cm^{-2}$, was used.  However, based on review of a subset of CSC~1.1 sources for which black body spectral fits were performed with reduced $\chi^2\le1.25$ ($N=755$), there is no support for such a low value.  Instead, there is a broad peak in the spectral fits at $N_{\rm H} \sim 7\times10^{21}\,\rm cm^{-2}$ and another, much smaller peak at $N_{\rm H} \sim 10^{15}\,\rm cm^{-2}$.  The latter value is unrealistic given that the expected lower bound to the column density is around $N_{\rm H} \sim 10^{19}\,\rm cm^{-2}$, as determined by the density of the ISM inside the local bubble \citep[{\em e.g.\/},][]{1987ARA&A..25..303C,1995A&A...294L..25E}. \citet{2010ApJ...725.1805B} has shown that these low $N_{\rm H}$ values are likely a result of fitting spectra with a simple model, while the true spectra are more complex and likely require multiple spectral components to describe the X-ray data. This leads to the $N_{\rm H}$ values being artificially low. 

\subsection{Spectral Hardness Ratios}
Spectral hardness ratios computed between the hard, medium, and soft energy bands are reported for all ACIS \addedtext{individual }observation detections\addedtext{, stacked observation detections,} and \addedtext{master} sources included in the catalog.  In CSC~2, the hardness ratio between two energy bands $x$ and $y$ is defined as $H_{xy} = {(F_x - F_y)/(F_x + F_y)}$, where $F_x$ and $F_y$ are the PSF 90\% ECF aperture photon fluxes in the bands $x$ and $y$, respectively, and we have chosen the convention where $x$ is always the higher energy band.  Note that this definition is {\em different\/} from the definition used in release~1 of the catalog, where the hardness ratio was defined as $H_{xy} = (F_x - F_y)/F_b$, where $F_b$ was the flux in the broad energy band.

While CSC release~1 used a Bayesian approach based on \citet{2006ApJ...652..610P} to compute hardness ratios from the total and background counts in each energy band, CSC~2 uses an alternative approach \citep{nowak_varhard} that makes use of the photon flux aperture photometry MPDFs, $P_x(F_x)$ for energy band $x$, computed in \S~\ref{sec:aperphot}.  This has the added advantage of ensuring consistency between the hardness ratios and the band photon fluxes, which was not always the case in the earlier catalog release.

Note that the hardness ratios reported at the per-observation detection, stacked observation detection, and master source level are determined identically, although different MPDFs are used in the computations.  At the per-observation detection level, the individual observation aperture photometry MPDFs are employed, whereas at the stacked observation detection and master source levels the combined aperture photometry MPDFs for the observations that comprise observation stack or Bayesian block are used, respectively.

$H_{xy}$ is defined by the relationships $F_x\equiv (1+H_{xy})F_{xy}/2$ and $F_y\equiv (1-H_{xy})F_{xy}/2$, where $F_{xy}$ is the sum of $F_x$ and $F_y$\null.  With this definition, 
\begin{displaymath}
P_{x,y}(F_x, F_y)\, dF_xd F_y = P_x(F_x)P_y(F_y){F_{xy}\over 2}\, dH_{xy} dF_{xy}.
\end{displaymath}
$P_{x,y}(F_x, F_y)$ is the two-dimensional probability distribution for the photon fluxes in the individual energy bands.  Since $P_x(F_x)$ and $P_y(F_y)$ are assumed to be independent in the aperture photometry calculations, $P_{x,y}(F_x, F_y)=P_x(F_x)P_y(F_y)$.  Although this requires the use of non-informative priors on the hardness ratio, informative priors on each of the individual flux bands could still be used.  

The joint distribution of the fluxes is now used to calculate the hardness ratio distribution. Specifically, the probability distribution for the hardness ratio becomes

\begin{eqnarray*}
P_{H_{xy}}(H_{xy})\,dH_{xy} &=& \int_{F_{xy}=0}^\infty dH_{xy} dF_{xy}\, P_x(F_x)P_y(F_y){F_{xy}\over 2} \\
&=& \int_{F_{xy}=0}^\infty dH_{xy} dF_{xy}\, P_x \left({(1+H_{xy})F_xy\over 2}\right) P_y \left({(1-H_{xy})F_xy\over 2}\right){F_{xy}\over 2}.
\end{eqnarray*}

The hardness ratio reported in the catalog is the value $H_{xy}$ that maximizes $P_{H_{xy}}(H_{xy})$.  The independent lower and upper confidence limits on $H_{xy}$ are determined by defining an amplitude cut on the probability distribution, $P^{cut}$, such that
\begin{displaymath}
\int_{P_{H_{xy}}\ge P^{cut}} dH_{xy}\, P_{H_{xy}}(H_{xy}) = 0.68.
\end{displaymath}
Since the values of $H_{xy}$ can run from $-1$ to $+1$, the lower confidence limit on $H_{xy}$ is the lowest value $\geq -1$ that
is included in the integration boundaries, and the upper confidence limit on $H_{xy}$ is the highest value $\leq +1$ that
is similarly included. 

\subsection{Estimating Source Variability}
\subsubsection{Intra-Observation Variability} \label{sec:intravar}
Temporal variability of a detection within a single observation is evaluated in the same way as in CSC release~1, as described in detail in \S~3.12.1 of \paperI\null.  The probability that a source is variable is estimated separately in each energy band using the commonly-used Kolmogorov-Smirnov \citep[K-S;][]{1951JASA...46...68M} and Kuiper \citep{kuiper1960tests} tests, as well as the Gregory-Loredo algorithm \citep{1992ApJ...398..146G}.

All three algorithms directly use the photon event arrival times to evaluate the variability probabilities.  Conceptually, a constant detection flux should be represented by a cumulative sum of photon events that increases linearly with time.   In practice, this null hypothesis must be modified to account for data gaps and also for variations of the effective areas of the \addedtext{modified} source and background region apertures due to the spacecraft dither motion during the observation.  These corrections ensure a detection that dithers across the edge of the detector or over a bad detector region is not erroneously classified as variable. 

The Gregory-Loredo test uses a Bayesian approach to detect variability, is entirely suitable for use with photon event data, and can work effectively in the presence of data gaps.  As described in \paperI, we have incorporated the capability to include temporal variation in the effective area.  Optimal resolution light curves generated by the Gregory-Loredo test for the detection and background, together with uncertainty and $3\sigma$ lower and upper confidence limits, are recorded in the CSC 2 light curve ({\tt lc3}) data product for the observation.  The power spectra of the light curves are evaluated for the presence of the fundamental spacecraft pitch and yaw dither frequencies or associated beat frequencies. If there is a peak in the power spectrum at one of these frequencies that is at least $5\times$ the RMS value, then a warning flag ({\tt dither\_warning\_flag}) is set in the catalog for the detection to indicate that the intra-observation variability properties are unreliable.

For each energy band, we combine the Gregory-Loredo probabilities, the odds ratio of obtaining the observed distribution of light curve time bins versus obtaining a flat distribution ({\em i.e.\/}, a non-variable detection), and the fractions of the light curve that fall within $3\sigma$ and $5\sigma$ of the average rate, to construct an intra-observation variability index defined in \paperI\null.  The variability index is a more robust measure of intra-observation variability, and is less sensitive to the number of net source counts, than variability probability.

At the stacked observation and master source levels, the intra-observation variability probabilities and indices for each energy band are the highest values across the set of individual observations that are linked to the stacked observation detection or master source, respectively.

\subsubsection{Inter-Observation Variability} \label{sec:intervar}
While the catalog intra-observation variability measures consider temporal variability {\em within\/} a single observation, inter-observation variability metrics evaluate the degree to which aperture photometry measurements from multiple individual observations of a single source are inconsistent with a constant flux.  In CSC 2, these metrics are based on analysis of the aperture photometry MPDFs computed in \S~\ref{sec:aperphot}\null.  At the stacked observation level, the inter-observation variability properties for a detection are computed from the set of observations included in the valid stack ({\em i.e.\/}, those for which the detection falls within the pixel mask for the observation).  Master source level inter-observation variability properties for a source are computed from the individual observation detections included in the valid stack of all stacked observation detections that have either unique or non-detection linkage to the master source, with the exception that individual ACIS observation detections for which the estimated photon pile-up fraction exceeds $\sim\!10\%$ are not considered unless {\em all\/} such detections exceed this threshold.  Non-detections are considered to be photometric upper limits at the epochs of the corresponding observations when evaluating inter-observation variability.

For a set of individual observation detections, the means, $\overline{F_x^i}$, and variances, ${\sigma_x^i}^2$, for each MPDF are
\begin{displaymath}
\overline{F_x^i} = \int\! dF_x^i\,F_x^iP_x^i(F_x^i); \quad
{\sigma_x^i}^2 = \int\! dF_x^i\,{(\overline{F_x^i}-F_x^i)}^2 P_x^i(F_x^i),
\end{displaymath}
where $P_x^i(F_x^i)$ is the MPDF of the PSF 90\% ECF aperture photon flux, $F_x^i$, in energy band $x$ for observation~$i$.

The inter-observation variability probability for energy band $x$ is evaluated using a likelihood ratio test by comparing the maximum log likelihood computed using the observed aperture photometry MPDFs with the maximum log likelihood computed with a constant aperture photometry flux.  For a set of $N$ observations, the former is computed by finding the set of fitted rates ${F_x^i}^{{\rm max}}$, that maximize
\begin{displaymath}
\log {{L^i}^{{\rm max}}} = \sum_i \log {P_x^i({F_x^i}^{{\rm max}})}.
\end{displaymath}
Similarly, the constant rate ${F_x}^{{\rm max}}$ that maximizes the log likelihood function is
\begin{displaymath}
\log {{L}^{{\rm max}}} = \sum_i \log {P_x^i({F_x}^{{\rm max}})}.
\end{displaymath}
In practice, the computation of the rates that maximize the log likelihoods are performed by minimizing the negative log likelihoods using Akima interpolations \citep{1970JACM...17..589A} of the logarithms of the MPDFs.  This procedure was found to yield better behavior in the wings of the probability distribution compared to other approaches such as spline fits to the distribution \citep{nowak_varhard}.

In the absence of inter-observation variability, the quantity
\begin{equation}
\label{eqn:intervarD}
D\equiv 2(\log {{L^i}^{{\rm max}}} - \log {{L}^{{\rm max}}})
\end{equation}
will be distributed as $\chi^2$ with $N-1$ degrees of freedom.  The probability, $p$, that the inter-observation variability is likely to be real is computed from the cumulative distribution for the $\chi^2$ statistic as
\begin{displaymath}
p={{\gammafunc{\left({{N-1}\over{2}}, {{D}\over{2}}\right)}}/{\Gammafunc{\left({{N-1}\over{2}}\right)}}},
\end{displaymath}
where $\gamma$ is the lower incomplete Gamma function and $\Gamma$ is the complete Gamma function.

An estimate of the highest level of variability is provided by computing the standard deviation of the inter-observation photon fluxes in each energy band as $F_x^\sigma = \max{(|{{F_x}^{{\rm max}}-{F_x^i}^{{\rm max}}}|/\sigma_x^i})$.

Finally, as in the previous release of the catalog, an inter-observation variability index is computed for each energy band, using the same definition as Table~7 of \paperI, with the exception that $D/(N-1)$ from equation~{\ref{eqn:intervarD}} is substituted for the reduced $\chi^2$ when computing the variability index.

\subsection{Source and Detection Codes and Flags} \label{sec:flags}
Similar to release~1, detections listed in the Per-Observation Detections and Stacked Observation Detections tables include sets of  {\em codes\/} and {\em flags\/} that identify specific circumstances for the detection in a form that is easily queryable by the catalog user.  Sources included in the Master Sources table include associated flags, but not codes.  The difference between the two is that the flags are Boolean quantities and indicate that the detection or source does or does not satisfy a specific condition, whereas codes are bit-encoded data values that provide an efficient way of recording multiple closely-related boolean quantities in a single datum.  

While some of the codes and flags are intended to provide a quick way to determine whether a detection or source satisfies a common query criteria, such as ``does the detection show intra-observation variability?'' or ``is the source extended?'', the majority warn the user of conditions that may impact the quality of the source or detection properties ({\em e.g.\/},, the detection is saturated), or that may limit the usefulness of the source or detection for some types of investigations ({\em e.g.\/}, the detection position was manually adjusted).  Source and detection flags, and detection codes and their associated bit encodings, are defined in Tables \ref{tab:masterproperties}--\ref{tab:perobsproperties}.

Multiple bits may be set simultaneously in a bit-encoded value, so the recorded integer equivalent value will be the sum of the appropriate values listed in Tables \ref{tab:stackobsproperties}--\ref{tab:perobsproperties}.  For example, if a detection is not consistent with a point source in the ACIS medium, hard, and broad energy bands then $\hbox{\tt extent\_code} = 4 + 8 + 16 = 28$ because the bits corresponding to integer values 4, 8, and 16 indicate \addedtext{extent} in the medium, hard, and broad bands, respectively.

Interpretation of the codes and flags, including the extent and variability codes and flags, confusion, saturated source and streak source flags, are discussed in \paperI\null.  Note that the interpretation of a flag or code is inherently qualified by the catalog table in which the datum appears.  For example, in the Per-Observation Detections table, a code or flag refers to the individual observation detection, whereas in the Stacked Observation Detections Table, the code or flag is a combination of the corresponding individual observation detection codes or flags for the set of observations that comprise the observation stack; whether the combination is performed as an {\em and\/} ({\em i.e.\/}, all) of the individual observation detection values or an {\em or\/} ({\em i.e.\/}, any) of the individual observation detection values is defined in Table~\ref{tab:stackobsproperties}.  Flags appearing in the Master Sources table are in general similarly comprised of combinations of the per-observation detection flags for the individual observations included in the longest flux-ordered Bayesian Block.  The exception is the variability flag, which is derived from {\em all\/} contributing observations irrespective of the Bayesian Block make-up.

Note that the bit encodings for three codes ({\tt conf\_code}, {\tt edge\_code}, and {\tt multi\_chip\_code}) have changed since release~1.  This was done so circumstances that may more highly impact data quality translate to higher numerical values for these codes.  For example, for {\tt edge\_code} a value of 4 (previously 1) now indicates that the detection position dithers off the detector boundary since that is more likely to impact to the data quality than just having some part of the background region dither off the detector boundary, which is now indicated by a value of 1 (previously 4)\null.

For CSC 2.0, the catalog variability flag, {\tt var\_flag},  in the Master Sources and Stacked Observation Detections tables, and the variability code, {\tt var\_code}, in the Per-Observation Detections table are defined similarly to release~1, {\em i.e.\/}, the flag or appropriate bit in the code is set only if the relevant {\em intra\/}-observation variability index is $\geq 3$.  Because of the significantly increased interest in time-domain astrophysics in recent years, in release 2.1 we have modified the definitions of {\tt var\_flag} in the Master Sources and Stacked Observation Detections tables to consider both {\em intra\/}- and {\em inter\/}-observation variability and have revised the variability index threshold for setting the flag from $\geq 3$ to $\geq 6$.  The latter change is also made to the definition of {\tt var\_code} in the Per-Observation Detections table.   These modifications change the meaning of the variability flag and code from ``the source (or detection) {\em may\/} be variable'' to ``the sources (or detection) {\em is definitely\/} variable'' and alters the focus from identifying variable sources that may need to be excluded from a study to identifying variable sources that are of interest for a study.

The manual intervention flags (4 at the stacked observation detections level and 5 at the master sources level) are set to true when a detection or source was subject to human quality assurance review {\em and\/} some parameters were changed manually as part of that review.  If there were no changes from the values determined by the automated catalog processing pipelines, then the manual intervention flags are not set.  Note in particular the difference between the {\tt man\_add\_flag} and the {\tt man\_inc\_flag}.  The former will be set if a source or detection is manually added to the catalog ({\em i.e.\/}, the source was not detected by the automated source detection process) {\em and\/} the source or detection meets all of the automated catalog inclusion criteria.  The {\tt man\_inc\_flag} is set if a source or detection is included in the catalog and {\em does not\/} meet all of the automated catalog inclusion criteria, irrespective of whether the source or detection was added manually or not.

\subsection{Quality Assurance}
A series of automated and human quality assurance review steps that follow the precepts of \S~3.14 of \paperI\ are used to guarantee the scientific integrity of the catalog, and all of the steps described in \paperI\ are performed similarly for this release.  Steps where human quality assurance review is or may be performed are identified elsewhere in the text.  \addedtext{Where significant scientific judgement may required to asses a dataset as part of a human review, the dataset is first sent to a primary reviewer for evaluation and any proposed changes are subsequently vetted by a secondary reviewer for concurrence.  This ``two-reviewer'' approach helps ensure both reliability and consistency between reviews as part of the overall quality assurance process.}

%This helps with spacing above the section header, and also avoids the subsequent subsection header being at the bottom of the page
\bigskip

\section{Preliminary Characterization}
Detailed statistical characterization of CSC source and detection properties is beyond the scope of this paper.\footnote{Additional characterization will be provided on the CSC website, see \url{https://cxc.cfa.harvard.edu/csc/char.html}.}  Here we limit the discussion to a preliminary evaluation of the CSC 2.0 false detection rate, completeness, and astrometric accuracy, which are provided to enable the user to better assess the reliability of the catalog.
\begin{deluxetable}{lccc}
\tabletypesize{\small}
\tablecolumns{4}
\tablecaption{False Detection Rates\label{tab:fssim}}
\tablehead{
\colhead{Stack} & \colhead{$T_{\rm stack}$} & \colhead{\truevalue\ FDR} & \colhead{\marginalvalue\ FDR}
}
\startdata
{\tt acisfJ0020335p283927\_001} & \phn\phn9.0 & 0.01 & 0.19 \\
{\tt acisfJ0152458p360906\_001} & 110.6 & 0.17 & 0.67 \\
{\tt acisfJ0025384m122430\_001} &132.7 & 0.16 & 0.70 \\
{\tt acisfJ0259013p133237\_001} & 135.1 & 0.01 & 0.31 \\
{\tt acisfJ0102415m491757\_001} & 291.7 & 0.05& 0.46 \\
{\tt acisfJ0839591p294814\_001} & 292.0 & 0.19 & 1.44 \\
{\tt acisfJ0839591p294814\_001}\tablenotemark{\scriptsize a} & 292.0 & 0.11 & 0.45 \\
\enddata
\tablenotetext{a}{Excluding false detections recorded on ACIS chip S4 (ACIS-8).}
\end{deluxetable}

\subsection{False Detection Rate}
False detection rates are estimated for observation stacks with exposure times ranging from $\sim\!10$--$300\,\rm ks$ by replacing the actual observation event lists blank-sky event lists derived from the background maps for the observation, randomized with Poisson noise.  Typically, between 150 and 200 realizations of the same simulation are  evaluated.  The \truevalue\ and \marginalvalue\ thresholds are set to achieve false detection rates of approximately 0.1 and 1 false detection per observation stack, respectively, for point-source detections brighter than the threshold (see \S~\ref{sec:lthresh}).  False detection rates for the simulated observation stacks in Table~\ref{tab:fssim} are in general consistent with desired rates, with the exception of stack {\tt acisfJ0839591p294814\_001}, which is a 4-observation stack that includes ACIS chip S4 (ACIS-8)\null.  As discussed in \paperI, this chip is affected by a pattern of linear streaks caused by a flaw in the serial readout \citep{houck_destreak}.  This results in an excess of detections in the vicinity of bad columns on chip S4 \citep{2011ApJS..194...37P}, even though the catalog pipelines so apply a correction for the presence of these ``streak'' events.  If the detections on chip S4 are excluded, the false detection rates for this simulated stack agree with the other simulations.

\subsection{Completeness}
\begin{figure}
\epsscale{1.0}
\plotone{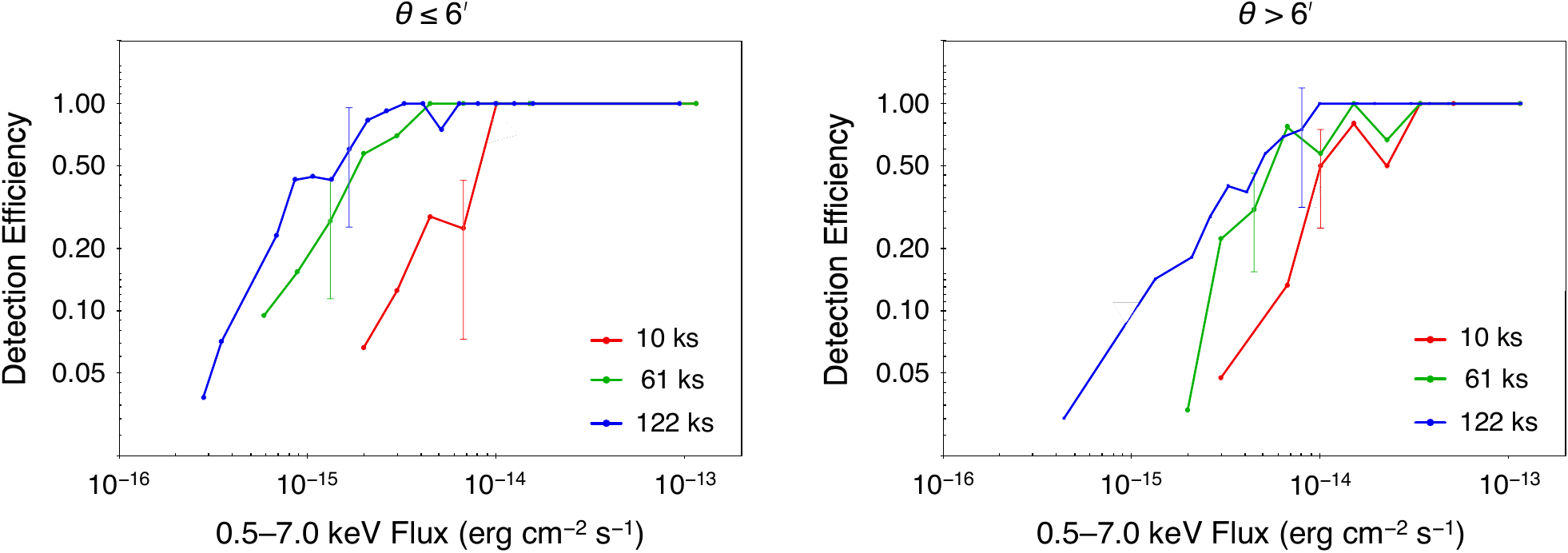}
\caption{\label{fig:deteff}
{\em Left:\/} Source detection efficiency estimated from ACIS observations with exposure times $\sim\!10$ (red), $61$ (green), and $122\,ks$ (blue) as a function of broad band energy flux, for detections with off-axis angles $\theta\le6'$.  Typical error bars are indicated for each curve. {\em Right:\/} Same, except for detections with off-axis angles $\theta>6'$.}
\end{figure}

Completeness of CSC 2.0 is evaluated by counting the number of sources detected in individual observations of the \Chandra\ Deep Field-South Survey (CDFS) and comparing those counts with the number of sources reported in the catalog of \citet{2017ApJS..228....2L}\null. The latter are derived from an analysis of stacked ACIS imaging observations totaling $\sim\!7\,Ms$, and so can be considered complete at the exposures of individual observations.  For CSC 2.0, three individual observations with exposure times $\sim\!10$, $61$, and $122\,ks$ (\obsid s 12047, 12054, and 17535, respectively) are considered, and a histogram of broad band fluxes is created for detections that would be classified as \marginalvalue\ or \truevalue\ using the catalog likelihood thresholds.  A similar histogram is constructed from the \citet{2017ApJS..228....2L} catalog using their ``full'' band fluxes ($0.5$--$7.0\,keV$, identical to the CSC broad band) for sources that fall within the fields-of-view of the 3 observations, and the ratio of the two distributions provide estimates of CSC 2.0 source detection efficiency (Figure~\ref{fig:deteff}).  Although the exposure time of the longest observation considered is short compared to the exposure times of longest duration observation stacks, we note nevertheless that $\sim\!95\%$ of CSC 2.0 observation stacks have exposure times $<100\,\rm ks$.

\subsection{Astrometric Accuracy}
\begin{figure}
\epsscale{0.75}
\plotone{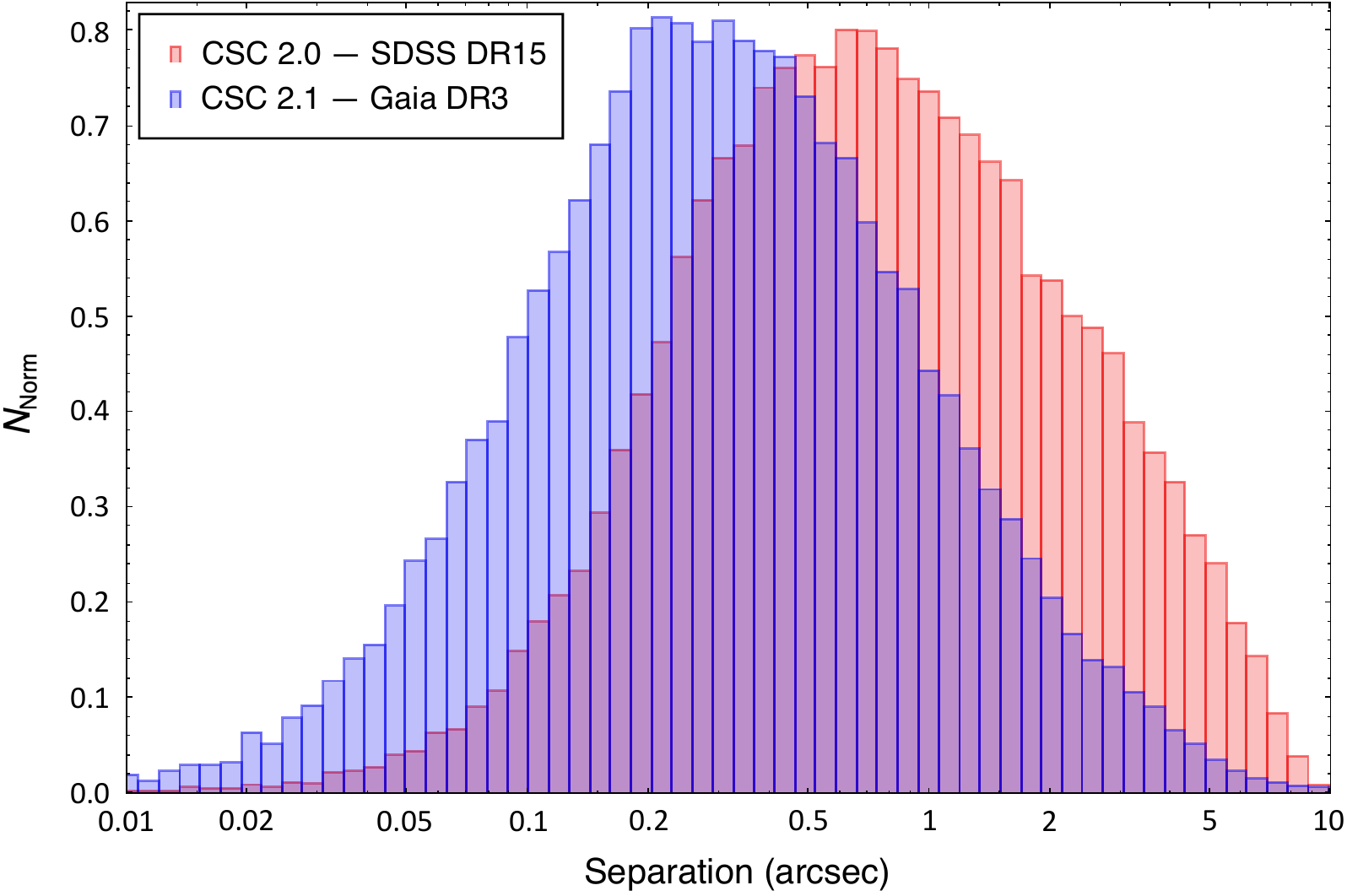}
\caption{\label{fig:CSC21_20astrom}
Normalized histograms of the astrometric angular separations between reference catalogs (SDSS DR15 for CSC 2.0 and Gaia DR3 for CSC 2.1) and cross-matched CSC master source positions for CSC 2.0 (red) and CSC 2.1 (blue), showing the overall improvement from matching release 2.1 astrometry to the Gaia-CRF3 reference frame. }
\end{figure}

The absolute astrometric accuracies of CSC 2.0 and CSC 2.1 were evaluated by cross-matching CSC master source positions with their optical stellar sources from SDSS DR15 \citep{2017AJ....154...28B}  for CSC2.0 and Gaia DR3 \citep{2023A&A...674A...1G} for CSC 2.1, using a technique similar to \citet{2011ApJS..192....8R}.  The measured angular separations between CSC master source positions and matched optical stellar sources are presented in Figure~\ref{fig:CSC21_20astrom}\null.  For separations larger than \addedtext{$\sim\!1\farcs0$--$2\farcs0$} the histogram fractions become increasingly overestimated due to an increased percentage of poor matches, typically due to invalid matches between X-ray and optical sources or due to matches with off-axis CSC detections that have large and asymmetric PSFs\null.   Because of the difficulty in unambiguously excluding poor matches {\em a priori\/} we choose not to remove them to avoid biasing the distribution.  \addedtext{For both catalog versions, the  master source positions are determined by combining stack detection positions as described in \S~\ref{sec:masterpos}\null.  For CSC 2.0, the stack detection positions are computed using the individual observation aspect solutions after applying the relative astrometric corrections in \S~\ref{sec:ofa}\null.  However, no absolute astrometric corrections are applied to the CSC 2.0 stack detection positions.  For CSC 2.1, the stack detection positions are corrected to the Gaia-CRF3 astrometric reference frame as described in \S~\ref{sec:absastrom}\null.}  Correcting the CSC astrometry to Gaia-CRF3 in CSC 2.1 reduces the median radial astrometric error from $0\farcs72$ to $0\farcs30$.

\section{CONCLUSIONS}
The \Chandra\ Source Catalog is a virtual X-ray astrophysics facility that provides to the astronomical community carefully-curated, high-quality, and uniformly calibrated and analyzed properties for X-ray sources detected in ACIS and HRC-I imaging observations obtained by the \Chandra\ X-ray Observatory.  Besides including more recent observations obtained since CSC release 1 was completed, the Release 2 Series improve upon the original catalog versions in several ways.  New background and source detection algorithms enable reliable detection of compact sources down to roughly 5 net counts over much of the field-of-view, yielding a significantly fainter catalog limiting sensitivity that was possible for release 1\null.  The new algorithms also allow the detection of bright, extended X-ray emission regions that were deliberately excluded from the earlier release.  Starting with release 2.1, catalog positions are corrected to the Gaia-CRF3 astrometric reference frame.

X-ray source and detection properties recorded in the catalog tables include position and position error, spatial extent compared to the local PSF, multi-band aperture photometry, hardness ratios, spectral fits using multiple models, and intra- and inter-observation temporal variability measures, typically with independent lower and upper confidence limits, extracted in 5 energy bands for ACIS observations and a single energy band for HRC data.  The catalog tables record the extracted properties for each source detected in the stacked observations, as well as separately for each observation, so that any temporal changes between observations are available for study.  Best estimates of properties derived from multiple detections of the same X-ray source on the sky are provided at the master source level.  The catalog includes 37 different types of source, detection, and field-based FITS format science-ready X-ray data products that can be used directly as a starting point for further analysis.

Release 2.0 includes 317,167 unique X-ray sources detected in ACIS and HRC-I imaging observations covering $559\,{\rm deg}^2$ of sky that were released publicly prior to the end of 2014, while release 2.1 includes $\sim\!407,806$ sources from observations released publicly prior to the end of 2021 with $730\,{\rm deg}^2$ sky coverage.

\begin{acknowledgments}
The authors wish to thank the many former and current \Chandra\ X-ray Center staff who have contributed to the development and production operations necessary to complete the \Chandra\ Source Catalog Release 2 Series.  In particular, Stephen Doe, Dan Nguyen, and Roger Hain contributed extensively to the Maximum Likelihood Estimator code.  Christopher Allen implemented the Bayesian X-ray aperture photometry algorithms.  Extensive additional software development support was provided by Jamie Budynkiewicz, Judy Chen, Danny Gibbs~II, Kenny Glotfelty, Diane Hall, Peter Harbo, Omar Laurino, Warren McLaughlin, Charles Paxson, and Panagoula Zografou.  Craig Anderson, Douglas Morgan, Amy Mossman, Joy Nichols, and Beth Sundheim all supported catalog human quality assurance reviews. Padmanabhan Ramadurai and the CXC Systems group provided extensive installation and operational support for the catalog high-performance cluster, and David W. Van Stone provided database administration services for the catalog databases. The authors acknowledge the support and guidance of the current and former \Chandra\ X-ray Center directors, Patrick Slane and Belinda Wilkes.  \addedtext{The authors further wish to thank the anonymous referee for a careful review of the manuscript, in addition to several recommendations and suggestions that improved the quality and readability of the paper.}

This paper makes use of data from the Sloan Digital Sky Survey (SDSS)\null.  Funding for the Sloan Digital Sky Survey has been provided by the Alfred P.\ Sloan Foundation, the Participating Institutions, the National Aeronautics and Space Administration, the National Science Foundation, the U.S.\ Department of Energy, the Japanese Monbukagakusho, and the Max Planck Society. The SDSS Web site is \url{http://www.sdss.org/}.  The SDSS is managed by the Astrophysical Research Consortium for the Participating Institutions. The Participating Institutions are The University of Chicago, Fermilab, the Institute for Advanced Study, the Japan Participation Group, The Johns Hopkins University, Los Alamos National Laboratory, the Max-Planck-Institute for Astronomy, the Max-Planck-Institute for Astrophysics, New Mexico State University, University of Pittsburgh, Princeton University, the United States Naval Observatory, and the University of Washington.

This publication makes use of data products from the Wide-field Infrared Survey Explorer, which is a joint project of the University of California, Los Angeles, and the Jet Propulsion Laboratory/California Institute of Technology, and NEOWISE, which is a project of the Jet Propulsion Laboratory/California Institute of Technology. WISE and NEOWISE are funded by the National Aeronautics and Space Administration.

This work has made use of data from the European Space Agency (ESA) mission {\em Gaia\/} (\url{https://www.cosmos.esa.int/gaia}), processed by the {\em Gaia\/} Data Processing and Analysis Consortium (DPAC, \url{https://www.cosmos.esa.int/web/gaia/dpac/consortium}). Funding for the DPAC has been provided by national institutions, in particular the institutions participating in the {\em Gaia\/} Multilateral Agreement.

Support for development of the \Chandra\ Source Catalog is provided by the National Aeronautics and Space Administration through the \Chandra\ X-ray Center, which is operated by the Smithsonian Astrophysical Observatory for and on behalf of the National Aeronautics and Space Administration under contract NAS\thinspace 8-03060.
\end{acknowledgments}

\vspace{5mm}
\facilities{CXO}
\software{CalDB4 \citep{2007ASPC..376..335E},
      CIAO \citep{2006SPIE.6270E..1VF}, 
      CXCDS \citep{2006SPIE.6270E..0NE}, 
      DS9 \citep{2003ASPC..295..489J},
      DTFE \citep{2011arXiv1105.0370C},
      HEALPIX \citep{2005ApJ...622..759G},
      MARX \citep{2012SPIE.8443E..1AD},
      PyMC3 \citep{2016PJCS....2..E55S},
      SAOTrace \citep{1995ASPC...77..357J},
      Sherpa \citep{2001SPIE.4477...76F},
      TOPCAT \citep{2005ASPC..347...29T},
      TreeCluster \citep{2004Bioinformatics.20.1453D},
      Triangle \citep{shewchuk1996triangle},
      XSPEC \citep{1996ASPC..101...17A}
      }

\bibliography{CSC_Paper2}{}
\bibliographystyle{aasjournal}

\end{document}